\documentclass{JINST}

\title{Evaluation of atomic electron binding energies for Monte Carlo particle transport}

\author{Maria Grazia Pia$^a$\thanks{Corresponding author.}, 
Hee Seo$^b$, 
Matej Bati\v{c}$^a$$^c$, 
Marcia Begalli$^d$, 
Chan Hyeong Kim$^b$, 
Lina Quintieri$^e$ and 
Paolo Saracco$^a$\\
\llap{$^a$}INFN Sezione di Genova,\\
  Via Dodecaneso 33, 16146 Genova, Italy\\
\llap{$^b$}Hanyang University, \\
17 Haengdang-dong, Seongdong-gu, Seoul, 133-791, Korea,\\
\llap{$^c$}Jo\v{z}ef Stefan Institute, \\
Jamova 39, 1000 Ljubljana,  Slovenia ,\\
\llap{$^d$}State University of Rio de Janeiro, \\
R. Sao Francisco Xavier, 524, 20550-013, Rio de Janeiro, RJ, Brazil\\
\llap{$^e$}INFN Laboratori Nazionali di Frascati, \\
Via Enrico Fermi 40, I-00044 Frascati (Rome), Italy\\
  E-mail: \email{MariaGrazia.Pia@ge.infn.it}}

\abstract{A survey of atomic binding energies used by general purpose Monte Carlo systems
is reported.
Various compilations of these parameters have been evaluated; their
accuracy is estimated with respect to experimental data.
Their effects on physics quantities relevant to Monte Carlo particle transport
are highlighted: X-ray fluorescence emission, electron and proton ionization
cross sections, and Doppler broadening in Compton scattering.
The effects due to different binding energies are quantified with respect to
experimental data.
The results of the analysis provide quantitative ground for the selection of
binding energies to optimize the accuracy of Monte Carlo simulation in
experimental use cases.
Recommendations on software design dealing with these parameters and on the
improvement of data libraries for Monte Carlo simulation are discussed.}


\usepackage{cite}
\usepackage{graphicx}
\usepackage{subfigure}
\usepackage{amsmath} 
\usepackage{multirow}

\begin{document}


\section{Introduction}
\label{sec_intro}
The
simulation of particle interactions in matter involves a number of atomic
physics parameters, whose values affect physics models applied to
particle transport and experimental observables calculated by the simulation.
Despite the fundamental character of these parameters, a consensus
has not always been achieved about their values, and different Monte Carlo codes use
different sets of parameters.

Atomic parameters are especially relevant to simulation scenarios that are
sensitive to detailed modeling of the properties of the interacting medium.
Examples include the generation of characteristic lines resulting from X-ray
fluorescence or Auger electron emission, and precision simulation studies, such
as microdosimetry, that involve the description of particle interactions with
matter down to energies comparable with the scale of atomic binding energies.

Simulation in these domains has been for an extended time the object of
specialized Monte Carlo codes; some general purpose Monte Carlo systems have
devoted attention to these areas, introducing functionality for the simulation
of fluorescence, PIXE (Particle Induced X-ray Emission) and microdosimetry.
In this context, emphasis has been placed on the development and validation of
the physics models implemented in the simulation systems, while relatively
limited effort has been invested into verifying the adequacy of the atomic
parameters used by general purpose Monte Carlo codes with regard to the requirements of new
application domains.

This paper surveys atomic binding energies 
used by well known Monte Carlo systems, including EGS \cite{egs5}, EGSnrc
\cite{egsnrc}, Geant4 \cite{g4nim,g4tns}, ITS (Integrated Tiger Series)
\cite{its5}, MCNP/MCNPX \cite{mcnp,mcnpx} and Penelope \cite{penelope}, and 
by some specialized physics codes.
These software systems use a variety of compilations of binding energies,
which are derived from experimental data or theoretical calculations; this paper
investigates their accuracy and their effects on simulations.

To date, to the best of our knowledge, no comprehensive review  
of the binding energies used by major Monte Carlo codes has been
documented in the literature yet.
The most recent review of atomic binding energies is included in the Handbook of
X-Ray Data \cite{zschornack}: it is limited to comparing two compilations of
binding energies \cite{bearden,siegbahn}, derived from experimental
data, with respect to a common reference \cite{sevier1979}; the comparison rests
on the visual appraisal of plots.
Out of the three compilations of binding energies examined, only two are used by major
Monte Carlo systems. 
We are not aware either of previously published objective estimates, based on
statistical methods, of the compatibility with experiment of the binding
energies used in Monte Carlo simulation, and of simulation observables that
depend on them.

A small subset of preliminary results of this study were summarized in a conference paper
\cite{binding_nss2010}.

%

\section{Compilations of electron binding energies}

The binding energies considered in this study concern neutral atoms in their
ground state; several compilations of their values, of experimental and
theoretical origin, are available in the literature.

Compilations based on experimental data are the result of the application of
selection, evaluation, manipulations (like interpolation and extrapolation) and
semi-empirical criteria to available experimental measurements to produce a set
of reference values covering the whole periodic system of the elements and the
complete atomic structure of each element.

Most of the collections of electron binding energies based on experimental data
derive from a review published by Bearden and Burr in 1967 \cite{bearden}.
Later compilations introduced further refinements in the evaluation of
experimental data and the calculation of binding energies for which no
measurements were available; they also accounted for new data taken after the
publication of Bearden and Burr's review.

Experimental atomic binding energies can be affected by various sources of
systematic effects; they originate not only from the use of different
experimental techniques in the measurements, but also from physical effects: for
instance, binding energies of elements in the solid state are different from
those of free atoms, and binding energy measurements can be affected by the 
chemical state of a solid.

The first attempt to calculate electron binding energies was reported by Slater
\cite{slater1955}; since then, various relativistic computations of neutral atom
binding energies have been performed \cite{crasemann}.
They exploit methods based on a Dirac-Hartree-Slater model, with
corrections for QED (quantum electrodynamics) effects and the nuclear charge
distribution.

\subsection{Selected compilations}
\label{sec_compilations}

This paper evaluates a selection of binding energy compilations, which are used
by general purpose simulation systems and some representative specialized codes:
\begin{itemize}
\item the compilation by Bearden and Burr \cite{bearden},
\item the compilation by Carlson \cite{carlson},
\item the tabulation included in Evaluated Atomic Data Library (EADL) \cite{eadl},
\item the compilation assembled by Sevier in 1979 \cite{sevier1979},
\item the compilations included in the seventh and eighth editions of the Table
of Isotopes \cite{toi1996,toi1978}, respectively published in 1978 and 1996,
\item the compilation by Williams included in the X-ray Data Booklet \cite{xbook}
and in the CRC Handbook of Chemistry and Physics \cite{crc90}.
\end{itemize}

Bearden and Burr performed a comprehensive evaluation of experimental X-ray
wavelength data; the techniques they used to establish a consistent energy scale
and to deal with elements with multiple or missing measurements are documented
in \cite{bearden}.
This compilation has been the basis for several other ones published
in the following years and is still used in some physics software systems.

Carlson's compilation reproduces the one by Lotz \cite{lotz} with a few
modifications and extensions, that concern the data for krypton and xenon,
the binding energies of elements with atomic number greater than 94 and the P
shell data of elements with atomic number between 87 and 95.
The compilation covers atomic numbers from 1 to 106; values are given 
for free atoms and are referenced to the vacuum potential.

The compilation by Lotz is based on Bearden and Burr's evaluated data,
complemented by other experimental measurements.
The tabulated binding energies were determined according to empirical criteria,
interpolation and extrapolation of available data.
Since values are listed for free atoms, the work function was taken into account
in converting experimental binding energies for solids.
According to \cite{lotz}, the uncertainties of the tabulated values are at most
2 eV for elements with atomic number up to 92; larger uncertainties, in some
cases greater than 10 eV, are reported for heavier elements.

The binding energies collected in the seventh edition of the Table of Isotopes
(identified in the following as ToI 1978) were taken from Shirley et al.
\cite{shirley1977} for elements with atomic number up to 30, and from a
compilation of experimental data by the Uppsala Group \cite{siegbahn_karlsson}
for heavier elements.
The tabulated binding energies derive mainly from photoelectron spectroscopic
measurements; data were taken from Bearden and Burr's compilation in cases
where experimental photoelectron measurements were not available.
Interpolation and extrapolation techniques were used to complement experimental
data.
The data are listed with reference to the Fermi level and concern 
elements with atomic number from 1 to 104.
Uncertainties are reported as about 0.1 eV for light elements and 1-2 eV for
most elements with atomic number greater than 30;  uncertainties
approaching 100 eV are mentioned for transuranic elements.
Shifts of the order of 10 eV in the binding energy of non-valence shells can
result from changes in the chemical state of the medium \cite{shirley1977}.

The binding energies collected in the eighth edition of the Table of Isotopes 
(identified in the following as ToI 1996) were taken from the compilation by
Larkins \cite{larkins}.
Binding energies are reported for solid systems referenced to the Fermi level,
except those for noble gases, Cl and Br, which are for vapor phase systems
referenced to the vacuum level.
Uncertainties may be as large as 10-20 eV for the inner orbitals in the
high-Z elements, and changes in chemical state can lead to substantial shifts in
the binding energies of non-valence shells \cite{shirley1977}.

The binding energies tabulated by Larkins are based on Sevier's 1972 compilation
\cite{sevier1972} for elements with atomic number up to 83 and on the
compilation by Porter and Freedman \cite{porter} for heavier elements;
with respect to these references, Larkins includes some updated values for Ar,
Ge, As, Se, Xe and Hg.
Sevier's 1972 tabulations were mainly an update to Bearden and Burr's ones to
include more recent measurements; a further extension was published by Sevier in
1979 \cite{sevier1979}.
Porter and Freedman combined a theoretical approach and experimental
measurements to interpolate data for heavy elements.

The eighth edition of the Table of Isotopes also includes a list of ionization
energies of the elements (concerning the least bound electron), which reflects the data available from NIST
(United States National Institute of Standards and Technology)
\cite{nist_ionipot}; these values differ in some cases from those in the
tabulation of electron binding energies in the same volume.

Williams' compilation is based on Bearden and Burr's data; some values are
taken from \cite{cardona_ley} with additional corrections, and some from
\cite{fuggle}. The energies are given relative to the vacuum level for the rare
gases and for H, N, O, F and Cl, relative to the Fermi level for metals and
relative to the top of the valence bands for semiconductors.
The tabulations concern elements with atomic number between 1 and 92.

The atomic subshell parameters collected in EADL are derived from theoretical
calculations by Scofield \cite{scofield1969, scofield1974}; besides these two
references, EADL documentation cites a ``private communication'' by Scofield,
dated 1988, as a source of the data.
Due to the scarcity of documentation about the origin of the binding energies
listed in EADL, it is difficult to ascertain how they were calculated, and what
assumptions and approximations may be underlying.
Binding energy values, although not for all elements and shells of the periodic
system, are reported in some publications by Scofield
\cite{scofield1978,scofield1990}; those in \cite{scofield1990} appear consistent
with EADL tabulations.
EADL data concern isolated, neutral atoms with atomic number up to 100.

\subsection{Binding energies used by physics software systems}

General purpose Monte Carlo systems and specialized codes use a variety of
binding energy compilations.

EGS5 uses the binding energies tabulated in the 1996 edition of the Table of
Isotopes, while EGSnrc uses the values of the earlier 1978
edition, as EGS4 \cite{egs4} did.

MCNP, MCNPX and ITS use the electron binding energies compiled by Carlson.

The Penelope 2008 version uses Carlson's compilation of binding energies;
earlier versions used the compilation included in the 1978 edition of the Table
of Isotopes.

The Geant4 toolkit uses various collections of binding energies.
The main reference for binding energies in Geant4 is the \textit{G4AtomicShells}
class in the \textit{materials} package;
according to comments in the code implementation, the binding energies values in
it derive from Carlson's compilation and the 73rd edition of the CRC
Handbook of Chemistry and Physics \cite {crc73}.


EADL values are used by the implementations of photon and electron interactions 
in Geant4 low energy electromagnetic package \cite{lowe_chep,lowe_nss}
based on the so-called Livermore Library, which
encompasses the Evaluated Electron Data Library (EEDL) \cite{eedl}, the
Evaluated Photon Data Library (EPDL97) \cite{epdl97} and EADL itself. 
EADL binding energies are also used in the calculation of proton ionization cross sections
described in \cite{haifa} and released in Geant4 9.4 in a modified version \cite{haifa2} to
address the drawbacks documented in \cite{tns_pixe}.
Proton ionization cross sections for Geant4 PIXE (Particle Induced X-ray
Emission) simulation described in \cite{tns_pixe} derive from ISICS \cite{isics}
tabulations using Bearden and Burr's binding energies.

Geant4 includes a C++ reimplementation of physics models originally
implemented in Penelope; Geant4 9.4 reimplements models from the 2008 version of
Penelope, while previous Geant4 releases included models equivalent to Penelope 2001.
The Geant4 9.4 reimplementation appears to use EADL values instead of the
binding energies used by Penelope 2008.
Binding energies corresponding to the values in the 1978 edition of the Table of
Isotopes are included in a Geant4 9.4 data set associated with Penelope.

Ionization energies consistent
with those reported by NIST \cite{nist_ionipot} are included in the Geant4
\textit{G4StaticSandiaData} class.

No reference to atomic binding energies can be retrieved in GEANT~3
documentation; however, according to comments embedded in the code, GEANT~3 used
Bearden and Burr's binding energies, with updated values for xenon derived from
\cite{grichine1991}.
Nevertheless, the \textit{GSHLIN} subroutine, where binding energies are
hard-coded, exhibits some discrepancies with respect to both Bearden and Burr's
tabulations and the values in \cite{grichine1991}; the origin of these values
could not be retrieved in the literature, nor in the software documentation.
Presumably, the code implementation and its comments went out of phase at some
stage of GEANT 3 evolution.

Atomic binding energies are relevant to PIXE calculations; two well known
software systems pertinent to this domain are GUPIX \cite{gupix1} and ISICS
\cite{isics}.
GUPIX uses Sevier's 1979 compilation of binding energies \cite{gupix3}, which
includes extensions to the 1972 collection by the same author.
ISICS uses Bearden and Burr's binding energies by default; the most recent
version of the code \cite{isics2011} offers the option of using the binding
energies assembled in Williams's compilation instead of Bearden and Burr's ones.

The authors of this paper could not retrieve track of the electron binding
energies used by FLUKA in the related software documentation and in the
literature, nor from direct inquiries with the maintainers of the code; 
it was not possible to ascertain them from the software implementation,
whose disclosure is subject to restrictions, as their presumed source file is in 
a binary encoded format.

\subsection{Comparison of binding energies compilations}

The binding energies collected in the various compilations exhibit some
differences, apart from those due to different references - the vacuum potential
or the Fermi level.

\begin{figure}[ht!]
    \begin{center}
        \subfigure[Carlson, ToI 1978-1996, Sevier 1979, Bearden and Burr]{%
            \label{fig_bek}
            \includegraphics[width=0.5\textwidth]{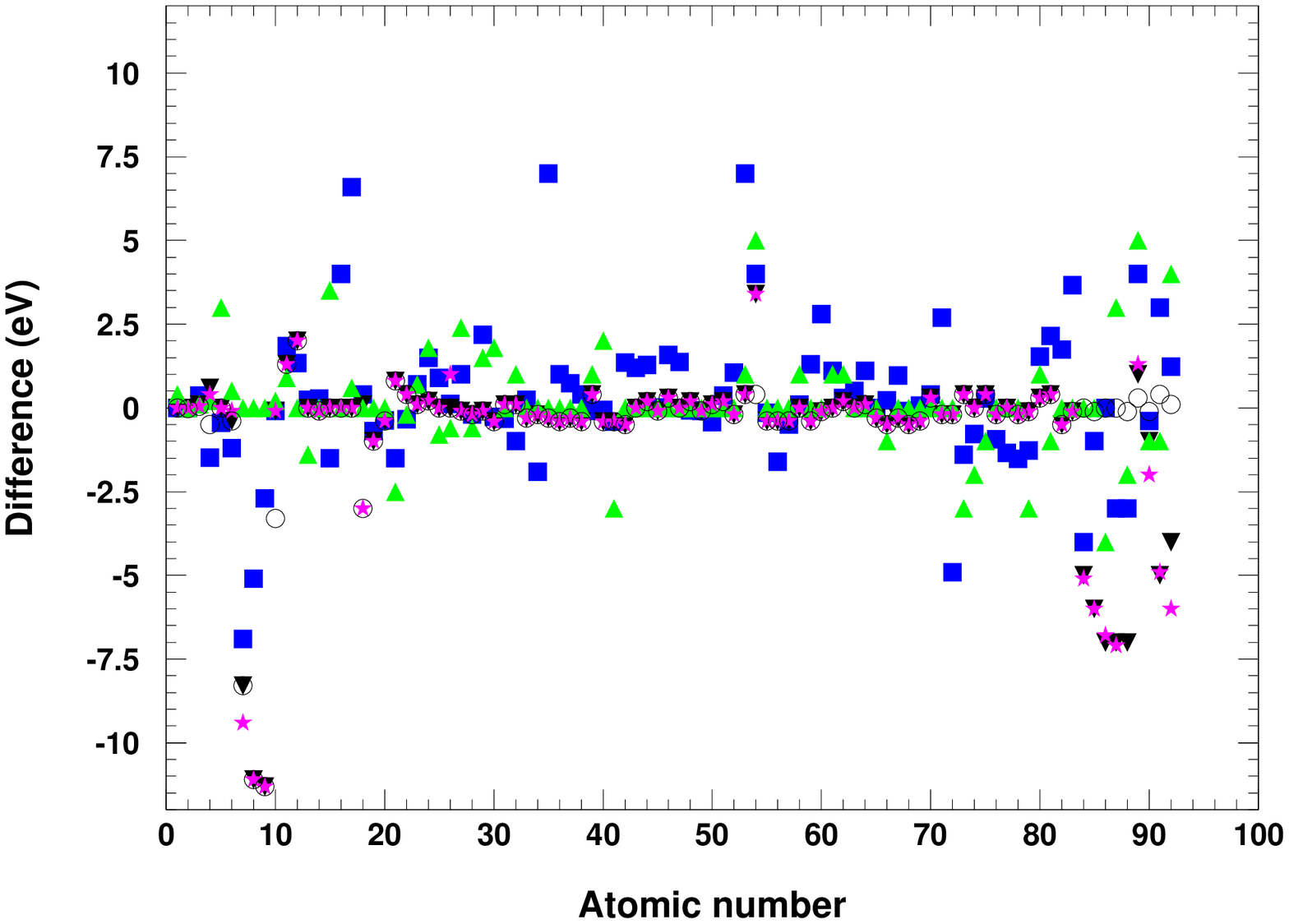}
        }%
        \subfigure[EADL]{%
           \label{fig_bek_eadl}
           \includegraphics[width=0.5\textwidth]{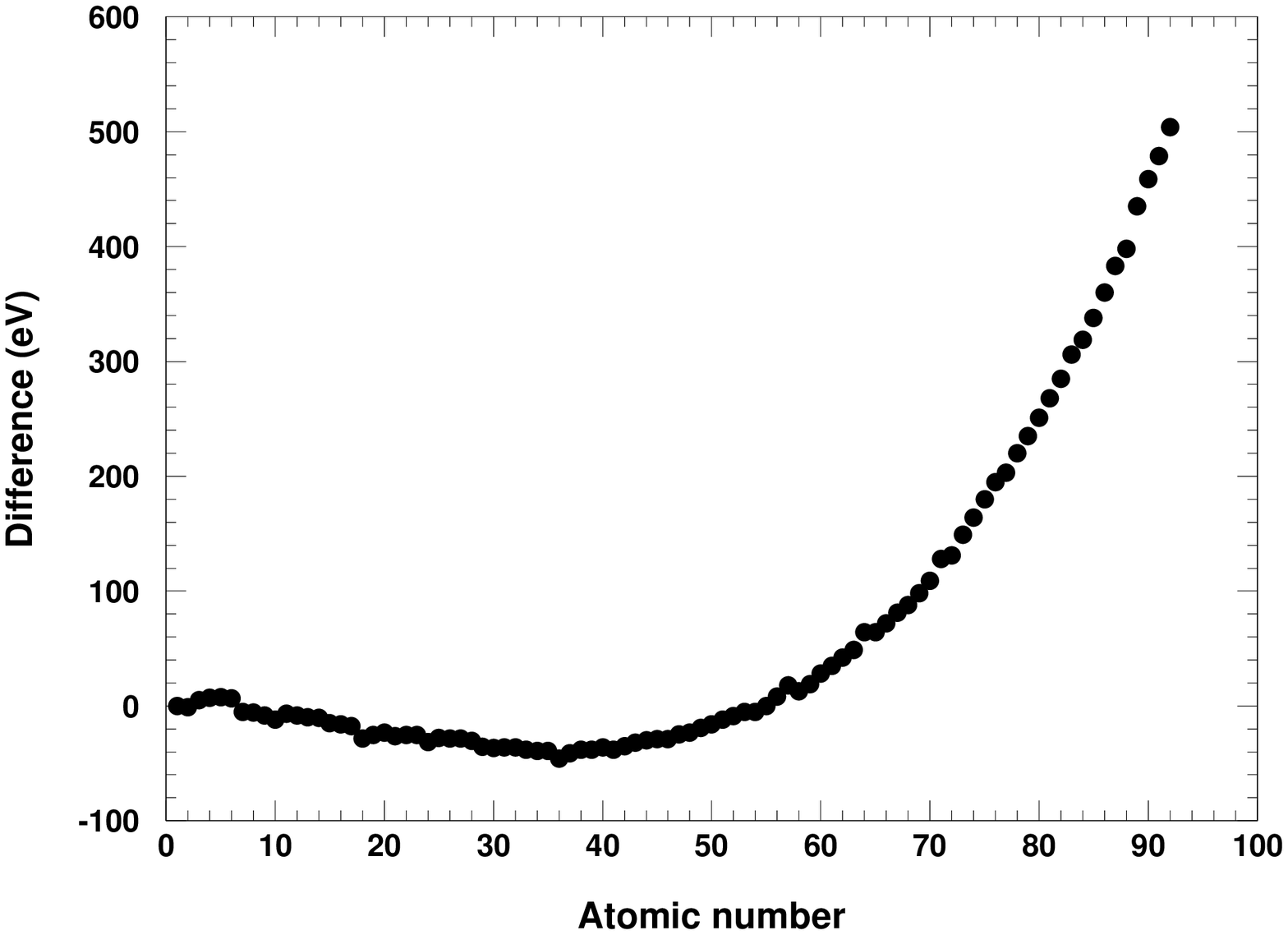}
        }\\ 
    \end{center}
    \caption{%
        Difference between K shell binding energies in various compilations and
binding energies in Williams' one versus atomic number: left, Carlson (blue
squares), Table of Isotopes 1996 (black down triangles), Table of Isotopes 1978
(green up triangles), Sevier 1979 (pink stars), Bearden and Burr (empty
circles); right, EADL (note the different scale).
     }%
    \label{fig_bek2}
\end{figure}

\begin{figure}[ht!]
    \begin{center}
        \subfigure[L$_{1}$]{%
            \label{fig_bel1}
            \includegraphics[width=0.5\textwidth]{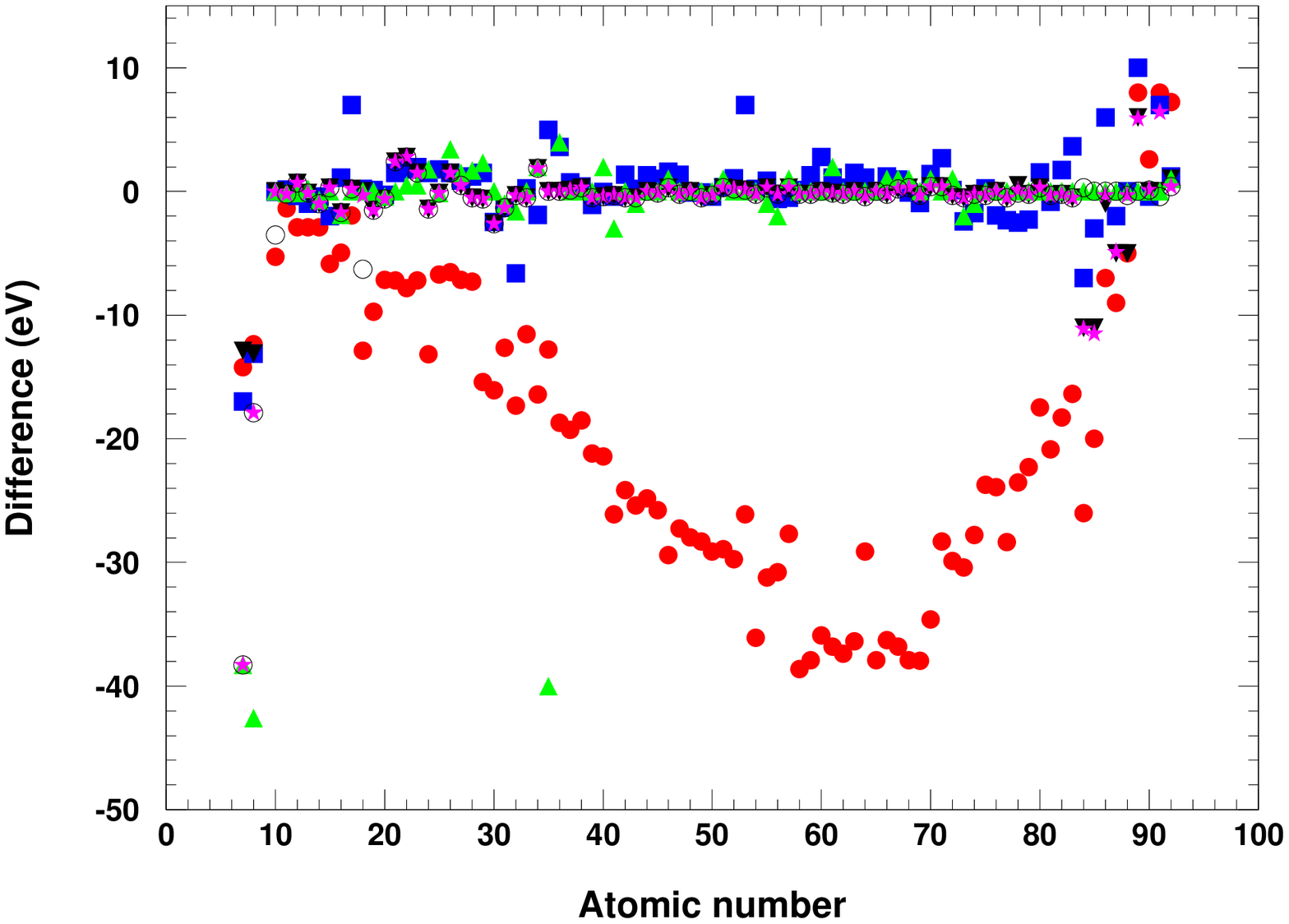}
        }%
        \subfigure[L$_{2}$]{%
           \label{fig_bel2}
           \includegraphics[width=0.5\textwidth]{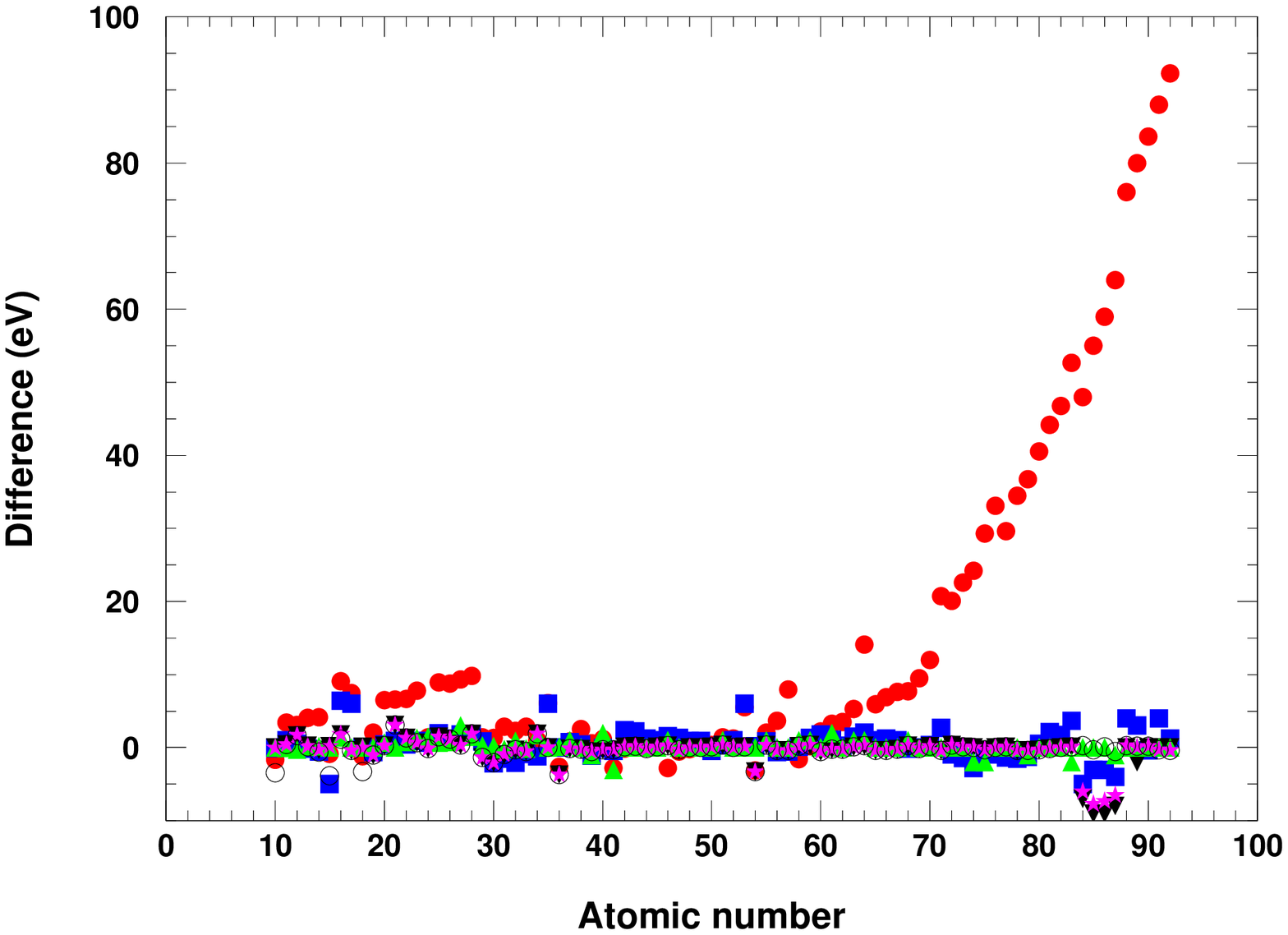}
        }\\ 
     \subfigure[L$_{3}$]{%
           \label{fig_bel3}
           \includegraphics[width=0.5\textwidth]{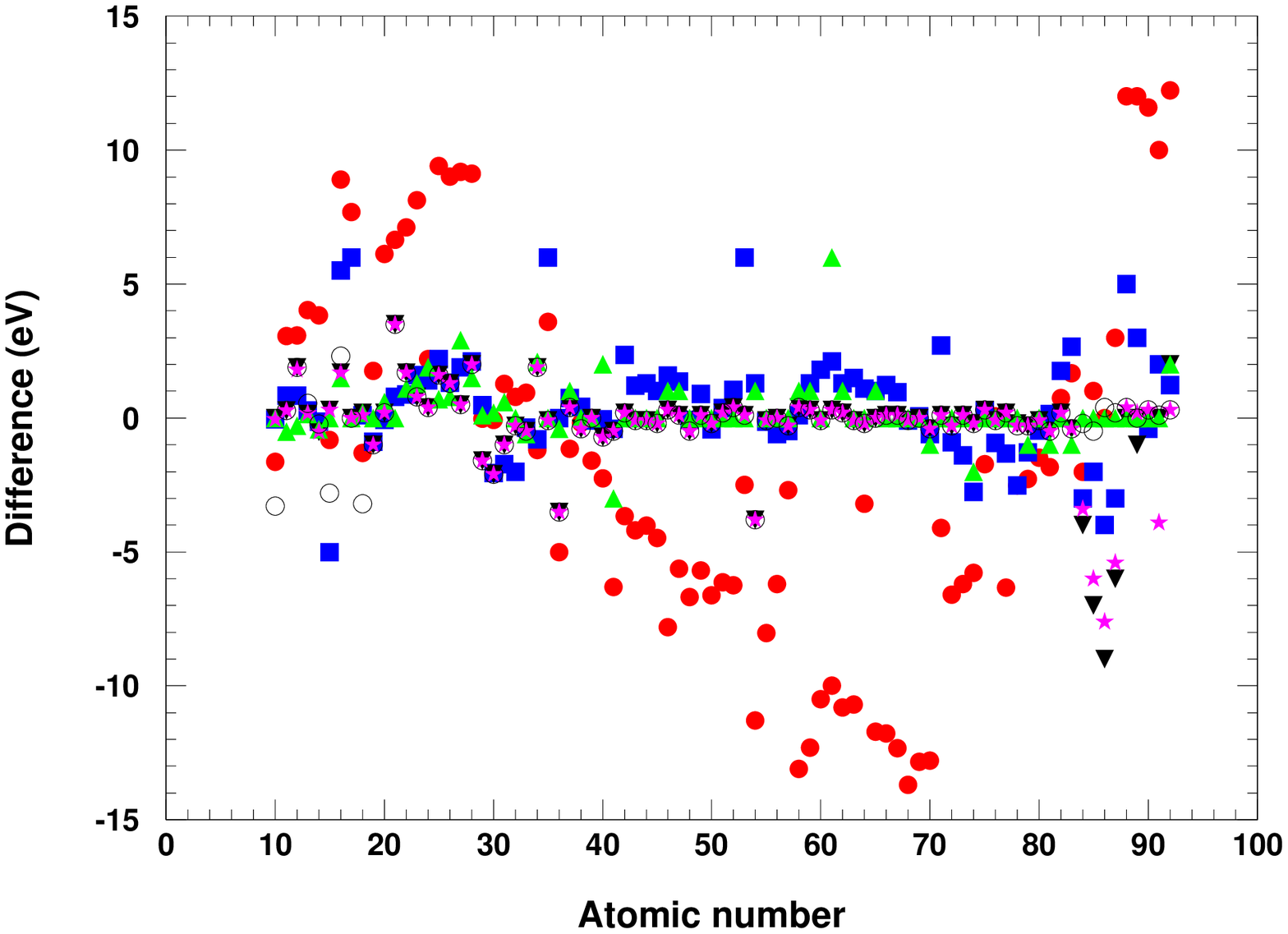}
	}%
    \end{center}
    \caption{%
        Difference between L sub-shell binding energies in various compilations and
binding energies in Williams' one versus atomic number: EADL (red
circles), Carlson (blue squares), Table of Isotopes 1996 (black down triangles),
Table of Isotopes 1978 (green up triangles), Sevier 1979 (pink stars), Bearden
and Burr (empty circles).
     }%
    \label{fig_bel}
\end{figure}

\begin{figure}[ht!]
    \begin{center}
        \subfigure[M$_{1}$]{%
            \label{fig_bem1}
            \includegraphics[width=0.5\textwidth]{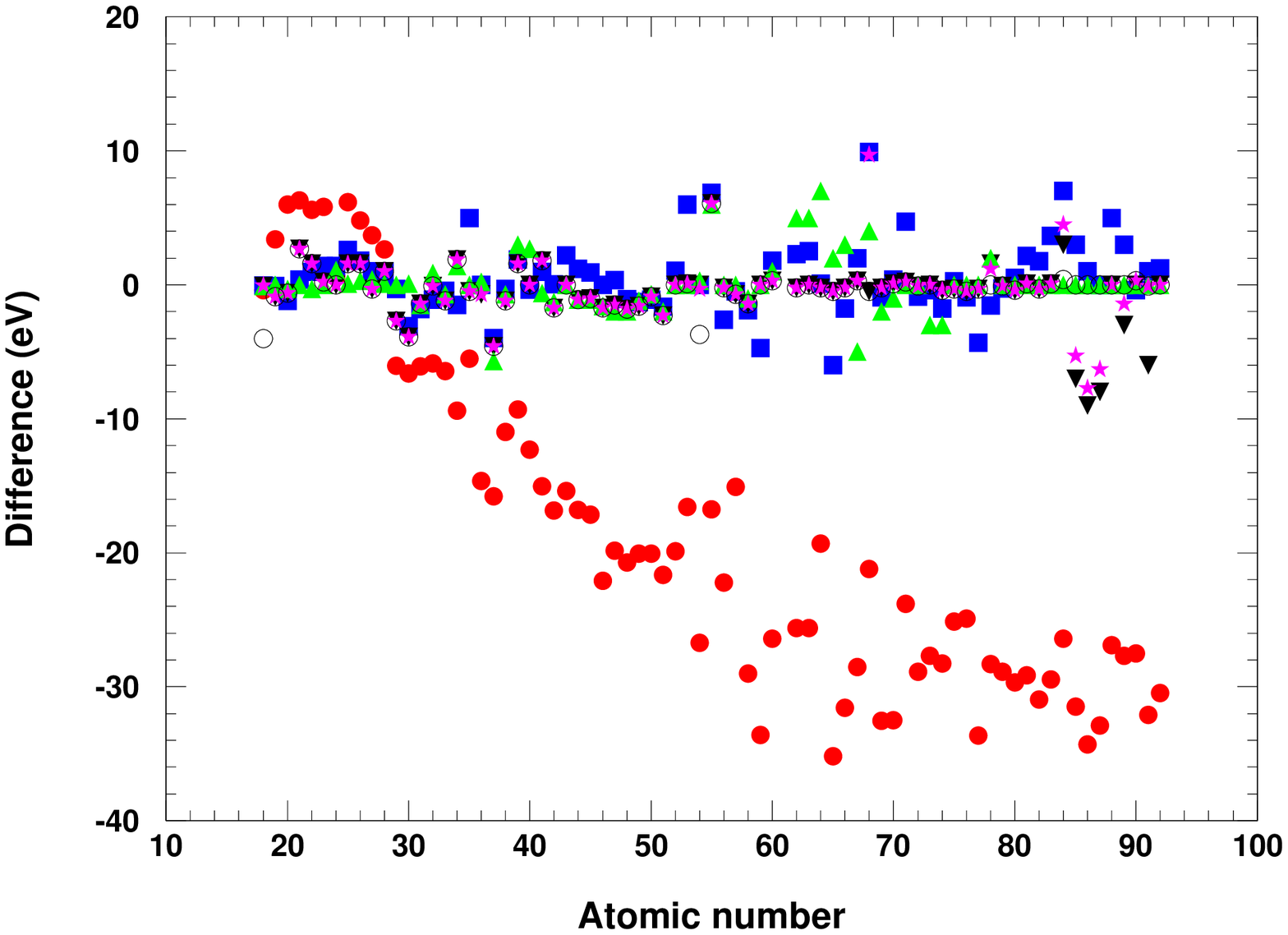}
        }%
        \subfigure[N$_{5}$]{%
           \label{fig_ben5}
           \includegraphics[width=0.5\textwidth]{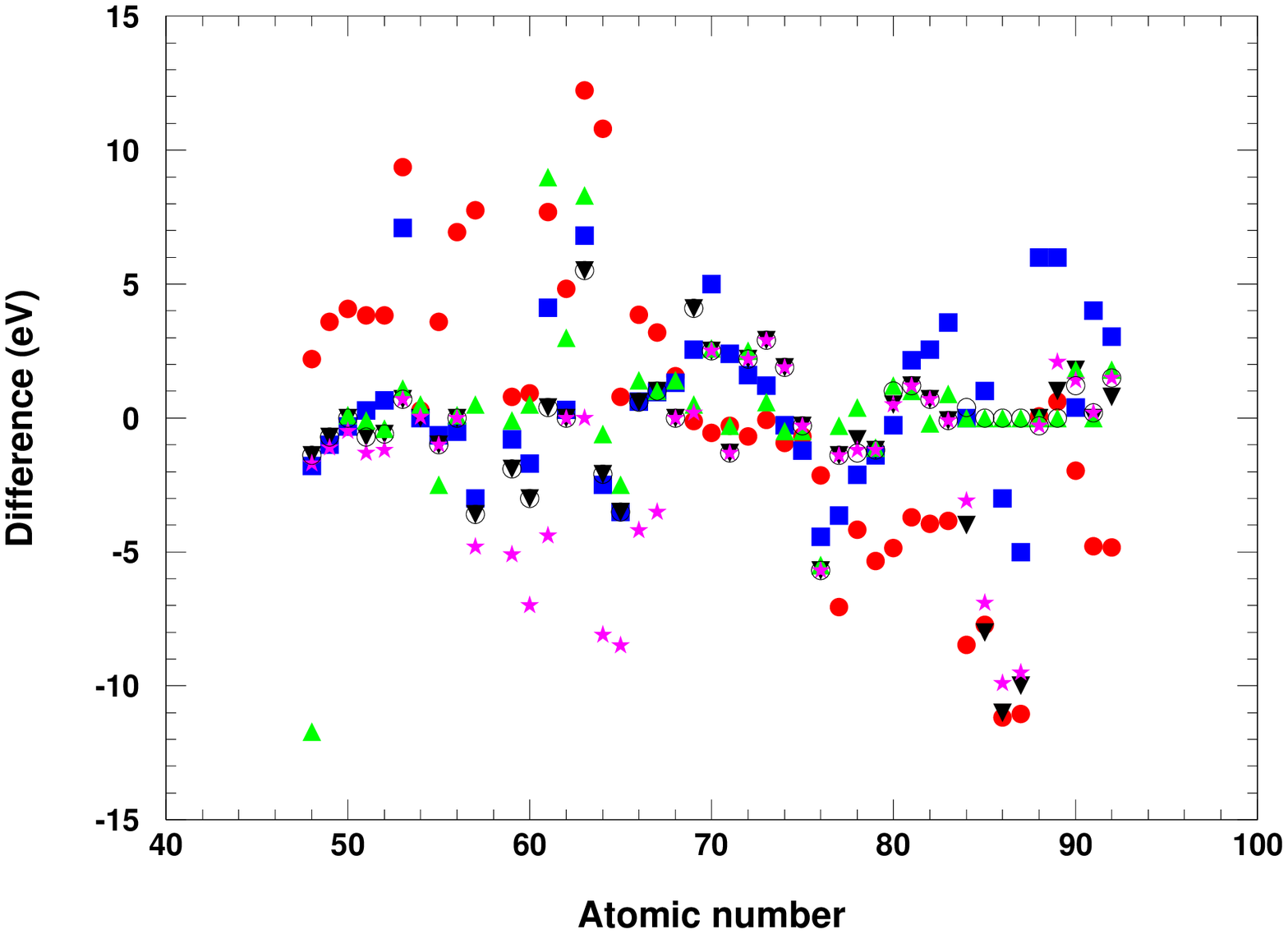}
        }\\ 
    \end{center}
    \caption{%
        Difference between representative M and N shell binding energies in various compilations and
binding energies in Williams' compilation versus atomic number: EADL (red
circles), Carlson (blue squares), Table of Isotopes 1996 (black down triangles),
Table of Isotopes 1978 (green up triangles), Sevier 1979 (pink stars), Bearden
and Burr (empty circles).
     }%
    \label{fig_bemn}
\end{figure}

A few examples of comparison are displayed in figures \ref{fig_bek2}-\ref{fig_bemn};
the plots show the difference between the binding energies in the various
compilations and the values in Williams' compilation.
The choice of Williams' compilation as a reference for plotting differences is
arbitrary; the main qualitative features of the plots are anyway equivalent, if
other empirical compilations are chosen as a reference instead of Williams' one.
The difference between EADL and Williams' K shell binding energies is plotted
separately from the other compilations, since the scale is approximately a
factor 30 larger.

The differences are of the order of a few electronvolts across the various empirical
compilations, as illustrated in figure \ref{fig_bek}, while they are larger
between EADL and the empirical compilations, especially for inner shells, as
shown in figure \ref{fig_bek_eadl}; they can reach a few hundred electronvolts for
the K shell of heavier elements.
The empirical compilations derive from a common source (Bearden and Burr's
review); therefore it is not surprising that they exhibit some similarities.


%

%
%

%

\section{Strategy of the study}

The study documented in this paper is driven by pragmatic motivations.
The analysis is focused on quantifying the accuracy of binding energy
compilations used in Monte Carlo systems, and their impact on physics models of
particle transport and on experimental observables produced by the simulation.
The evaluation aims at identifying one or more optimal options for Monte Carlo 
applications in experimental practice.
A comprehensive review of the physical ground, experimental measurements and
methods of calculations of electron binding energies is outside the scope of
this paper.

Two complementary approaches are adopted: direct validation of tabulated
electron binding energy values and the evaluation of effects on related physics
quantities, like ionization cross sections and X-ray fluorescence emission.
Both analyses involve comparisons with experimental data and a
comparative appraisal of the accuracy of the various compilations.

The evaluations of effects on dependent simulated quantities represent a
significant set of scenarios to characterize the problem domain, although they
are not intended to exhaust all the possible present and future uses of
atomic binding energies in radiation transport codes.
The test cases are chosen to address simulation observables that depend directly 
and explicitly on electron binding energies; this requirement is essential to 
avoid the risk of introducing a possible bias in the evaluation process due to 
other physics modeling features than the binding energies subject to assessment.
The test cases concern quantities that can be directly measured, so that 
the simulation outcome can be compared with experimental data to evaluate
the contribution of different atomic binding energies to the achieved accuracy.
Theoretical quantities that cannot be directly measured are explicitly not
considered, since possible effects due to electron binding energies would be
difficult to univocally identify and quantify in the analysis.

The study is articulated over a variety of test cases, which involve different
physics issues and reference data; the analysis methods are tailored to the
physical and experimental features of each test case.
Various statistical tools are exploited to quantify the accuracy of the
distributions examined in this study and the difference (or equivalence) of the
various binding energy compilations; 

Goodness-of-fit tests mentioned in the following sections utilize the
Statistical Toolkit\cite{gof1,gof2}.
Whenever applicable, multiple goodness-of-fit tests are  applied to mitigate
the risk of systematic effects in the conclusions of the analysis due to
peculiarities of the mathematical formulation of the various methods.

A combination of Student's t-test and F-test is applied to study the distribution of 
differences between the data subject to evaluation and reference values,
when goodness-of-fit tests do not exhibit adequate discriminant power over
some analyzed data samples.
The t-test is utilized to estimate the compatibility with null mean difference,
while, once the sample exhibiting the narrowest distribution of differences (i.e. the
lowest variance) has been singled out, the F-test is used to identify the data
samples whose variance is statistically equivalent to the narrowest distribution.

The binding energies listed in the various compilations are given with respect
to different references (vacuum or Fermi level).
In the following comparisons the original values are corrected to account for
the work function as appropriate to ensure a consistent reference.
Values of the work function are taken from the compilation of the CRC Handbook
of Physics and Chemistry \cite{crc90}, which is considered an authoritative
source for these data in experimental practice; they are complemented by data
from \cite{halas} and \cite{drummond} for elements not included in the 
compilation of \cite{crc90}.

The analyses reported in the following sections concern elements with atomic
number between 1 and 92, unless differently specified.
This range ensures uniform treatment of the various compilations in their
comparative appraisal, since all the examined compilations cover these elements,
while only a subset of them deal with transuranic elements.
Moreover, established experimental references of transuranic elements suitable
for the analysis of binding energies are scarce.

\section{Evaluation of reference binding energies}

Comparison with experimental data is the prime method to evaluate the accuracy
of simulation models; this validation method requires reliable experimental 
measurements as a reference.
Three authoritative collections of experimental binding energies (two of which
are partially overlapping) are used for this purpose; they include values for a
limited number of elements and shells, therefore they can validate only part of the
content of the compilations mentioned in the previous sections.


\subsection{Comparison with high precision reference data}

Experimental values of elemental binding energies reported in the literarature
exhibit significant discrepancies \cite{powell1991}; they can affect the validation 
of binding energy compilations.

Inconsistencies in the measurements are mostly due to inadequacies in the
calibration of binding energy scales of the various instruments, and are often
visible when comparing measurements performed by different laboratories.

Binding energy measurements may differ also for physical reasons: elemental
binding energies differ in the vapour and condensed states, and a chemical shift
is present when atoms are investigated in chemical compound states.
Moreover, binding energies for atoms implanted by ion bombardment into a metal foil substrate are
shifted with respect to those for a foil of the pure element.
Measurements on different surfaces of a crystal can result in different
ionization energy values.

These effects may be sources of systematic errors, which can be 
significant when comparing the binding energies collected in the various
compilations with experimental values.

A further source of uncertainties derives from the conversion between binding
energy values of solids referenced to the Fermi level and to the vacuum level;
this operation involves adding, or subtracting, the value of the work function.


A collection of high precision binding energies \cite{powell1995} was assembled
by Powell for the purpose of constituting a reference for the NIST X-ray
Photoelectron Spectroscopy Database \cite{nist_xps}.
Data published by different laboratories were subject to a retroactive
calibration procedure; the original experimental values were corrected to
produce a set of 61 binding energy values, concerning elements with
atomic numbers between 4 and 84 and shells from K to N.
The uncertainty of the reference energies is reported to be 0.061 eV
\cite{powell1995}.
These high precision data have been used to evaluate the accuracy
of the binding energy compilations examined in this study.

\begin{figure}[ht!]
     \begin{center}
        \subfigure[Difference]{%
            \label{fig_powell}
            \includegraphics[width=0.5\textwidth]{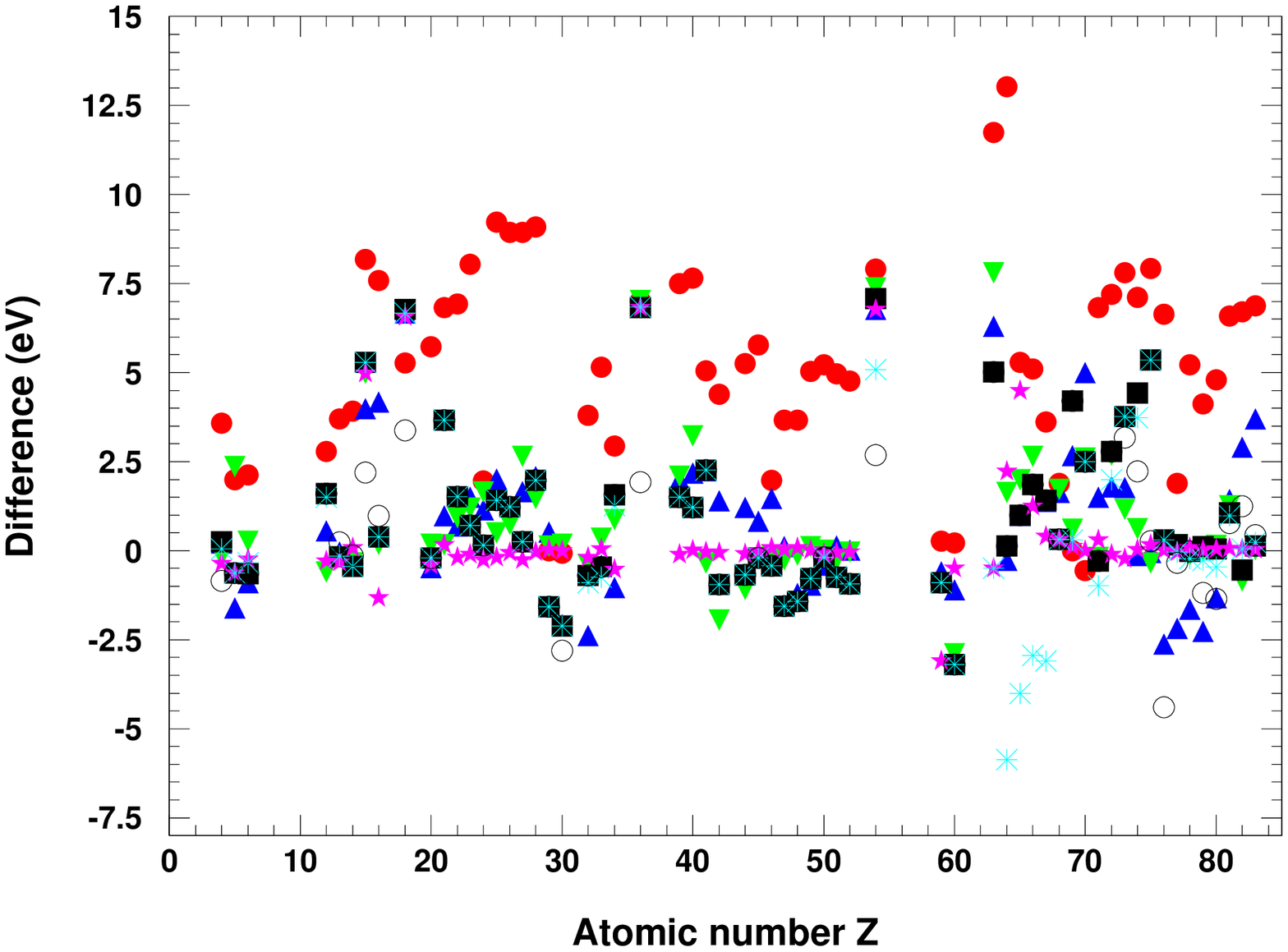}
        }%
        \subfigure[Relative difference]{%
           \label{fig_powellrel}
           \includegraphics[width=0.5\textwidth]{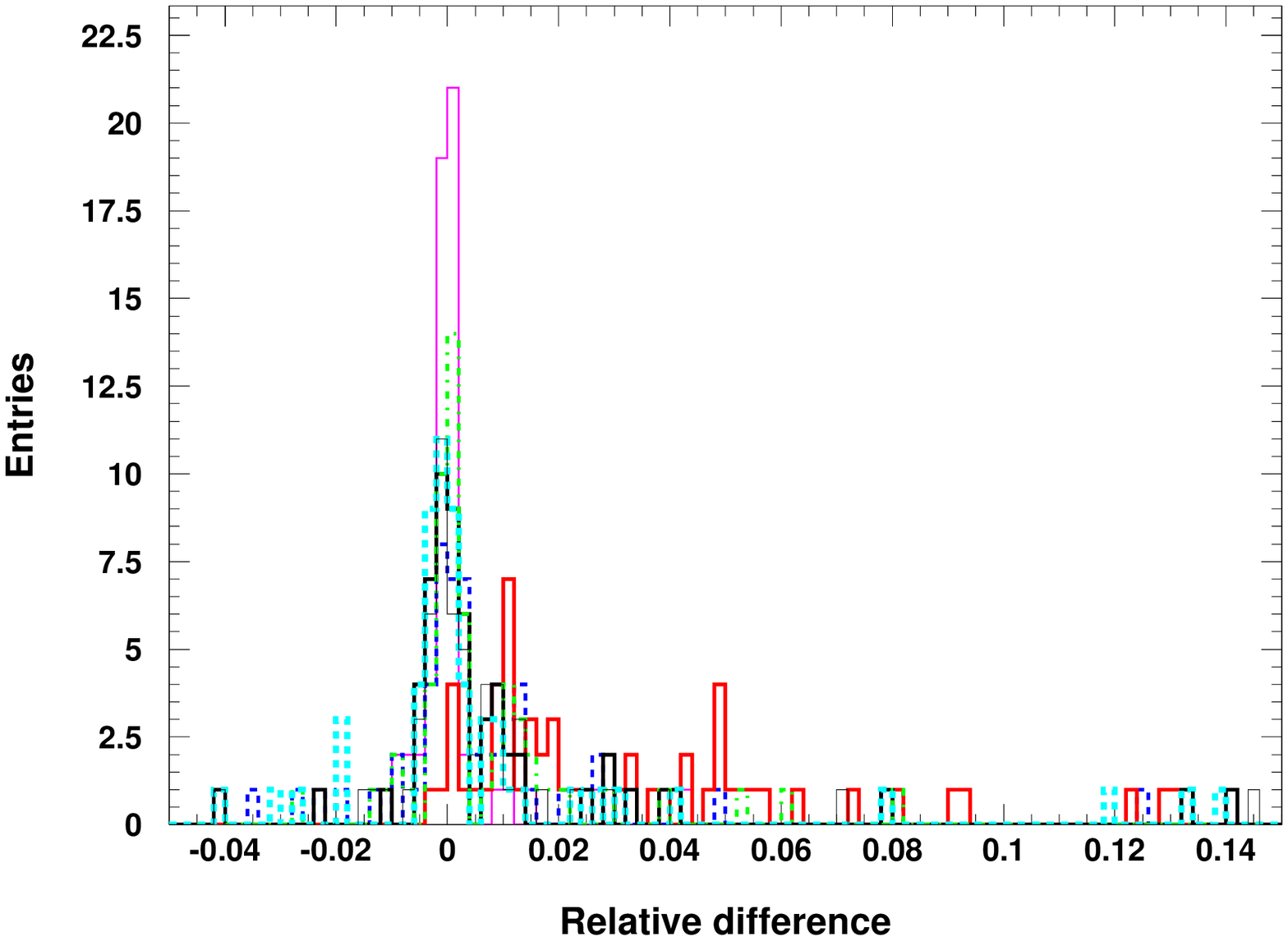}
        }\\ 
    \end{center}
    \caption{%
       Comparison of binding energies in various compilations and
reference data from \cite{powell1995} versus atomic number: EADL (red circles),
Carlson (blue up triangles), Table of Isotopes 1996 (black squares), Table of
Isotopes 1978 (green down triangles), Williams (pink stars), Sevier
1979 (turquoise asterisks).
     }%
   \label{fig_powell2}
\end{figure}

The difference of the binding energies in the various compilations with respect
to the reference values of \cite{powell1995} is shown in figure \ref{fig_powell};
the relative difference with respect to the same data is shown in figure
\ref{fig_powellrel}.
EADL binding energies appear less accurate than the samples from other
compilations and exhibit a systematic shift with respect to the reference
values.

It is worthwhile to remark that no quantitative, systematic validation of EADL
binding energies has been reported yet in the literature.
EADL documentation \cite{eadl} states in a section devoted to the accuracy of
the data that ``by comparing subshell parameters from a number of different
sources, it can be seen that there is still a disagreement of about 1\% between
the binding energies''; nevertheless this statement is not supported by any
objective demonstration, either directly in the documentation itself or through
references to the literature.
EADL documentation does not specify whether the other sources considered in the
above statement were other theoretical calculations, or empirical compilations, 
or experimental data.

%

Goodness of fit tests, which are commonly applied in statistical analysis to
compare data distributions, do not appear adequate to discriminate the
compatibility of the various compilations with respect to the reference data of
\cite{powell1995}.

Goodness of fit tests based on the empirical distribution function
(Kolmogorov-Smirnov\cite{kolmogorov1933,smirnov1939}, Anderson-Darling
\cite{anderson1952,anderson1954} and Cramer-von Mises
\cite{cramer1928,vonmises1931}) result in p-values greater than 0.999 for all
the data samples: this means that the differences between the binding energies
of the various compilations and the reference data are small with respect to the
sensitivity of these tests to detect discrepancies in the distributions subject
to comparison.
It is worthwhile to remark that the power of goodness-of-fit tests is still a
subject of research in statistics.

On the other hand, the $\chi^2$ statistic \cite{bock} based on the uncertainties
of the reference data reported in \cite{powell1995} (0.061 eV) results in
p-values much smaller than 0.001 for all the binding energy samples; therefore
the $\chi^2$ test would reject the hypothesis of compatibility of any
compilation with the reference data sample with 0.001 significance.


The outcome of the $\chi^2$ test depends critically on the correct estimate of
the uncertainties of the data subject to test.
The procedure applied in \cite{powell1995} to build the reference data sample
mitigates the risk of possible systematic effects due to instrumental
calibration, which may affect raw experimental data; nevertheless, other
sources, independent from the intrinsic precision of the measurement, may
contribute to the overall uncertainty.

Previously mentioned physical and chemical shifts of the experimental data,
associated with the conditions of the measurements, may introduce systematic
effects.
In this context, one should take into account that, while the reference values
in \cite{powell1995} reflect the experimental configuration and instrumental energy resolution for each
measured material, the data tabulated in the various empirical compilations
are the result of manipulations, such as 
interpolations, extrapolations and fits, over large collections of heterogeneous
experimental data from multiple sources:
the generic binding energy estimates deriving from these procedures may not
adequately account for the peculiarities of experimental mesurements performed
in specific physical and chemical configurations.
The calculation of the $\chi^2$ test statistic includes only the uncertainties
associated with Powell's reference data; it does not account for errors
associated with the binding energies of the various compilations.

A further source of uncertainty is associated with the work function in cases
where a conversion between the Fermi and vacuum reference level should be
applied for consistency between the distributions subject to comparison.
Moreover, experimental values of the work function are affected by the technique of
measurement and the cleanliness of the surface.
The CRC compilation does not report the uncertainties of the work functions;
therefore this additional error cannot be included in the computation of the
$\chi^2$ test statistic concerning Carlson's and EADL binding energies.

Due to these considerations, caution should be exercised in interpreting the
outcome of the $\chi^2$ test as physically significant,
as the nominal uncertainties values involved in the calculation may not
realistically represent the actual uncertainties associated with the tested data
samples.

Other statistical methods than goodness-of-fit testing were exploited to
quantify the compatibility between the various binding energies compilations and
the reference data of \cite{powell1995}.

A Student's t-test was performed to estimate whether the 
differences between the binding energies of the various compilations and the
corresponding reference values are compatible with a true mean of zero.
The p-values resulting from this test are summarized in Table \ref{tab_tpowell}.
The binding energies of Williams' and 1979 Sevier's compilations are compatible
at 0.05 level of significance with null mean difference with respect to the
reference data; the t-test rejects the hypothesis of compatibility with zero
mean difference with 0.01 significance for all the other compilations.
It should be stressed that these tests, as well as similar ones
reported in the following, do not compare the compilations with each other, 
but how well they reproduce the set of precision reference data.

All the binding energies compilations exhibit 
rather large differences with respect to the reference data for rare gases 
(Ar, Xe and Kr) reported in \cite{powell1995}, that derive from
implants in other media.
If one excludes these data from the t-test, also Bearden and Burr's  binding energies 
are compatible with zero mean difference with respect to the
reference data at 0.05 significance level.

\begin{table}
\begin{center}
\caption{Student's t-test applied to the difference with respect to Powell's reference data.}
\label{tab_tpowell}
\begin{tabular}{|l|cc|cc|}
\hline
				& \multicolumn{2}{|c|}{All data} 	& \multicolumn{2}{|c|}{Excluding Ar, Xe, Kr} \\
{\bf Compilation} 		&{\bf Mean (eV)} 	& {\bf p-value}	&{\bf Mean (eV)} 	& {\bf p-value}	\\
\hline
Bearden and Burr		&0.51		& 0.026		&0.40		& 0.083 		\\
Carlson			&0.95		& 0.001		&0.65		& 0.008		\\
EADL				&4.85		& $<$0.0001	&4.75		& $<$0.0001	\\
Sevier 1979			&0.41		& 0.181		&0.11		& 0.673		\\
ToI 1978			&1.00		& 0.0004		&0.69		& 0.002		\\
ToI 1996			&0.99		& 0.001 		&0.69		& 0.006		\\
Williams			&0.41		& 0.076 		&0.09		& 0.545		\\

\hline
\end{tabular}
\end{center}
\end{table}

\begin{table}
\begin{center}
\caption{F-test applied to the differences with respect to Powell's reference data.}
\label{tab_sigmapowell}
\begin{tabular}{|l|c|c|}
\hline
{\bf Compilation} 		& {\bf Standard deviation (eV)}	&{\bf p-value}\\
\hline
Bearden and Burr		& 1.71	&0.0005		\\
Carlson			& 1.80	&0.0001		\\
EADL				& 1.56	&0.005		\\
Sevier 1979			& 2.01 	&$<$0.0001		\\
ToI 1978			& 1.61	&0.002		\\
ToI 1996			& 1.84	&0.0001		\\
Williams			& 1.07	& -			\\
\hline
\end{tabular}
\end{center}
\end{table}

The binding energies of Williams' compilation exhibit the narrowest
distribution of differences with respect to the reference data of
\cite{powell1995}, as can be seen in figure \ref{fig_powellrel}.
The standard deviations related to the various compilations are listed in Table
\ref{tab_sigmapowell}, excluding the data for argon, xenon and krypton, which 
are treated as outliers. 
The table also reports the p-values of the F-test to evaluate the hypothesis of
equality of variance associated with the various compilations with respect to
the binding energies of Williams' compilation; the distributions subject to the
F-test concern the difference between the binding energies in the compilations
and the reference data of \cite{powell1995}.
The statistical analysis confirms the qualitative evidence of figure
\ref{fig_powellrel}, since the null hypothesis is rejected with 0.01 significance level
for all the test cases.

\subsection{Comparison with NIST recommended binding energies}

A similar analysis has been performed with respect to the collection of recommended
binding energies for principal photoelectron lines assembled by NIST \cite{nist_principal}.
This collection consists of 85 values; it includes most of the reference binding
energies discussed in \cite{powell1995}, along with additional data, mainly
concerning outer shells than those reported in \cite{powell1995}.
The data for noble gases listed in \cite{powell1995} are not included in this
set of recommended values.

\begin{figure}
\centerline{\includegraphics[angle=0,width=11cm]{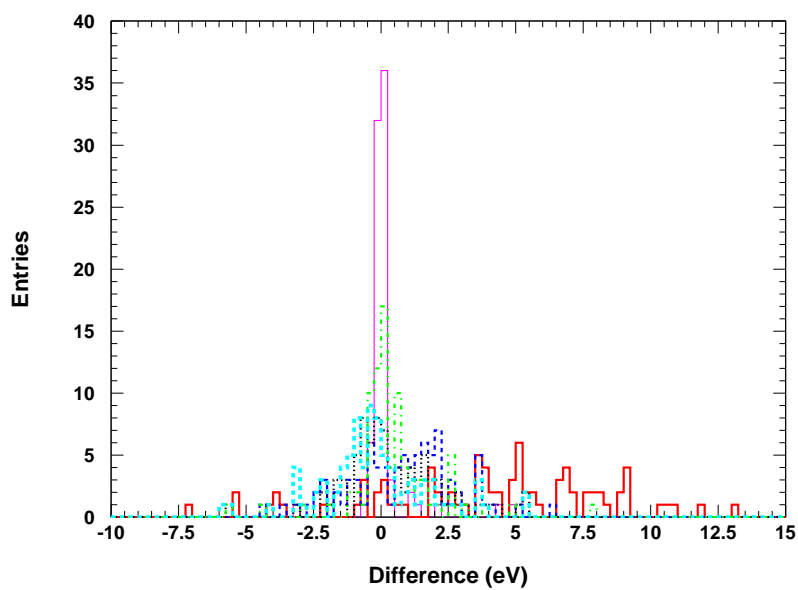}}
\caption{Difference between binding energies in various compilations and
reference data from \cite{nist_principal}: Carlson (dashed blue line), Table of
Isotopes 1996 (dotted black line), Table of Isotopes 1978 (dotted-dashed
green line), Williams (thin pink solid line), Sevier 1979 
(thick turquoise dashed line) and EADL (thick solid red
line).}
\label{fig_principal}
\end{figure}

Figure \ref{fig_principal} shows the difference between the binding energies of
the various compilations and the NIST recommended values.

The comparison with these reference data adopts a similar method to the
one described in the previous section.

The p-values resulting from Student's t-test for compatibility with zero mean 
difference with respect to the reference data are reported in Table \ref{tab_tprincipal}.
The binding energy samples extracted from the Williams' and 
Sevier's 1979 compilations are compatible with zero mean difference at 0.05 level
of significance; those from the two editions of the Table of Isotopes are
compatible at 0.01 level.
Similarly to the previous case, the distribution of differences with the lowest
standard deviation is the one associated with Williams' compilation.

The standard deviations of the distributions of differences from the reference
data are reported for all the data samples in Table \ref{tab_sigmaprincipal},
together with the p-values of the F-test for the equality of variance with
respect to the distribution associated with Williams' compilation.
Consistently with the qualitative features of figure \ref{fig_principal}, the
variance associated with the other compilations is incompatible with the 
variance related to  Williams' binding energies.

Based on this statistical analysis, one can conclude that Williams' compilation best
reproduces experimental reference binding energies.
It should be stressed, however, that the NIST reference sample represents a
small subset of the periodic system of elements: approximately 6\% of the total
number of shells of elements with atomic number up to 92.

\begin{table}
\begin{center}
\caption{Student's t-test applied to the difference with respect to NIST recommended binding energies.}
\label{tab_tprincipal}
\begin{tabular}{|l|c|c|}
\hline
{\bf Compilation} 		&{\bf Mean (eV)}	& {\bf p-value}	\\
\hline
Bearden and Burr		&0.35			&0.070 		\\
Carlson			&0.75			& 0.0003		\\
EADL				&3.96			& $<$0.0001	\\
Sevier 1979			&-0.19			& 0.387		\\
ToI 1978			&0.44			& 0.017		\\
ToI 1996			&0.52			& 0.012		\\
Williams			&0.11			& 0.246 		\\
\hline
\end{tabular}
\end{center}
\end{table}

\begin{table}
\begin{center}
\caption{F-test applied to the difference with respect to NIST recommended binding energies.}
\label{tab_sigmaprincipal}
\begin{tabular}{|l|c|c|}
\hline
{\bf Compilation} 		& {\bf Standard deviation (eV)}	&{\bf p-value}\\
\hline
Bearden and Burr		& 1.74	&$<$0.0001	\\
Carlson			& 1.94	&$<$0.0001	\\
EADL				& 4.14	&$<$0.0001	\\
Sevier 1979			& 2.03	&$<$0.0001	\\	
ToI 1978			& 1.66	&$<$0.0001	\\
ToI 1996			& 1.86	&$<$0.0001	\\
Williams			& 0.88 	& -			\\
\hline
\end{tabular}
\end{center}
\end{table}

\subsection{Evaluation of ionization energies}
\label{sec_ionipot}
The ionization energy (in the past referred to as ionization potential), is the least
energy that is necessary to remove an electron from a a free, unexcited, neutral
atom, or an additional electron from an ionized atom.
In the following analysis it is considered to be equal to the binding energy of
the least bound electron in the atom.

A compilation of reference experimental ionization potentials is available from
NIST \cite{nist_ionipot}; the same values are also reported in the Table of
Isotopes \cite{toi1996} in a table distinct from the compilation of electron
binding energies and in the CRC Handbook of Chemistry and Physics \cite{crc90}.
This compilation does not list the uncertainties of the ionization energies it collects,
but NIST web site comments that they range from less than
one unit in the last digit of the given values to more than 0.2 eV.

The lowest binding energies for each element in the various compilations have
been compared to the reference ionization energies collected by NIST.
The compilations of the 1978 edition of the Table of Isotopes and Williams 
do not include many outer-shell binding energies; this limitation may be
related to the emphasis of these compilations on experimental effects related to
inner shells, like measurements concerning X-ray fluorescence or Auger electron
emission.
Therefore the following analysis was restricted to the compilations of Bearden 
and Burr, Carlson, EADL, Sevier 1979 and 1996 edition of the Table of Isotopes, 
which list a full set of electron binding energies.

\begin{figure}
\centerline{\includegraphics[angle=0,width=11cm]{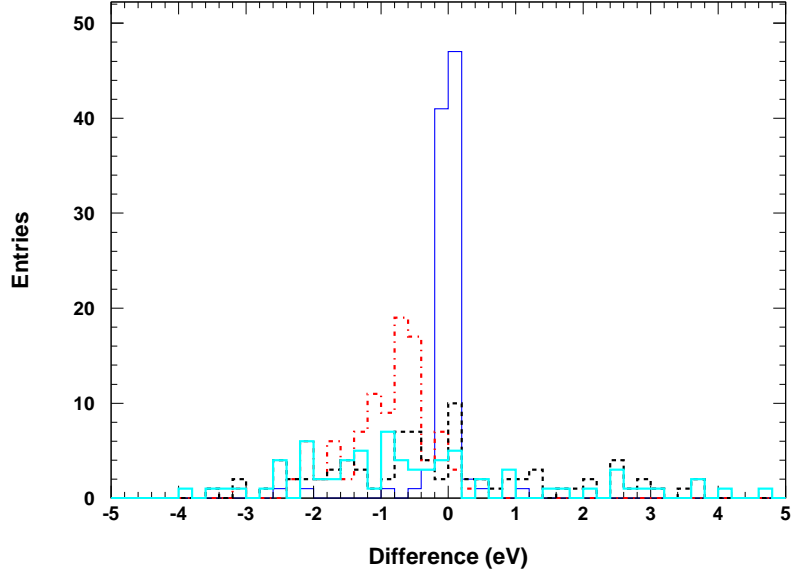}}
\caption{Difference between ionization energies in various compilations and
NIST reference experimental data: Carlson (thin solid blue line), Table of
Isotopes 1996 (dashed black line),  Sevier 1979 
(thick solid turquoise line) and EADL (dot-dashed red
line).}
\label{fig_ionipot}
\end{figure}

The difference between ionization energies derived from the various compilations
and NIST reference data is shown in figure \ref{fig_ionipot}.
Carlson's binding energies appear to be in closest agreement with NIST ionization energies.

Goodness of fit tests are sensitive to the differences exhibited by the various
compilations with respect to the NIST reference collection.
The results of the Kolmogorov-Smirnov, Anderson-Darling and Cramer-von Mises
tests are listed in Table \ref{tab_gof_ionipot}. 
The hypothesis of compatibility with NIST reference data is rejected by all the
tests with 0.001 significance for EADL and Bearden and Burr's data.
Carlson's compilations and the Table of Isotopes 1996 are compatible with the
reference data at 0.05 significance level according to all the tests.
Regarding Sevier's 1979 compilation, the Anderson-Darling test rejects the
hypothesis of compatibility at 0.05 significance level, while the
Kolmogorov-Smirnov and Cramer-von Mises test do not; the different response of
these tests near the critical region of 0.05 significance could be explained by
the greater sensitivity of the Anderson-Darling test statistic to fat tails.

\begin{table}
\begin{center}
\caption{P-values from goodness-of-fit tests concerning NIST reference ionization potentials.}
\label{tab_gof_ionipot}
\begin{tabular}{|l|c|c|c|}
\hline
{\bf Compilation} 		& {\bf Kolmogorov}	& {\bf Anderson}	& {\bf Cramer}	\\
                         		& {\bf Smirnov}		& {\bf Darling}	& {\bf von Mises} 	\\
\hline
Bearden and Burr		& $<$0.001		& $<$0.001	& $<$0.001	\\
Carlson			& 0.670			& 0.995		& 0.963 	\\
EADL				& $<$0.001		& $<$0.001	& $<$0.001	\\
Sevier 1979			& 0.061			& 0.023	   	& 0.067	\\
ToI 1996			& 0.096			& 0.099		& 0.213	\\
\hline
\end{tabular}
\end{center}
\end{table}

\section{Effects on fluorescence X-ray energies}
%

Compilations of characteristic X-ray energies are available, at least for lines of
experimental interest, which in principle could be used in Monte Carlo
simulation to determine the energy of secondary products of atomic relaxation.
Nevertheless, these experimental tabulations can hardly satisfy the requirements
of general purpose Monte Carlo codes, which require the ability of generating
any atomic transition for any element.
The energies of X-rays and Auger electrons resulting from atomic relaxation are
often computed by Monte Carlo codes as the difference between the binding
energies of the shells involved in the transition; in this approximation the
binding energies of the atom in an ionised state are assumed to be the same as
in the ground state.
Therefore the accuracy of the simulation of the secondary products of atomic
relaxation is determined by the accuracy of the binding energies implemented in
the Monte Carlo system (apart from the physical approximation of neglecting the
difference between the binding energies of an ionised atom and a neutral one in
the ground state).

The accuracy of the examined binding energy compilations to reproduce the energy
of atomic relaxation products has been estimated with respect to the
experimental X-ray energies reported in the review by Deslattes et al.
\cite{deslattes}, which concerns K and L transitions.

A comparison of X-ray energies calculated by Geant4, based on EADL binding
energies, with respect to the same experimental data is documented in
\cite{relax_nist}.
That study showed that, according to the outcome of the Kolmogorov-Smirnov test,
all the X-ray energies simulated by Geant4 are compatible with the experimental
data with 0.1 significance for all transitions and all elements; the relative
difference between simulated and experimental values is approximately 1-2\% for
most individual transitions.
The present study is extended to  binding energy compilations other than EADL.

\begin{figure}[ht!]
    \begin{center}
        \subfigure[KL$_2$]{%
            \label{fig_kl2}
            \includegraphics[width=0.5\textwidth]{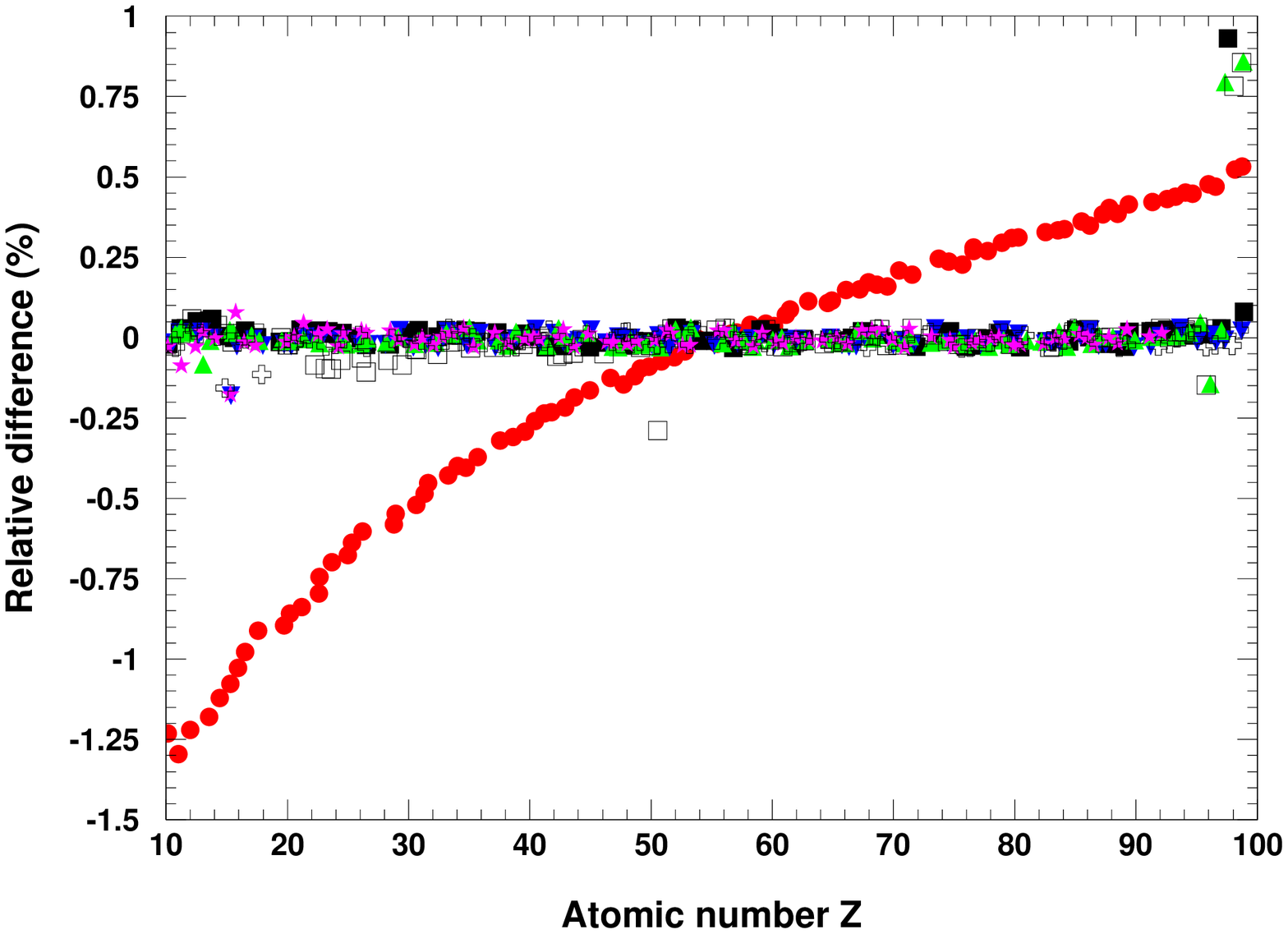}
        }%
        \subfigure[KM$_2$]{%
           \label{fig_km2}
           \includegraphics[width=0.5\textwidth]{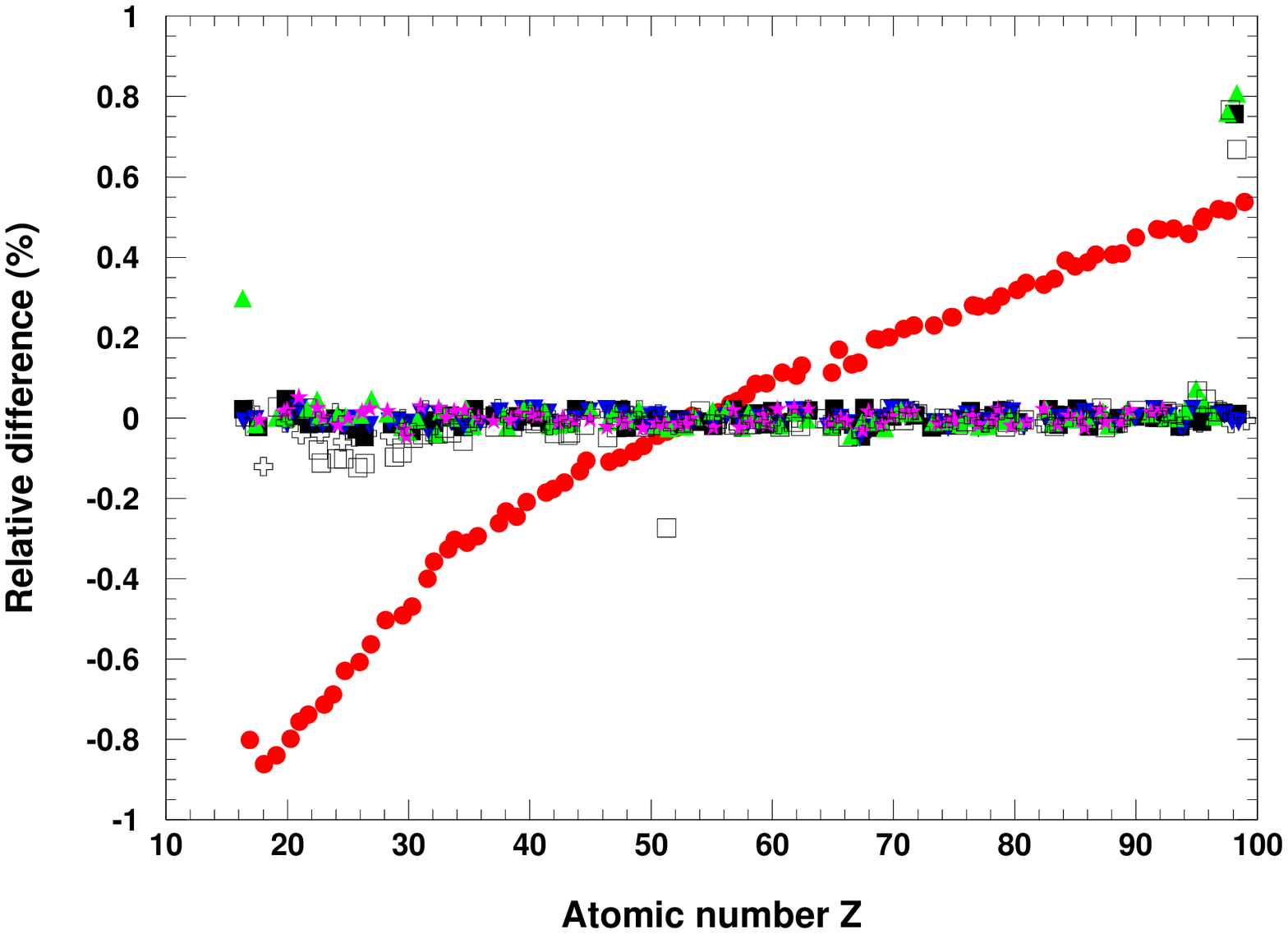}
        }\\ 
     \subfigure[KN$_2$]{%
           \label{fig_kn2}
           \includegraphics[width=0.5\textwidth]{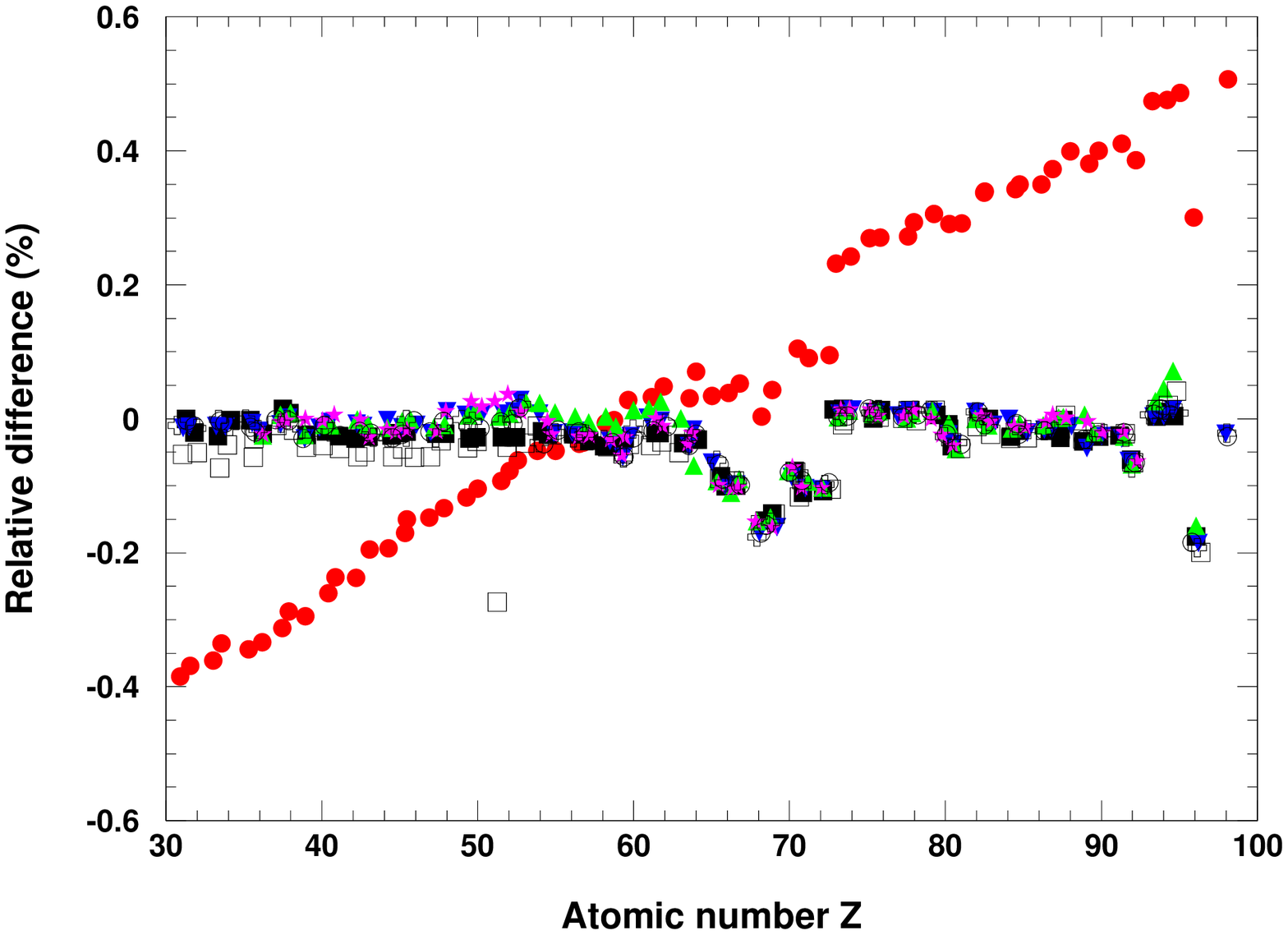}
	}%
    \end{center}
    \caption{%
        K-shell transitions, 
relative difference between binding energies in various compilations and
experimental data from \cite{deslattes} versus atomic number: EADL (red circles),
Carlson (blue up triangles), Table of Isotopes 1996 (black squares), Table of
Isotopes 1978 (green down triangles), Williams (pink stars), Sevier
1979 (turquoise asterisks) and G4AtomicShells (empty squares).
     }%
    \label{fig_trans_k}
\end{figure}
\begin{figure}[ht!]
     \begin{center}
        \subfigure[L$_1$M$_3$]{%
            \label{fig_l1m3}
            \includegraphics[width=0.5\textwidth]{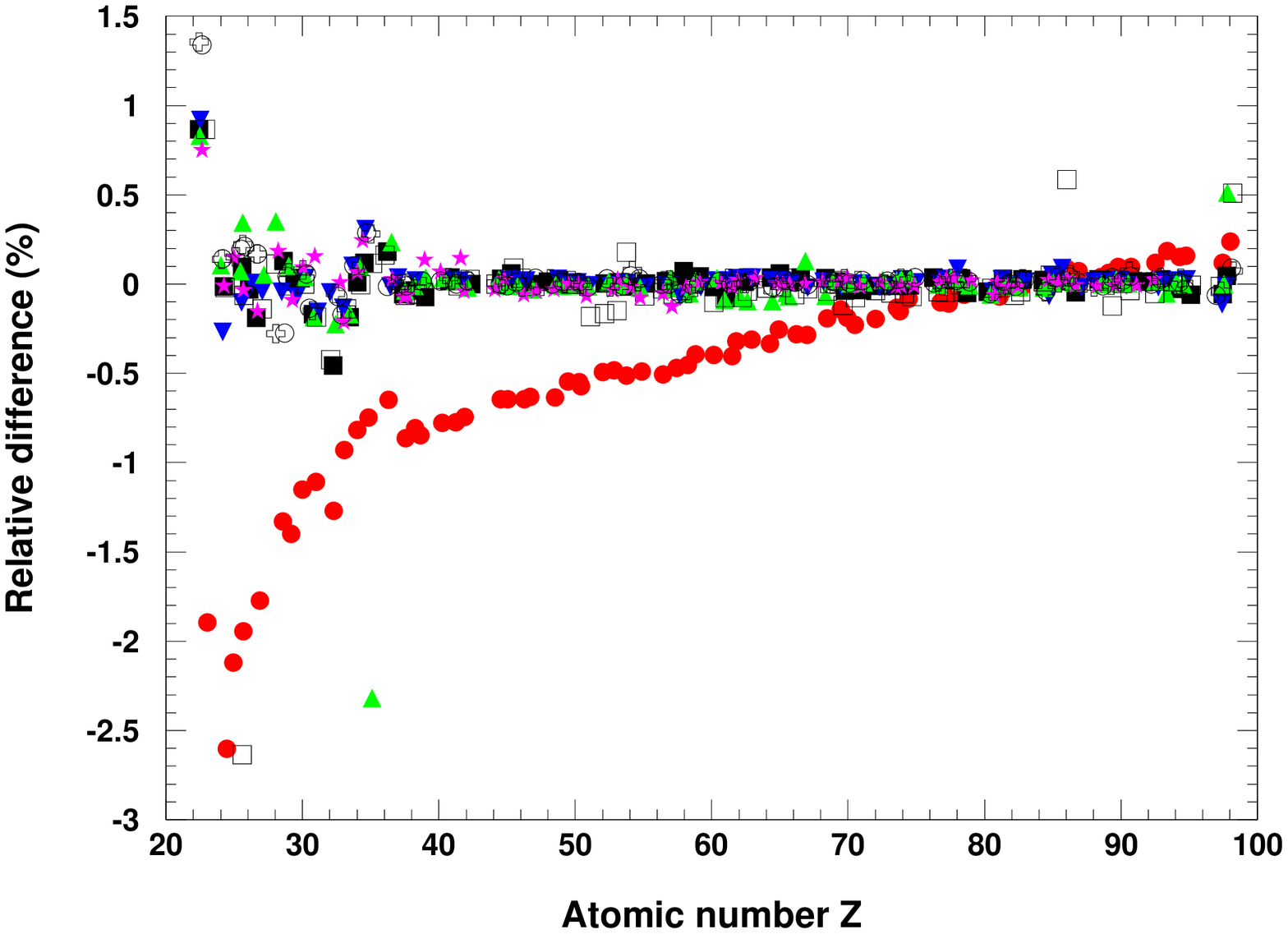}
        }%
        \subfigure[L$_1$N$_2$]{%
           \label{fig_l1n2}
           \includegraphics[width=0.5\textwidth]{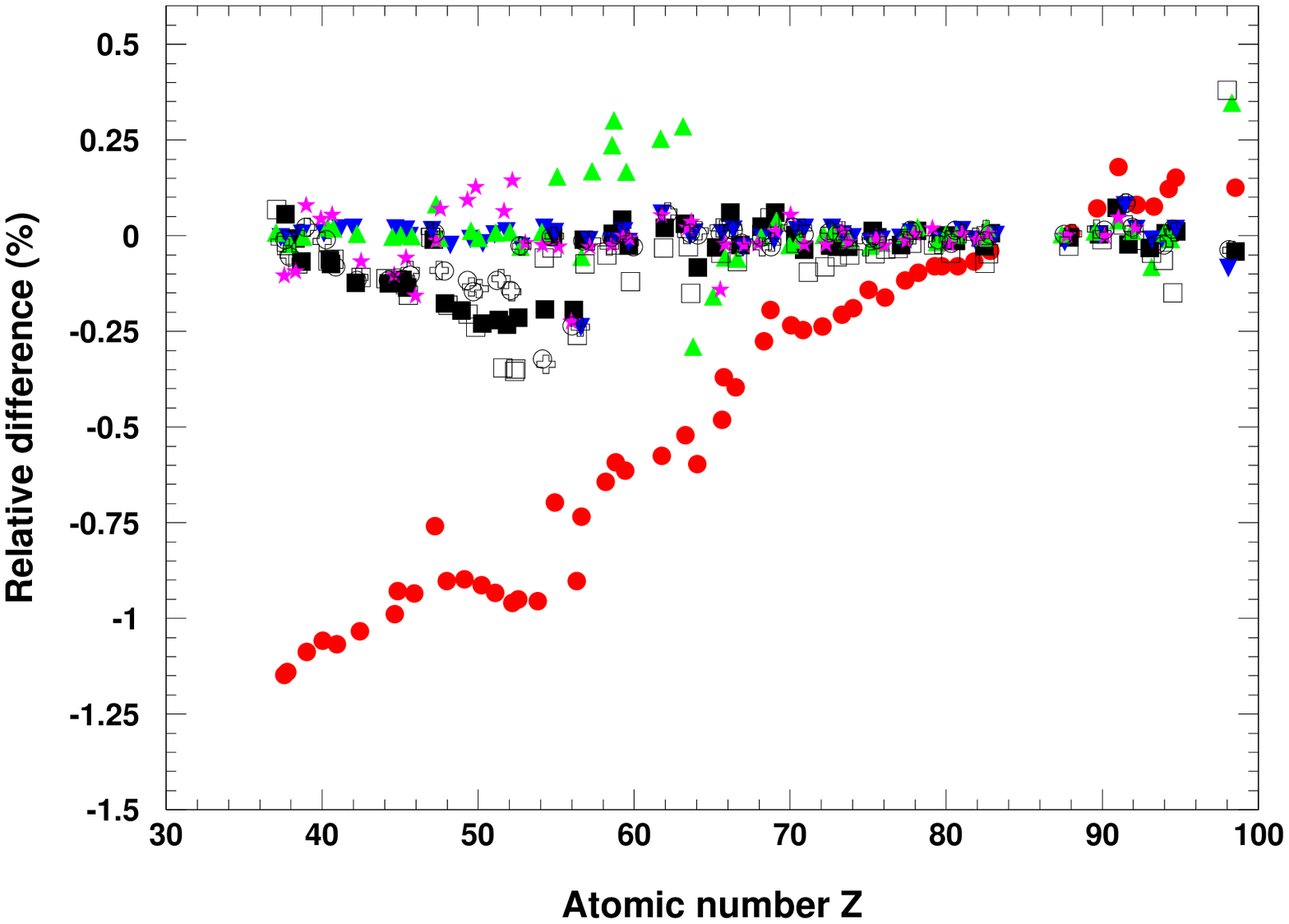}
        }\\ 
    \end{center}
    \caption{%
       L$_1$-shell transitions, 
relative difference between binding energies in various compilations and
experimental data from \cite{deslattes} versus atomic number: EADL (red circles),
Carlson (blue up triangles), Table of Isotopes 1996 (black squares), Table of
Isotopes 1978 (green down triangles), Williams (pink stars), Sevier
1979 (turquoise asterisks) and G4AtomicShells (empty squares).
     }%
   \label{fig_trans_l1}
\end{figure}
\begin{figure}[ht!]
    \begin{center}
        \subfigure[L$_2$M$_1$]{%
            \label{fig_l2m1}
            \includegraphics[width=0.5\textwidth]{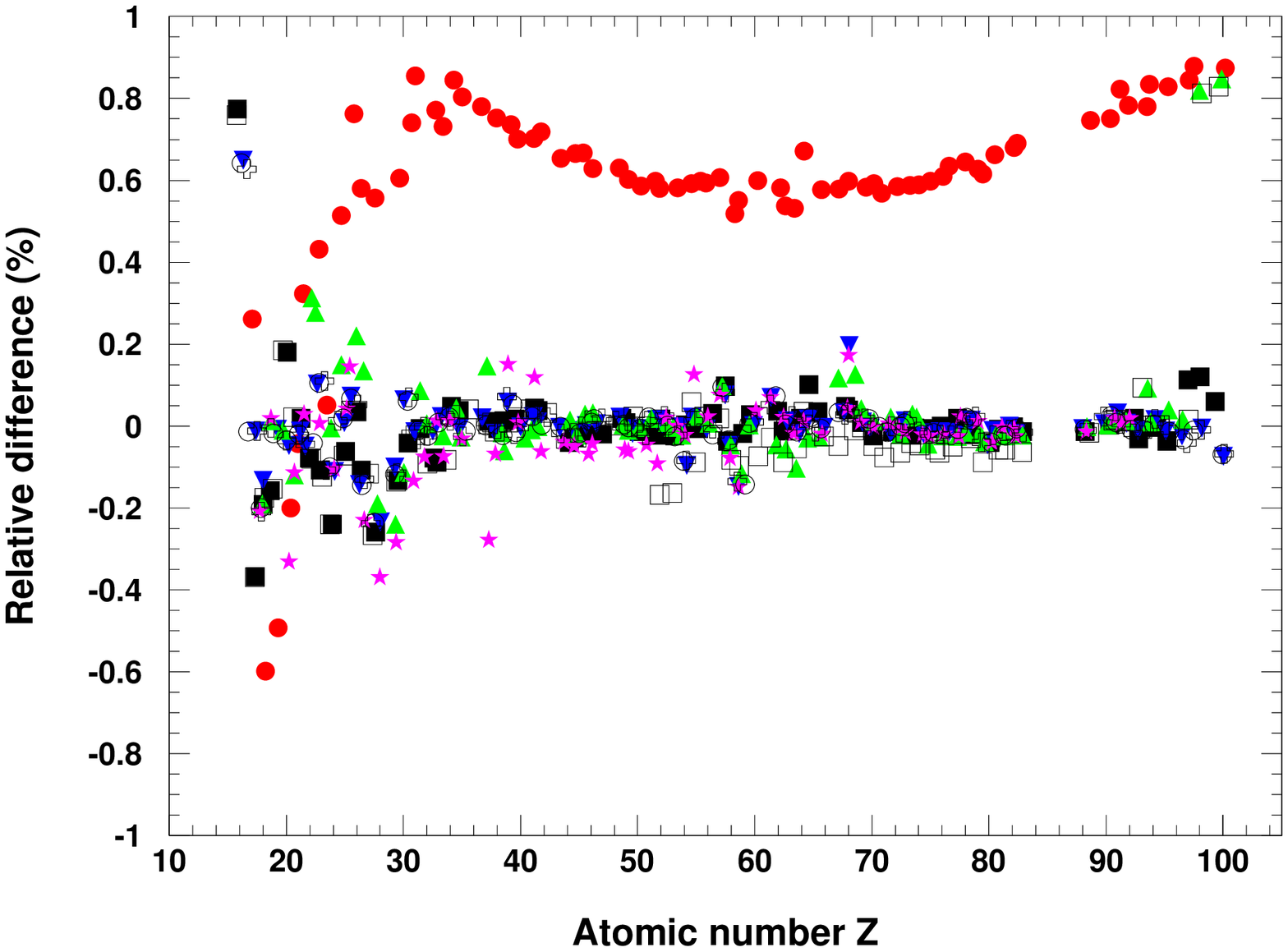}
        }%
        \subfigure[L$_2$N$_4$]{%
           \label{fig_l2n4}
           \includegraphics[width=0.5\textwidth]{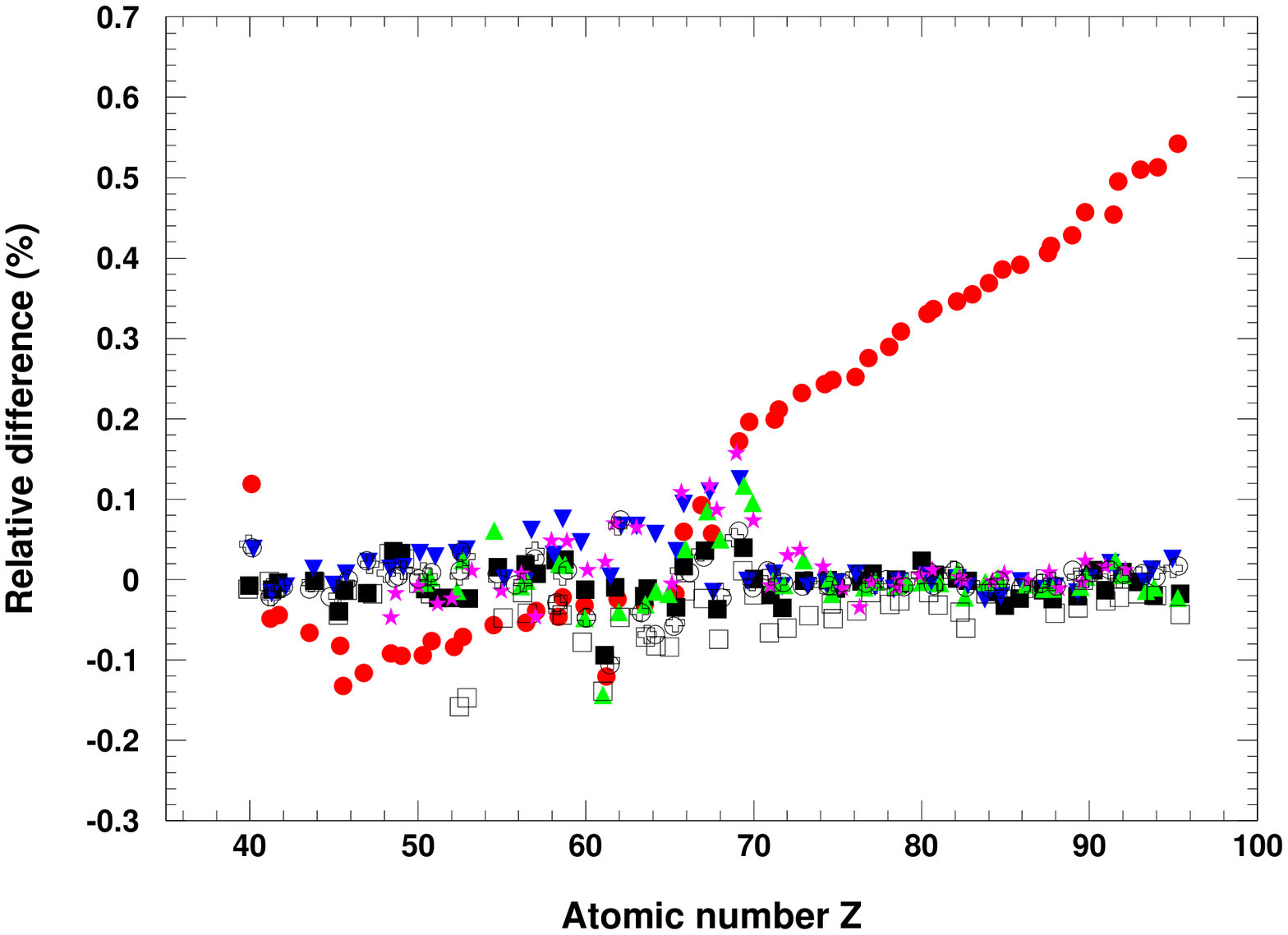}
        }\\ 
    \end{center}
    \caption{%
       L$_2$-shell transitions, 
relative difference between binding energies in various compilations and
experimental data from \cite{deslattes} versus atomic number: EADL (red circles),
Carlson (blue up triangles), Table of Isotopes 1996 (black squares), Table of
Isotopes 1978 (green down triangles), Williams (pink stars), Sevier
1979 (turquoise asterisks) and G4AtomicShells (empty squares).
     }%
    \label{fig_trans_l2}
\end{figure}
\begin{figure}[ht!]
    \begin{center}
        \subfigure[L$_3$M$_5$]{%
            \label{fig_l3m5}
            \includegraphics[width=0.5\textwidth]{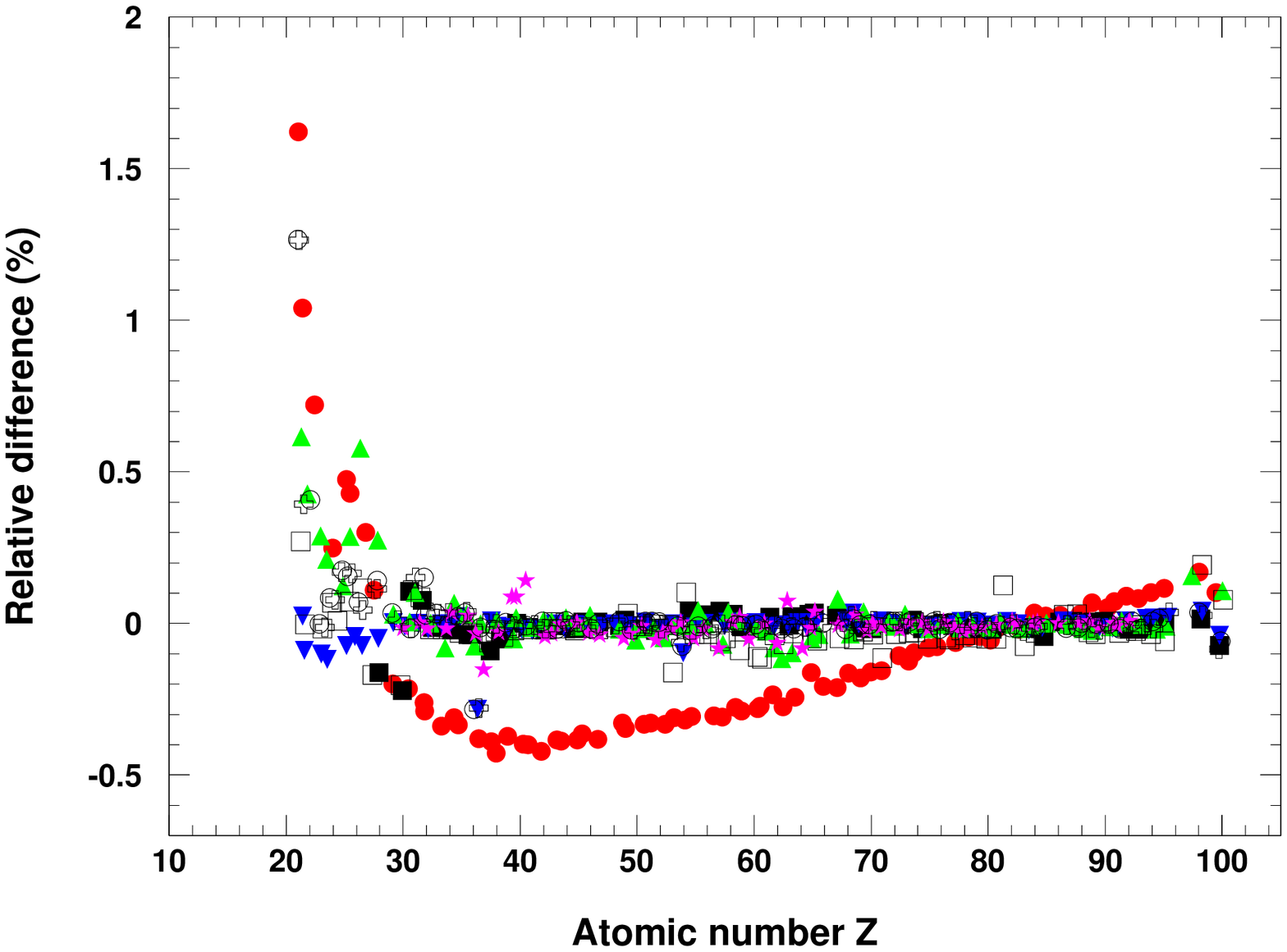}
        }%
        \subfigure[L$_3$N$_1$]{%
           \label{fig_l3n4}
           \includegraphics[width=0.5\textwidth]{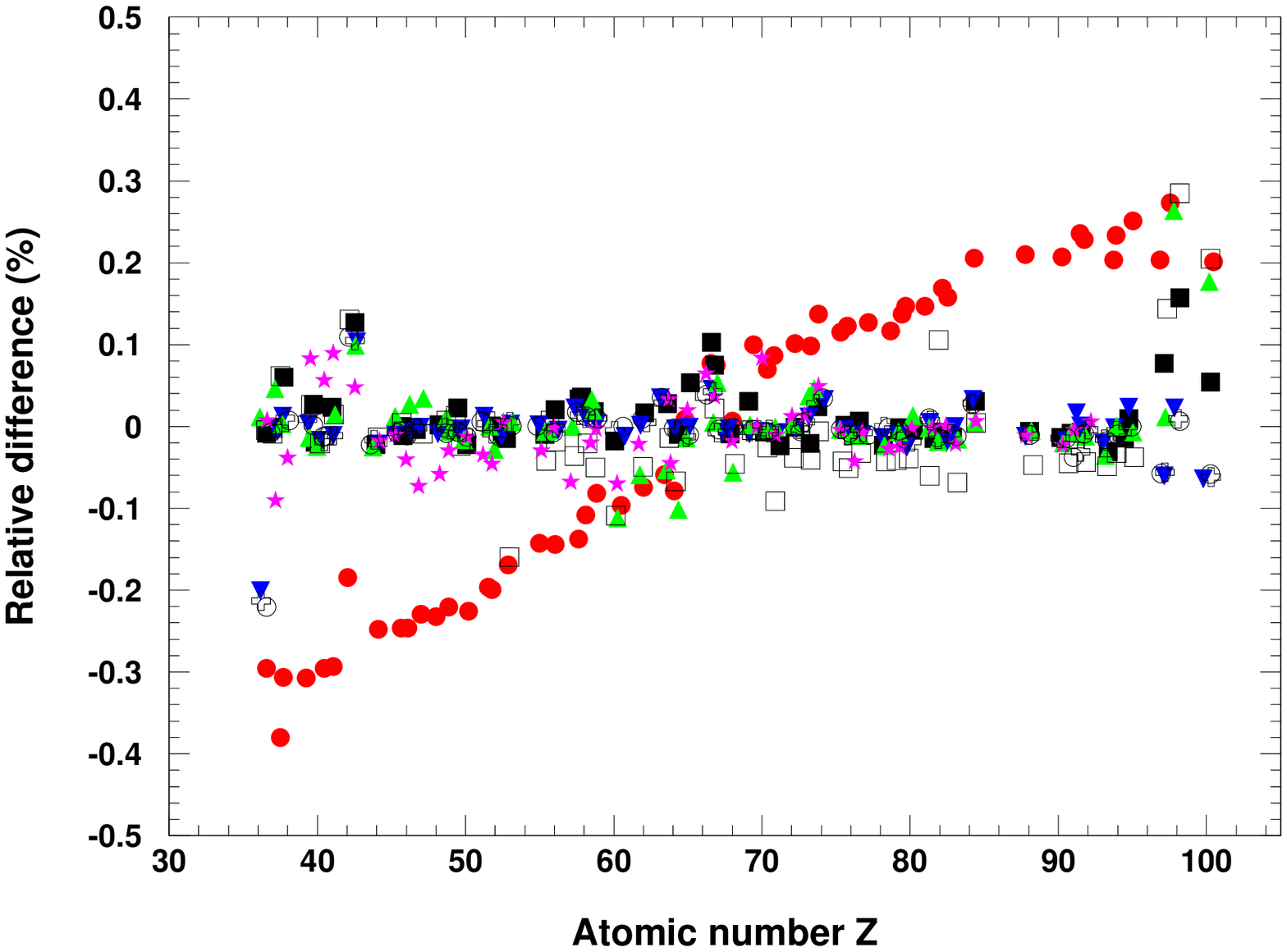}
        }\\ 
    \end{center}
    \caption{%
       L$_3$-shell transitions, 
relative difference between binding energies in various compilations and
experimental data from \cite{deslattes} versus atomic number: EADL (red circles),
Carlson (blue up triangles), Table of Isotopes 1996 (black squares), Table of
Isotopes 1978 (green down triangles), Williams (pink stars), Sevier
1979 (turquoise asterisks) and G4AtomicShells (empty squares).
     }%
    \label{fig_trans_l3}
\end{figure}

A selection of representative plots of the relative difference between
calculated X-ray energies and the experimental data of \cite{deslattes} is shown
in figures \ref{fig_trans_k}-\ref{fig_trans_l3}; X-ray energies are calculated from the
various compilations of binding energies.
It is evident from the plots that the energies calculated by EADL appear less
accurate than those based on the other compilations.
Nevertheless, as already found in \cite{relax_nist}, the discrepancies of the
energies deriving from EADL with respect to measurements are quite small (less
than 2\% in general).

Similarly to what has been discussed in the previous section, goodness-of-fit
tests based on the empirical distribution function are not sensitive to such
small differences: the hypothesis of compatibility between experimental data and
X-ray energies based on EADL (the compilation that is evidently responsible for
the largest discrepancies) is not rejected at 0.1 level of significance
\cite{relax_nist}.

The $\chi^2$ test has limited discriminant power as well, due to the small
uncertainties of the experimental reference data in \cite{deslattes} (less than
0.1 eV for some transitions), which lead to the rejection of the null hypothesis
of compatibility between calculated and experimental X-ray energies
in a large number of test cases.
It is hard to ascertain whether this result of the $\chi^2$ test is due to
underestimated uncertainties for some transitions and elements, or reflects a
realistic conclusion that X-ray energies calculated from binding energy
differences do not achieve the same accuracy by which X-ray energies are
experimentally measured.

Similarly to what was described in the previous section, a t-test was applied to
evaluate whether the distribution of differences between calculated and
experimental X-ray energies is compatible with a true mean of zero.
For each transition the t-test was performed over all the elements for which
experimental values are reported in\cite{deslattes}; the fraction of test cases
for each transition for which the hypothesis of compatibility is not rejected
with 0.05 significance is listed in Table \ref{tab_xttest}.

\begin{table}
\begin{center}
\caption{Number of test cases compatible at 0.05 significance level with mean null
difference between calculated and experimental X-ray energies.}
\label{tab_xttest}
\begin{tabular}{|l|c|ccc|c|c|}
\hline
{\bf Compilation} 	& {\bf K}	& {\bf L$_1$}	& {\bf L$_2$}	& {\bf L$_3$} 	& {\bf K$+$L}	& {\bf Fraction}\\
\hline
Bearden and Burr	& 10		& 12			& 11			& 11			& 44 			& 0.92 $\pm$ 0.04 \\
Carlson		& 7		& 11			& 11			& 9			& 38  		& 0.79 $\pm$ 0.06 \\ 
EADL			& 8		& 4			& 1			& 3 			& 16  		& 0.33 $\pm$ 0.07 \\
Sevier 1979		& 10		& 9	   		& 12			& 9 			& 40  		& 0.83 $\pm$ 0.05 \\
ToI 1978		& 7		& 11			& 8			& 9 			& 35  		& 0.76 $\pm$ 0.06 \\
ToI 1996		& 9		& 12			& 11			& 12			& 44  		& 0.92 $\pm$ 0.04 \\
Williams		& 8		& 12			& 11			& 9 			& 40  		& 0.85 $\pm$ 0.05 \\
\hline
\end{tabular}
\end{center}
\end{table}

The largest number of test cases where the hypothesis of compatibility with null
average difference is not rejected with 0.05 significance is achieved by the
compilations of the 1996 Table of Isotopes and Bearden and Burr's review (44
out of 48 test cases).

The hypothesis whether the compatibility of the other binding energy
compilations with zero mean is equivalent to the one achieved by the 1996 Table
of Isotopes and Bearden and Burr's review was tested by means of contingency
tables.
Contingency tables were built by counting in how many t-test cases the
rejection of the null hypothesis occurs, or does not occur; these counts are
respectively identified as "fail" or "pass".
The results concerning K and L shells are summed to obtain a larger sample size.
They were analyzed by means of Fisher's exact test \cite{fisher}, Pearson's
$\chi^2$ test (whenever the number of entries in each cell justifies the use of
this test) and the $\chi^2$ test with Yates continuity correction \cite{yates}.
The contingency table concerning the comparison of EADL and the 1996 Table of
Isotopes is reported in Table \ref{tab_contx}, along with the p-values of the
three tests applied to it.

\begin{table}
\begin{center}
\caption{Contingency table associated with the t-test :
applied to X-ray energies derived from EADL and from the 1996 Table of Isotopes.}
\label{tab_contx}
\begin{tabular}{|l|c|c|c|c|}
\hline
{\bf $\chi^2$ test outcome}		&{\bf ToI 1996} 		& {\bf EADL}	\\
\hline
Pass					& 44		& 16 \\
Fail					& 4		& 32 \\
\hline
p-value Fisher test			& \multicolumn{2}{|c|}{$< 0.0001$} \\
p-value Pearson $\chi^2$		& \multicolumn{2}{|c|}{not applicable} \\
p-value Yates $\chi^2$		& \multicolumn{2}{|c|}{$< 0.0001$} \\
\hline
\end{tabular}
\end{center}
\end{table}

The hypothesis of equivalence with respect to the results of the t-test is
rejected with 0.05 significance for EADL; it is not rejected for Carlson's,
Sevier's and Williams' compilations.
The outcome of the tests is controversial for the contingency table concerning
the 1996 and 1978 editions of the Table of Isotopes: the p-values are 0.050 for
Fisher's exact test, 0.039 for Person's $\chi^2$ test and 0.075 for the $\chi^2$
with Yates' continuity correction.
The compatibility between EADL and the the 1996 Table of Isotopes is excluded
even if a looser 0.01 significance for the rejection of the null hypothesis is
set both in the t-test and in the contingency tables.


%
%
%
%

%

%

%

\begin{figure}[ht!]
    \label{fig_kdiff}
    \begin{center}
        \subfigure[KL$_3$]{%
            \label{fig_kl3diff}
            \includegraphics[width=0.5\textwidth]{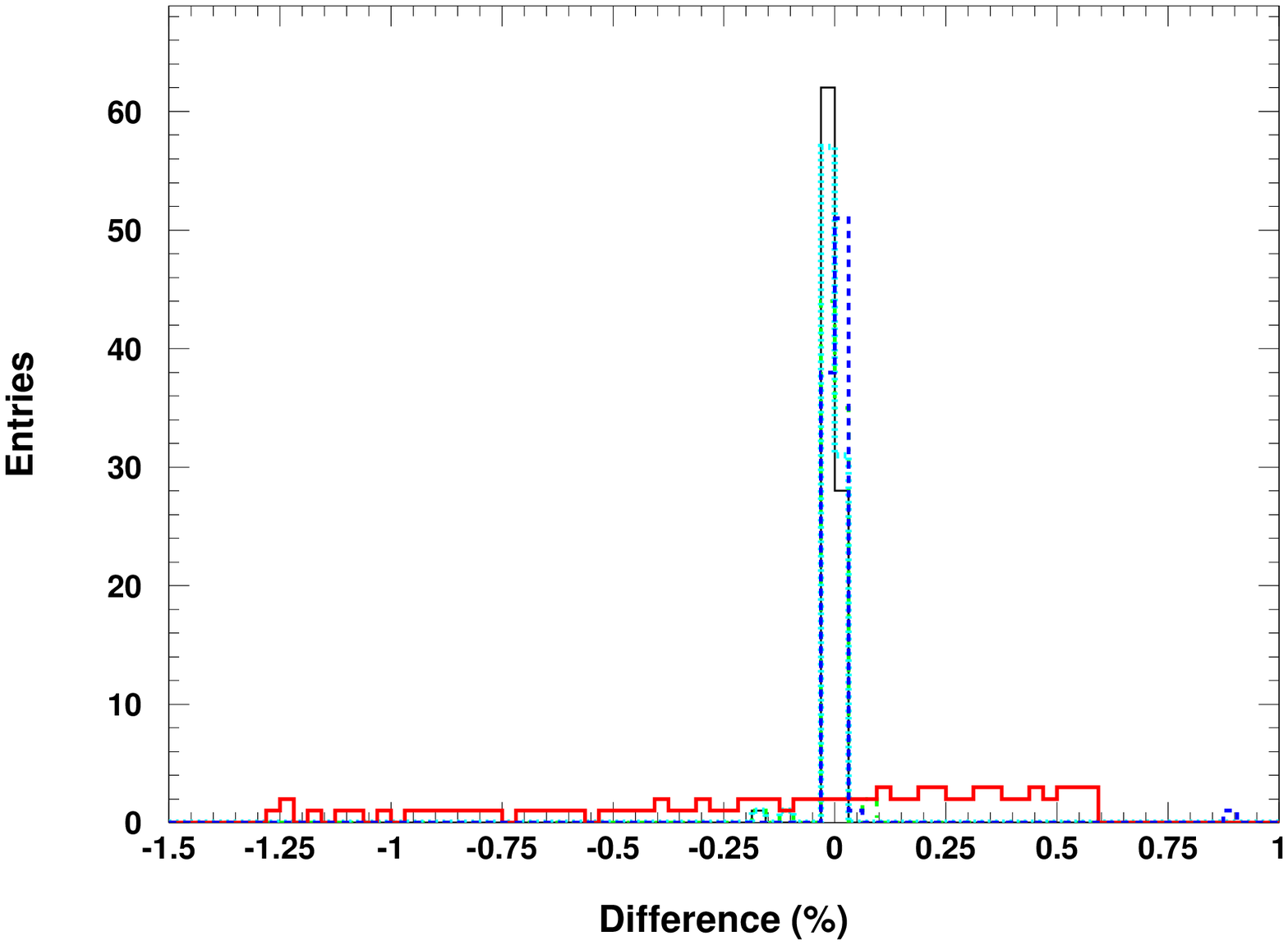}
        }%
        \subfigure[KL$_3$]{%
           \label{fig_km3diff}
           \includegraphics[width=0.5\textwidth]{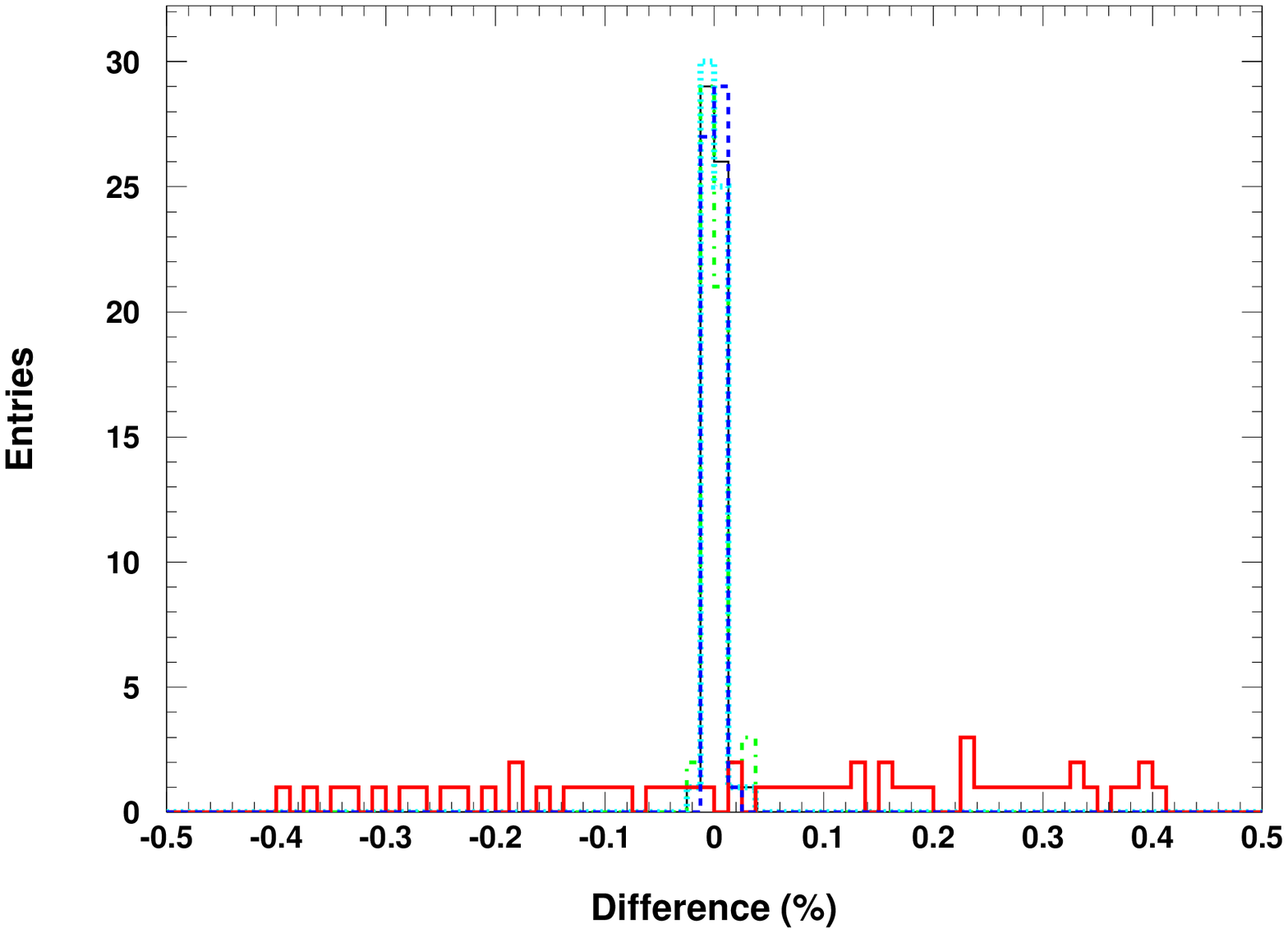}
        }\\ 
    \end{center}
    \caption{%
       -shell transitions, 
relative difference between binding energies in various compilations and
experimental data from \cite{deslattes}: EADL (thick solid red line),
Carlson (dashed blue line), Table of Isotopes 1996 (thin solid black line), 
Williams (dash-dotted green line), Sevier 1979 (dotted turquoise
line); the results of the other compilations considered in this study, which are
not shown, exhibit a narrow distribution similar to the other compilations,
except EADL.
     }%
\end{figure}

\begin{figure}[ht!]
     \begin{center}
        \subfigure[L$_1$M$_2$]{%
            \label{fig_l1m2diff}
            \includegraphics[width=0.5\textwidth]{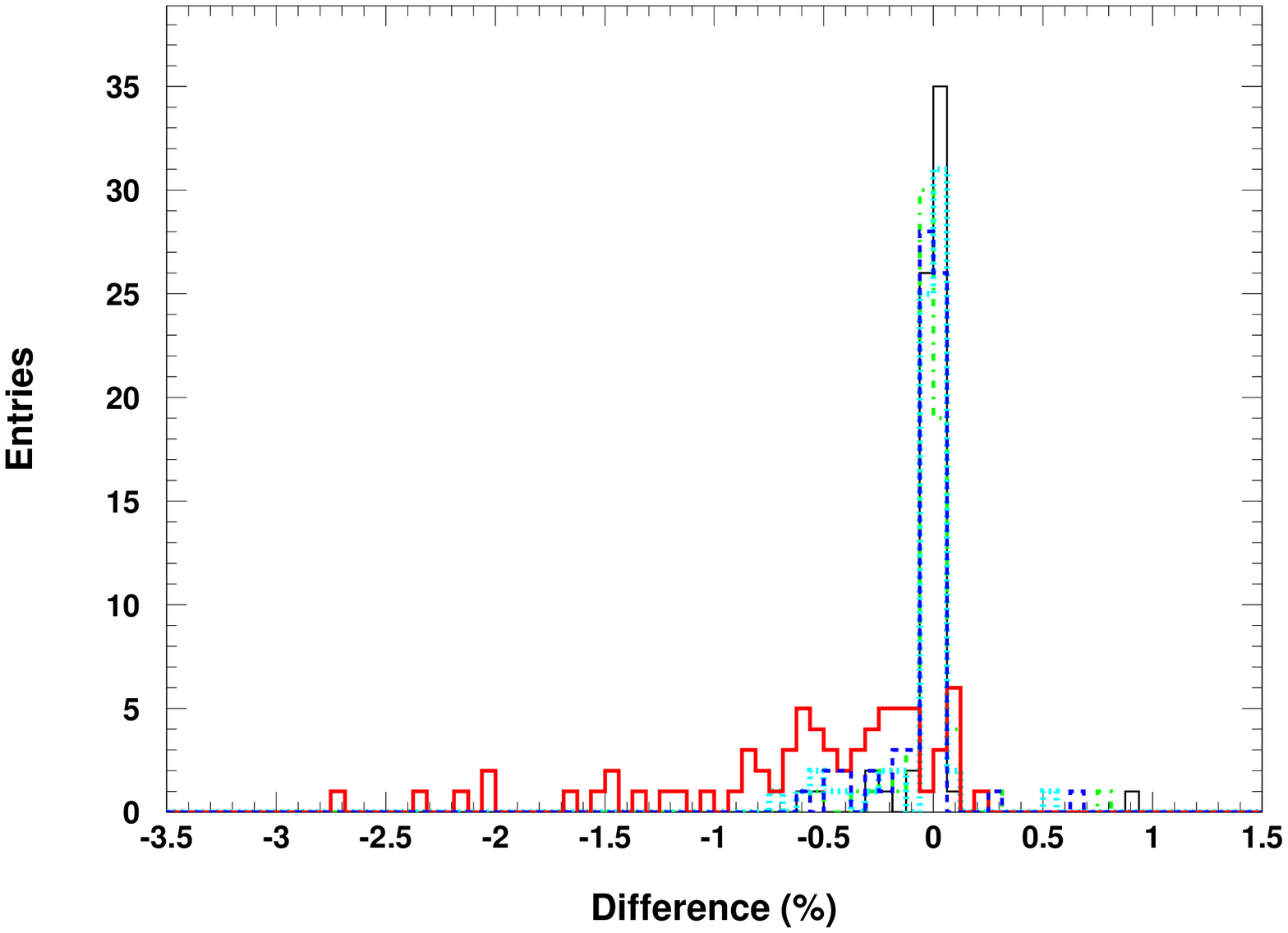}
        }%
        \subfigure[L$_2$M$_1$]{%
           \label{fig_l2m1diff}
           \includegraphics[width=0.5\textwidth]{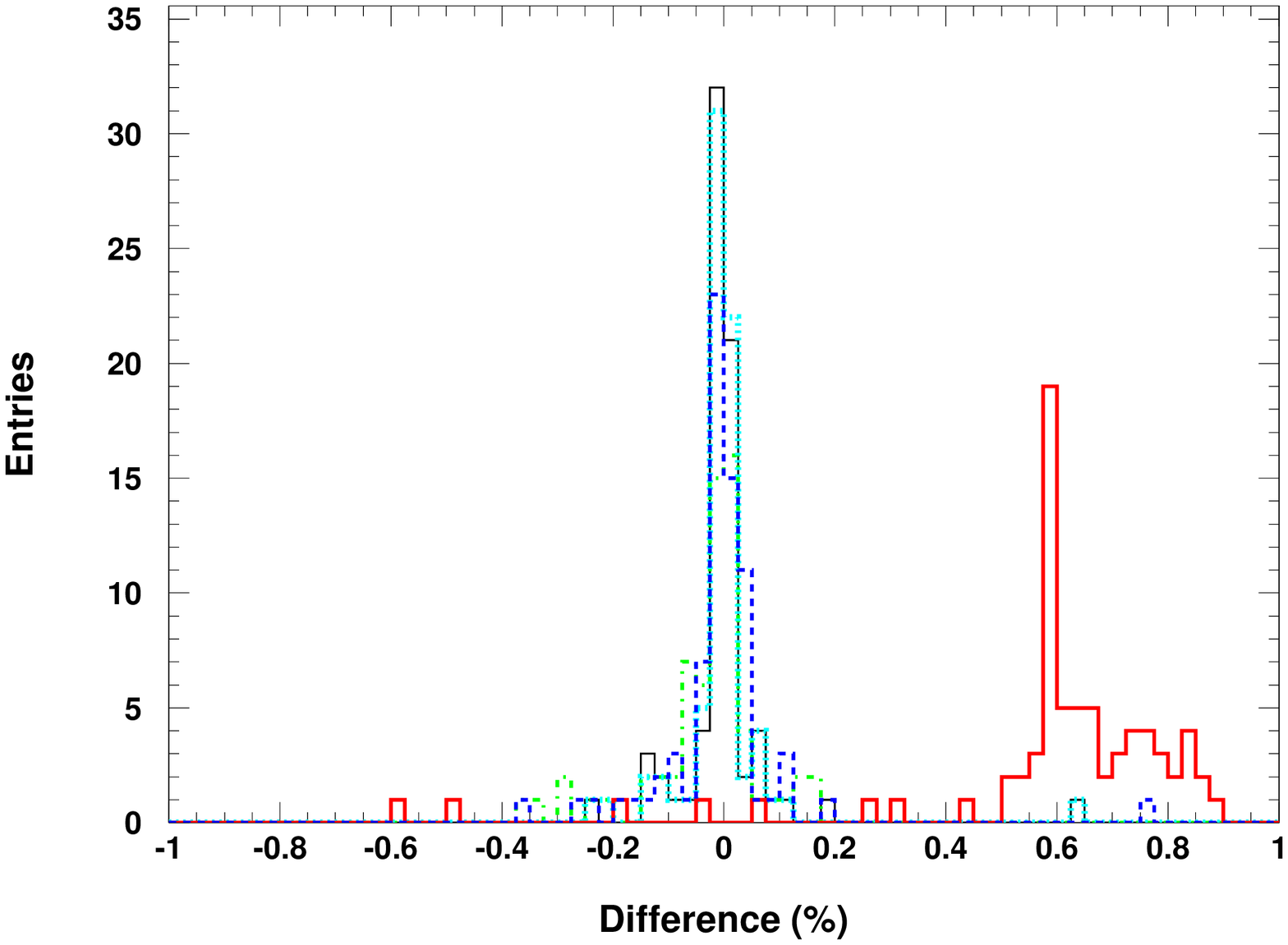}
        }\\ 
       \subfigure[L$_3$N$_5$]{%
            \label{fig_l3n5diff}
            \includegraphics[width=0.5\textwidth]{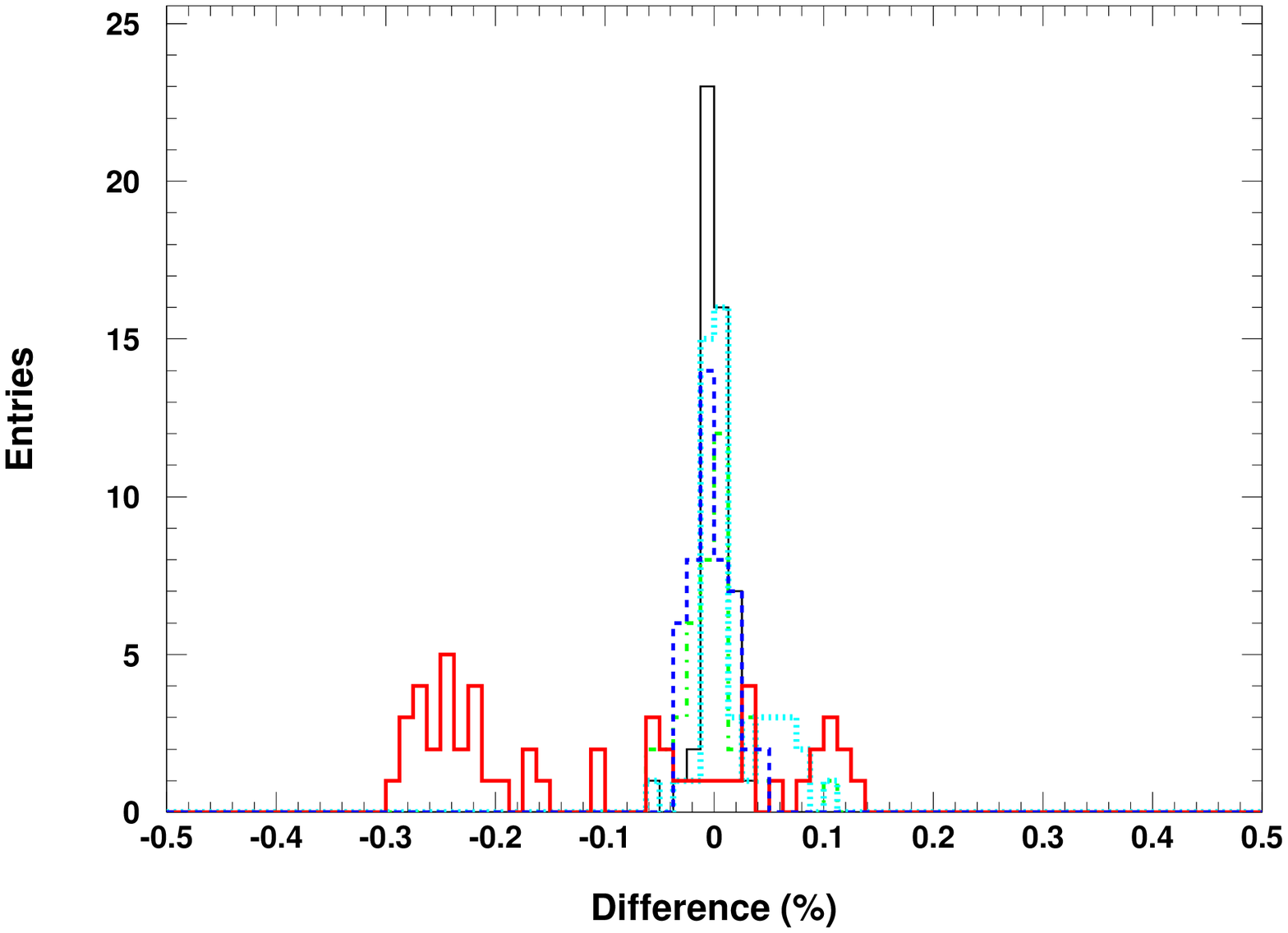}
        }%
    \end{center}
    \caption{%
       L-shell transitions, 
relative difference between binding energies in various compilations and
experimental data from \cite{deslattes}: EADL (thick solid red line),
Carlson (dashed blue line), Table of Isotopes 1996 (thin solid black line), 
Williams (dash-dotted green line), Sevier 1979 (dotted turquoise
line); the results of the other compilations considered in this study, which are
not shown, exhibit a narrow distribution similar to the other compilations,
except EADL.
     }%
   \label{fig_ldiff}
\end{figure}

The distribution of the difference between the X-ray energies calculated from
binding energy tabulations and the experimental values of \cite{deslattes} is
wider for EADL than for all the other compilations; this result can be
appreciated in a few representative plots (figures
\ref{fig_kdiff}-\ref{fig_ldiff}).

%

The equivalence of the variance of the differences between calculated and
experimental X-ray energies was estimated by means of the F-test.
For each transition, the variance of the corresponding data sample was compared
to the variance associated with the 1996 Table of Isotopes, which exhibits the
narrowest distribution of differences between calculated and experimental X-ray
energies.
The hypothesis of equivalence of the variances under test was rejected if the
p-value from the F-test was smaller than 0.01.
The fraction of transitions for which the outcome of the F-test indicates that
there is no significant difference in the respective variances is listed in
Table \ref{tab_ftest}.

\begin{table}
\begin{center}
\caption{Fraction of K and L transitions for which the variance 
of the difference between calculated and experimental
X-ray energies is equivalent to the variance associated 
with the 1996 Table of Isotopes.}
\label{tab_ftest}
\begin{tabular}{|l|c|}
\hline
{\bf Compilation}			&{\bf Fraction of transitions}		\\
\hline
Bearden and Burr			&0.60	$\pm$ 0.07 \\
Carlson				&0.69	$\pm$ 0.07	\\				
EADL					&0.25	$\pm$ 0.06 \\
Sevier 1979				&0.79	$\pm$ 0.06 \\
ToI 1978    				&0.65	$\pm$ 0.07 \\
Williams				&0.70	$\pm$ 0.07 \\
\hline
\end{tabular}
\end{center}
\end{table}

The results of the F-test are consistent with the qualitative appraisal of the
accuracy of the distributions in figures \ref{fig_kdiff}-\ref{fig_ldiff}.
It is worthwhile to recall that the F-test is sensitive to the normality of the
distributions to which is applied; although the differences between calculated
and experimental data are expected to be normally distributed, the results
reported in Table \ref{tab_ftest} may be affected by some details of the
distributions subject to comparison.

The analysis of X-ray energies suggests that better accuracy in the reproduction
of K and L transition energies can be achieved by binding energy
compilations other than EADL.
A comment on the accuracy of EADL transition energies of the M series, mentioning
deviations from experiment up to 10 MeV, is reported in the X-Ray Data Handbook
\cite{zschornack} with the recommendation of using other experimental data
preferably; nevertheless, no comprehensive, quantitative demonstration of EADL
accuracy is reported, nor references are cited in support of the appraisal of
EADL accuracy and consequent recommendation.

The original design of Geant4 atomic relaxation described in \cite{tns_relax}
would easily accommodate the improvement of the accuracy of the simulated
energies through alternative binding energy options: the software implementation 
would handle the process of atomic relaxation transparently, if a different
tabulation of binding energies is supplied as an external file.

\section{Effects on ionization cross sections}

Some analytical formulations of cross sections for the ionization of atoms by
charged particle impact involve atomic binding energies.
Two of these models are considered in this study to ascertain whether different
binding energy compilations would produce significant differences in the cross
section values: the Binary-Encounter-Bethe model (BEB) \cite{beb1994} for
electron impact ionization and the ECPSSR (Energy Loss Coulomb Repulsion
Perturbed Stationary State Relativistic) model \cite{ecpssr} for proton impact
ionization.
For both models the effects on the accuracy of the cross section calculations
are quantitatively estimated through a comparison with experimental data.

\subsection{Electron impact ionization cross sections}

Two models of electron impact cross sections, the Binary-Encounter-Bethe 
\cite{beb1994} model and the Deutsch-M\"ark model \cite{dm1987}, have been
designed, implemented and validated in view of extending and improving Geant4
simulation capabilities in the energy range below 1 keV.
Their features, verification and validation are briefly summarized in
\cite{beb_mc2010,beb_nss2010} and extensively documented in a dedicated paper
\cite{tns_beb}.
The first software development cycle has been focussed on modeling total
ionization cross sections; the validation process and the analysis of the effect
of atomic binding energies concern these calculations, although the BEB model
has the capability of calculating cross sections for the ionisation of
individual shells.

The BEB cross section for the ionization of subshell \textit{i} is given by:
\begin{equation}
 \sigma_{i} = \frac{S}{t+(u+1)/n}\left[\frac{\text{log}(t)}{2}\left(1-\frac{1}{t^2}\right)
+1-\frac{1}{t}-\frac{\text{log}(t)}{t+1} \right]
\label{eq_beb}
\end{equation}
where:
\begin{equation}
 t=\frac{T}{B}, ~~~~~ u=\frac{U}{B}, ~~~~~ S=4\pi a^2_0 N \left(\frac{R}{B}\right)^2
\label{eq_bebsub}
\end{equation}
In the above equations \textit{B} is the electron binding energy, \textit{N} is
the the occupation number, \textit{T} is the incident electron energy,
\textit{U} is the average electron kinetic energy, \textit{t} and \textit{u} are
normalized incident and kinetic energies, $n$ is the principal quantum number
(only taken into account when larger than 2), \textit{a$_0$} is the Bohr radius
and \textit{R} is the Rydberg constant.
The sum over all the subshells \textit{i} of an atom gives the total (counting)
cross section; in practice, only the valence shell and a few outer subshells
contribute significantly to determine the cross section value.

The original BEB cross sections \cite{beb1994} used binding energy values 
calculated by the authors of the model.
Only a few of those values are documented in \cite{beb1994}; they were utilized in
the software verification process to assess the correctness of the
implementation, but such a small set is inadequate for using the model in a
general purpose simulation system, which must be able to calculate cross
sections for any target atoms.
The BEB model developed for use with Geant4 utilizes a full set of binding
energies and provides the option of accessing alternative compilations.

The analysis addressed two issues: the sensitivity of cross sections to the
values of the binding energies used in the calculation, and the evaluation of
the accuracy with respect to experimental data.

\begin{figure}[ht!]
    \begin{center}
        \subfigure[Nitrogen]{%
            \label{fig_beb7}
            \includegraphics[width=0.5\textwidth]{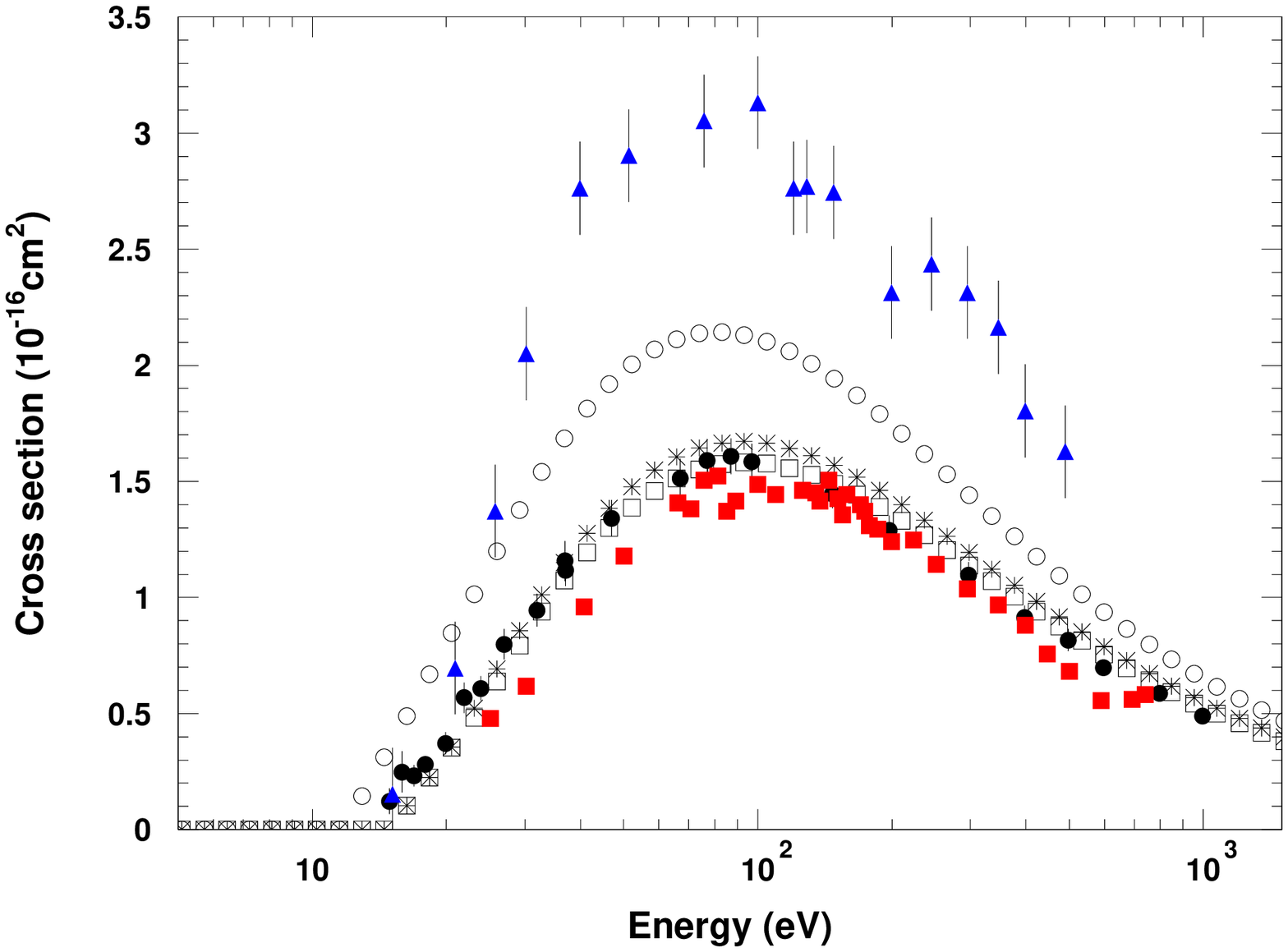}
        }%
        \subfigure[Germanium]{%
           \label{fig_beb32}
           \includegraphics[width=0.5\textwidth]{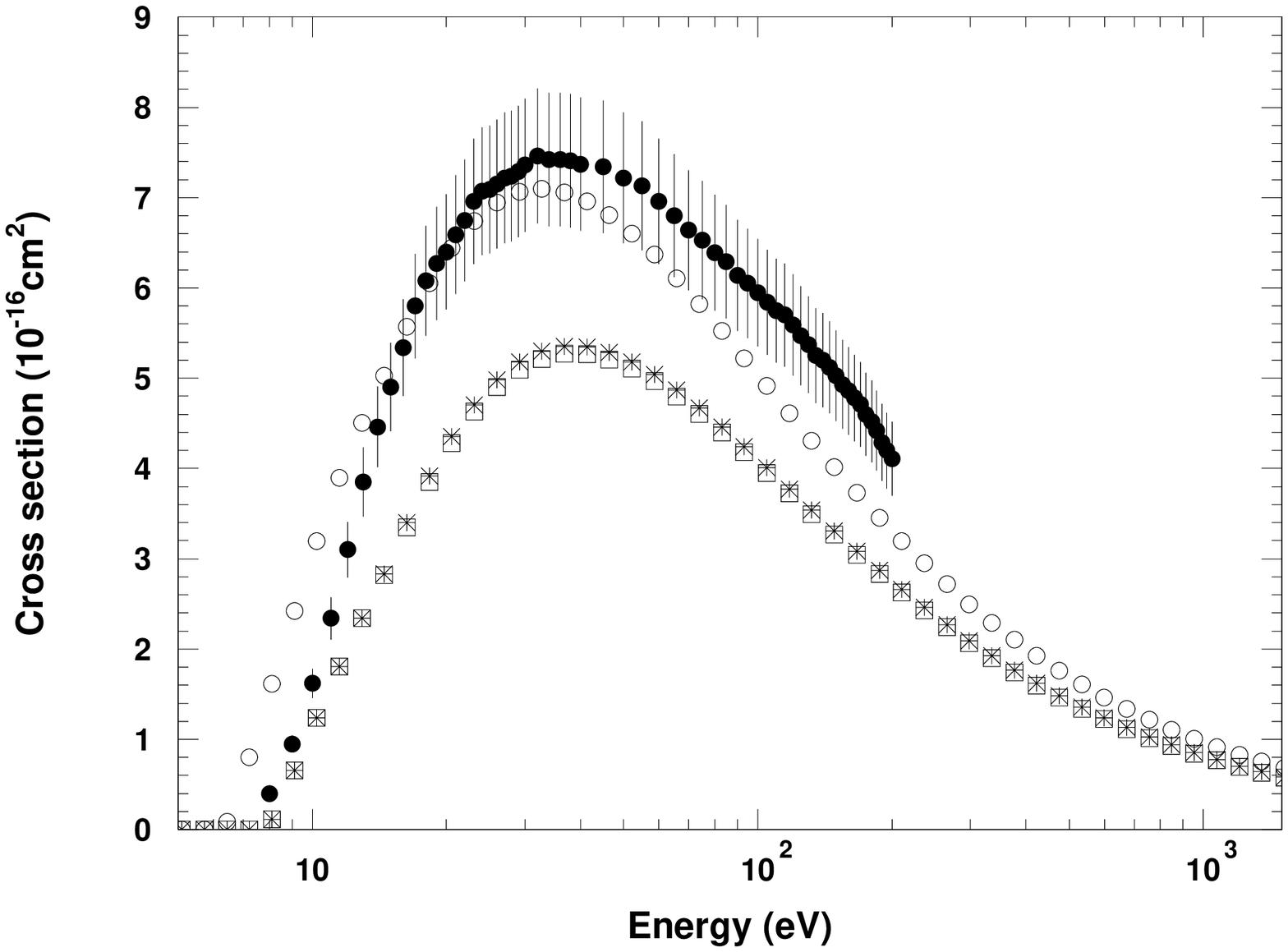}
        }\\ 
    \end{center}
    \caption{%
       BEB electron impact ionization cross section,
BEB model with EADL binding energies for all shells (empty circles), 
BEB model with EADL binding energies except for ionization energies
replaced by NIST values (empty squares), 
BEB model with Lotz binding energies (asterisks)
and experimental data:
(a) from 
\cite{expBrook1978} (red squares),
\cite{smith} (black circles) and
\cite{expPeterson} (green stars),
(b) from \cite{expFreund} (black circles). 
     }%
    \label{fig_beb}
\end{figure}

Two examples of the effects of different binding energies on the calculated
cross sections are shown in figure \ref{fig_beb}.
They illustrate three options of binding energies: Lotz's compilation, which is
also used by the Deutsch-M\"ark model, EADL data for all shells and EADL with
ionization potentials replaced by NIST values \cite{nist_ionipot} (identified in the following as
"modified EADL").
Lotz's compilation is identical to Carlson's apart from a few exceptions;
according to the analysis in section \ref{sec_ionipot}, Carlson's compilation
appears the most accurate with respect to NIST ionization potentials, while EADL
exhibits the largest differences with respect to other compilations in both
inner and outer-shell binding energies.
Significant differences are visible in the cross sections, when different
ionization energies are used in the calculation, while different inner-shell
binding energies appear to have relatively small effects.

%



The effect of these three options, which can be considered as extreme
alternatives in the BEB calculation, has been quantified through a statistical
analysis.

First, cross sections calculated with different binding energies were compared
via goodness-of-fit tests; the test concerned all elements with atomic number
between 1 and 92, and incident electron energies from 1 eV to 10 keV, 
divided in two ranges: those up to 100 eV, and those above.
The results of the comparisons are summarized in Tables \ref{tab_beblotz} and
\ref{tab_bebnist}.
The Kolmogorov-Smirnov test appears the most sensitive to differences in the
cross sections deriving from the considered binding energy options.
The hypothesis of compatibility between cross sections calculated with
Lotz's binding energies and with modified EADL is rejected with 0.05
significance only for a few heavy elements (with Z$>$80) in the lower energy
range (below 100 eV): this result indicates that total ionization cross sections 
are marginally affected by inner-shell binding energies.
The hypothesis of compatibility between the cross sections based on EADL and
modified EADL is rejected in a larger number of test cases, especially
in the lower energy range: this result shows that
the cross sections are sensitive to the values of the ionization potential.

\begin{table}
\begin{center}
\caption{Test cases in which the hypothesis of compatibility of BEB cross sections
based on Lotz and modified EADL binding energies is not rejected.}
\label{tab_beblotz}
\begin{tabular}{|l|c|c|}
\hline
{\bf Test}			&{\bf Fraction} 	&{\bf Fraction} \\
				&(E $<$ 100 eV) 			&(E $>$ 100 eV) 	\\
\hline
Kolmogorov-Smirnov	&0.91	$\pm$ 0.05 	&1.00	         $-$ 0.02 \\
Anderson-Darling		&0.97	$\pm$ 0.03	&1.00	         $-$ 0.02 \\				
Cramer-von Mises  	&0.97	$\pm$ 0.03 	&1.00	         $-$ 0.02 \\
\hline
\end{tabular}
\end{center}
\end{table}

\begin{table}
\begin{center}
\caption{Test cases in which the hypothesis of compatibility of BEB cross sections
based on EADL and modified EADL is not rejected.}
\label{tab_bebnist}
\begin{tabular}{|l|c|c|}
\hline
{\bf Test}			&{\bf Fraction} 	&{\bf Fraction} \\
				&(E $<$ 100 eV) 			&(E $>$ 100 eV) 	\\
\hline
Kolmogorov-Smirnov	&0.70	$\pm$ 0.05 	&0.98	$\pm$ 0.02 \\
Anderson-Darling		&0.83	$\pm$ 0.04	&0.98	$\pm$ 0.02 \\				
Cramer-von Mises  	&0.95	$\pm$ 0.02 	&1.00	$-$ 0.02 \\
\hline
\end{tabular}
\end{center}
\end{table}

The following analysis evaluated whether different ionization
potentials would significantly affect the accuracy of the calculated cross
sections with respect to experimental data.
The effects of different ionization energies on the accuracy of the calculation
are not straightforward to ascertain from a qualitative appraisal of the data:
in fact, within the data sample one can identify test cases where either
configuration - with NIST values or with EADL original values - appears to
better reproduce the experimental data as shown, for instance, in figure
\ref{fig_beb}.
Therefore a statistical analysis was performed, examining the compatibility with
experiment of two sets of BEB cross sections, which use respectively EADL
binding energies for all shells, or the modified EADL with NIST ionization
potentials.
The two sets of cross sections were compared to the same experimental measurements
\cite{dmDeutsch2005}-\cite{exp92}, consisting of more than 120 individual data
sets and concerning more than 50 elements.
The comparison with experimental data exploits goodness-of-fit tests
(Kolmogorov-Smirnov, Anderson-Darling, Cramer-von Mises and $\chi^2$); their
significance was set to 0.05.
The test was articulated over five distinct energy ranges below 1 keV to
appraise in detail the accuracy of the calculated cross sections.


The number of test cases for which the null hypothesis of compatibility between
calculated and measured cross sections is rejected, or not rejected, is reported
in Table \ref{tab_bebeadl} for the two examined binding energy options.
Below 1 keV the hypothesis of compatibility with experimental data is
always rejected in a smaller number of test cases when the cross section
calculation utilizes NIST ionization energies instead of EADL original ones.
No difference is observed above 1 keV.

The analysis by means of contingency tables does not reject the hypothesis of
equivalence between the two cross section categories at reproducing experimental
data in any of the considered energy ranges.
Nevertheless, the probability that the better performance associated with NIST
ionization energies in all five trials could be due to chance only is 0.03.

\begin{table}
\begin{center}
\caption{Test cases for which the hypothesis of compatibility of BEB
cross sections with experimental data is rejected, or not rejected, with 0.05
significance.}
\label{tab_bebeadl}
\begin{tabular}{|c|ccc|ccc|}
\hline
{\bf Energy}	&\multicolumn{3}{|c|}{\bf EADL}		&\multicolumn{3}{|c|}{\bf Modified EADL}			\\

(eV)		& Pass	& Fail	&Pass Fraction  	& Pass  & Fail 	&Pass Fraction      \\
\hline					                	                
$<20 $		& 67	& 40	& 0.63$\pm$0.05		& 79    & 28 	&0.74$\pm$0.04 \\
20-50		& 61	& 68	& 0.47$\pm$0.04		& 81    & 48  	&0.63$\pm$0.04 \\
50-100		& 40	& 84	& 0.32$\pm$0.04 	& 49    & 75 	&0.40$\pm$0.04 \\
100-250		& 47	& 80 	& 0.37$\pm$0.04 	& 56    & 71  	&0.44$\pm$0.04 \\
250-1000	& 45	& 31	& 0.59$\pm$0.06 	& 47    & 29  	&0.62$\pm$0.06 \\
$>$1000		& 14 	& 11 	& 0.56$\pm$0.10 	& 14    & 11  	&0.56$\pm$0.10 \\
\hline
\end{tabular}
\end{center}
\end{table}


\subsection{Proton impact ionization cross sections}

A similar study was performed on proton ionization cross sections.
Several cross section models for the computation of inner-shell ionization by
proton and $\alpha$ particle impact have been released in Geant4 version 9.4
\cite{tns_pixe}; they include calculations based on the plane wave Born
approximation (PWBA) \cite{pwba}, the ECPSSR (Energy-loss Coulomb Perturbed Stationary State Relativistic) model \cite{ecpssr} in a number of
variants and a collection of empirical models, deriving from fits to
experimental data.
The PWBA and ECPSSR cross sections (in all their variants) exploit tabulations
produced by the ISICS  (Inner-Shell Ionization Cross Sections) code \cite{isics} for K, L and M shells.

The formulation of the PWBA and ECPSSR cross sections involves atomic binding
energies.
For a given shell the PWBA cross section is given by
\begin{equation}
\sigma_{\text{PWBA}} = \sigma_0 \theta^{-1}  F \left( \frac{\eta}{\theta},\theta \right)
\label{eq_pwba}
\end{equation} 
where:
\begin{equation}
\sigma_{0} = 8 \pi a_{0}^{2} \left( \frac{Z_{1}^{2}}{Z_{2}^4} \right)
\end{equation}
$a_0$ is the Bohr radius, $Z_1$ is the projectile atomic number, $Z_{2}$ is the
effective atomic number of the target atom, $F$ is the reduced universal cross
section, with the reduced atomic electron binding energy $\theta$ and reduced projectile
energy $\eta$ given by
\begin{equation}
\label{eq_theta}
\theta = 2 n^2 \frac{U_2}{Z_2^2}
\end{equation}
and
\begin{equation}
\label{eq_eta}
\eta = 2 \frac{E_1}{M_1 Z_2^2}
\end{equation}
respectively.
In equations  \ref{eq_theta} and \ref{eq_eta}
$E$, $M$ and $U$ represent the energy, mass and 
atomic binding energy.
In the above formulae the indices 1 and 2 refer respectively to the projectile
and the target.
The analytical formulation of the reduced universal cross section $F$ can be
found in \cite{isics}; it involves the reduced atomic electron binding energy.
The ECPSSR cross section for a given shell is expressed in terms of the PWBA value:
\begin{equation}
\sigma_{\text{ECPSSR}} = C_B^E(dq{_0}^{B}\zeta)\sigma_{\text{PWBA}}\left(\frac{m_R\left(\frac{\xi}{\zeta}\right)\eta}{(\zeta\theta)^2},\zeta\theta\right)
\end{equation}
where $C_E^B$ is the Coulomb deflection correction, $\zeta$ is the correction
factor for binding energy and polarization effects, $m_R$ is the relativistic
correction, $q_0$ is the minimum momentum transfer and \begin{equation}
\xi = v_1 \frac{Z_2}{U_2}
\end{equation}
$v_1$ being the projectile velocity.

The PWBA and ECPSSR cross section tabulations distributed with the Geant4 code
were produced with the  ISICS 2008 version, which uses the Bearden and Burr binding
energies.
Recent updates to ISICS \cite{isics2011} offer the option of
using Williams' compilation of binding energies as alternative values to the default
Bearden and Burr's ones;
a further evolution of ISICS \cite{isics_linux} lets the user specify an
arbitrary source of atomic binding energies, thus providing access to any of the
options analyzed in this paper.
This version, which was used to produce the data for this paper, 
involves new implementations of some parts of the ISICS code, which contribute 
to the numerical correctness and computational robustness of the software.
The new features and verification of this new version of ISICS are documented 
in \cite{isics_linux}.
The experimental validation of cross sections generated by this new version of
ISICS produces consistent results with those reported in \cite{tns_pixe}, when
the two code versions are run in the same configuration.

\begin{figure}[ht!]
    \begin{center}
        \subfigure[Carbon]{%
            \label{fig_crossk_6}
            \includegraphics[width=0.5\textwidth]{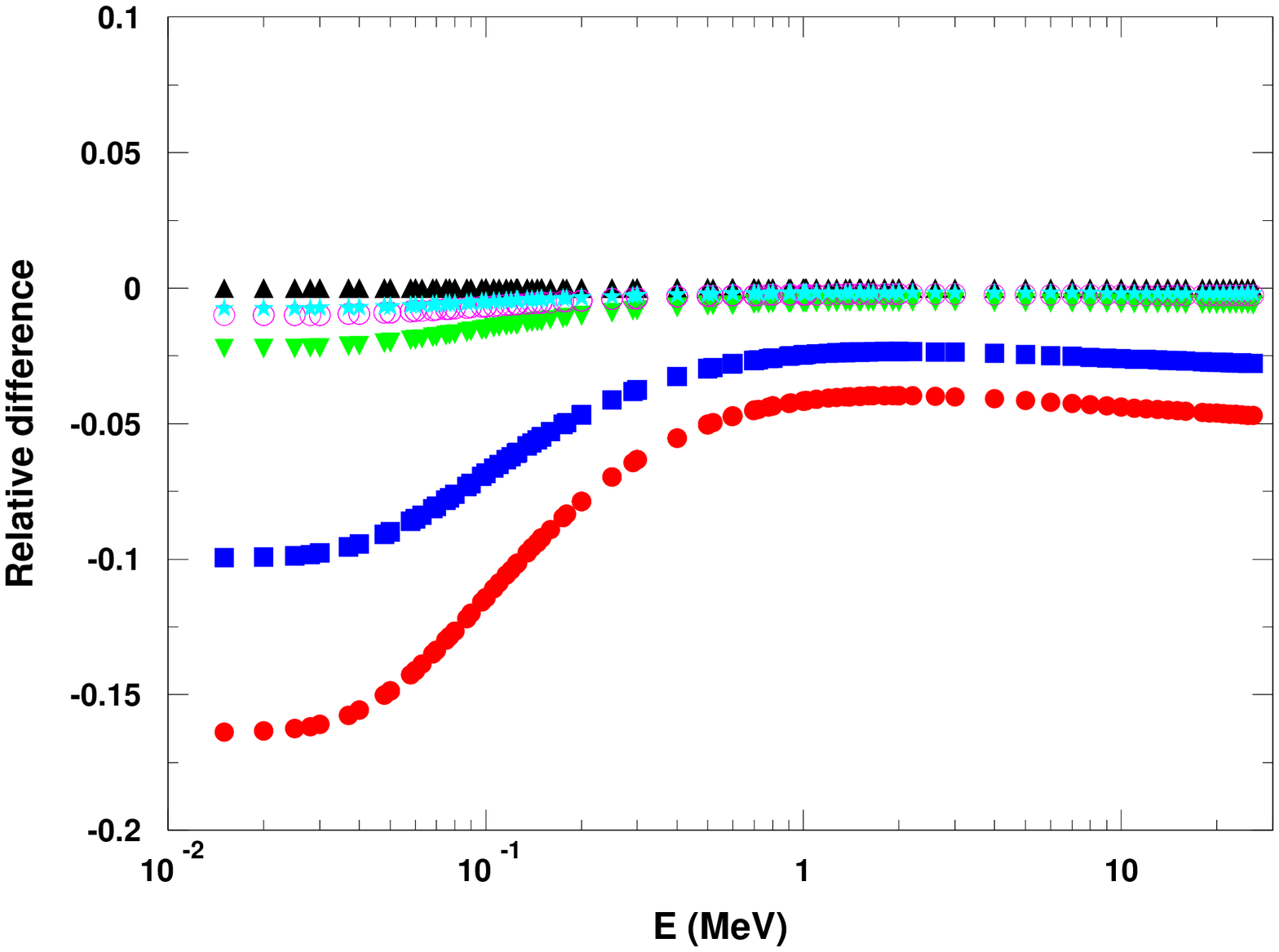}
        }%
        \subfigure[Copper]{%
           \label{fig_crossk_14}
           \includegraphics[width=0.5\textwidth]{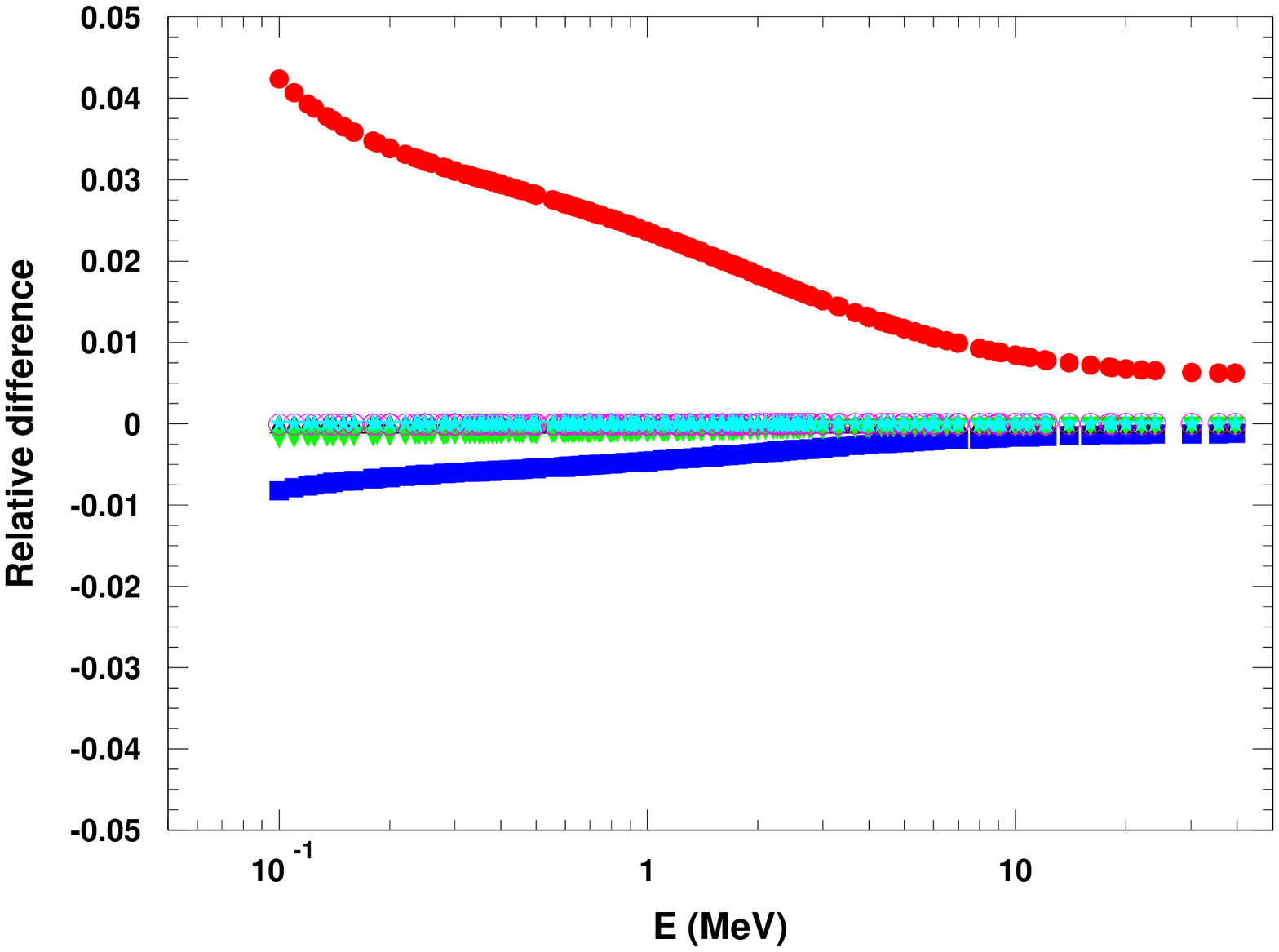}
        }\\ 
    \end{center}
    \caption{%
       K shell ionization cross sections by proton impact on carbon (a) and copper (b) calculated by the
ECPSSR model with different binding energies: the plot shows the relative
difference with respect to the values calculated with Bearden and Burr's binding
energies used by default by ISICS; the symbols identify cross sections calculated with EADL
(red circles), Carlson (blue squares), 1996 Table of Isotopes (black up
triangles), 1978 Table of Isotopes (green down triangles), Williams (pink empty
circles) and Sevier 1979 (turquoise stars) binding energies. Some symbols are
not visible in the plot due to the close values of some cross sections. 
     }%
    \label{fig_crossk}
\end{figure}

\begin{figure}[ht!]
    \begin{center}
        \subfigure[L$_1$]{%
            \label{fig_crossl1_48}
            \includegraphics[width=0.5\textwidth]{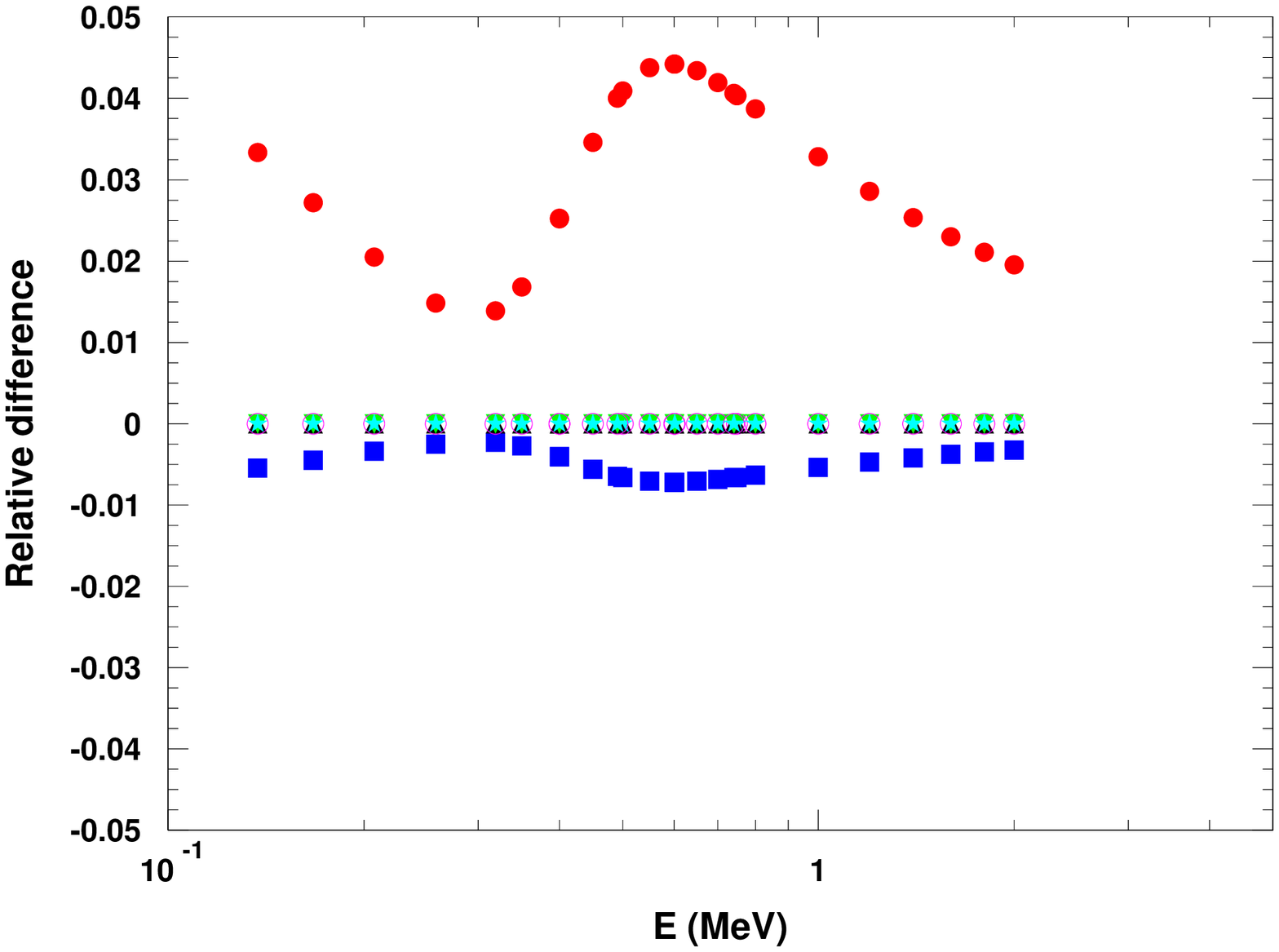}
        }%
        \subfigure[L$_2$]{%
           \label{fig_crossl2_48}
           \includegraphics[width=0.5\textwidth]{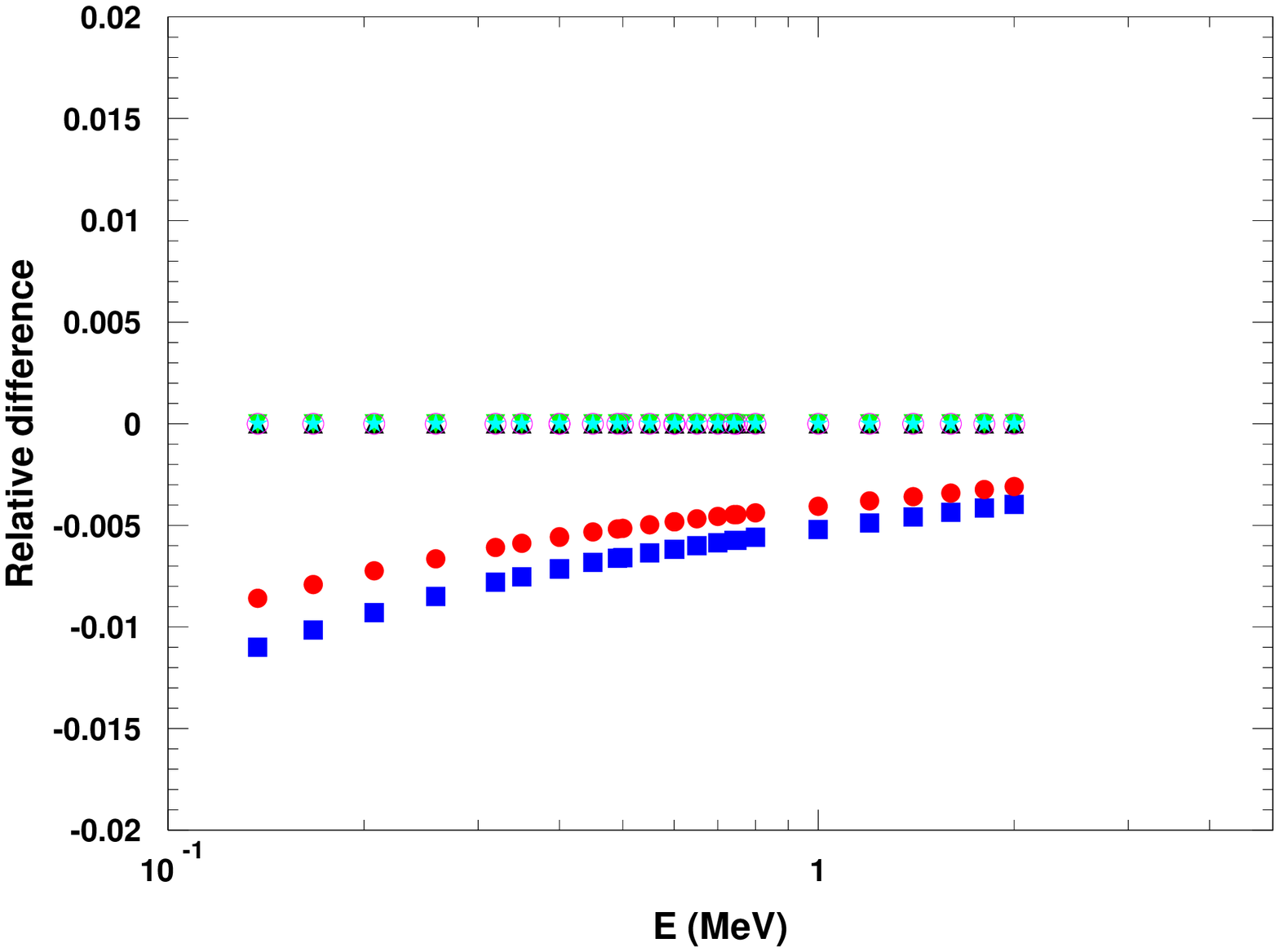}
        }\\ 
        \subfigure[L$_3$]{%
            \label{fig_crossl3_48}
            \includegraphics[width=0.5\textwidth]{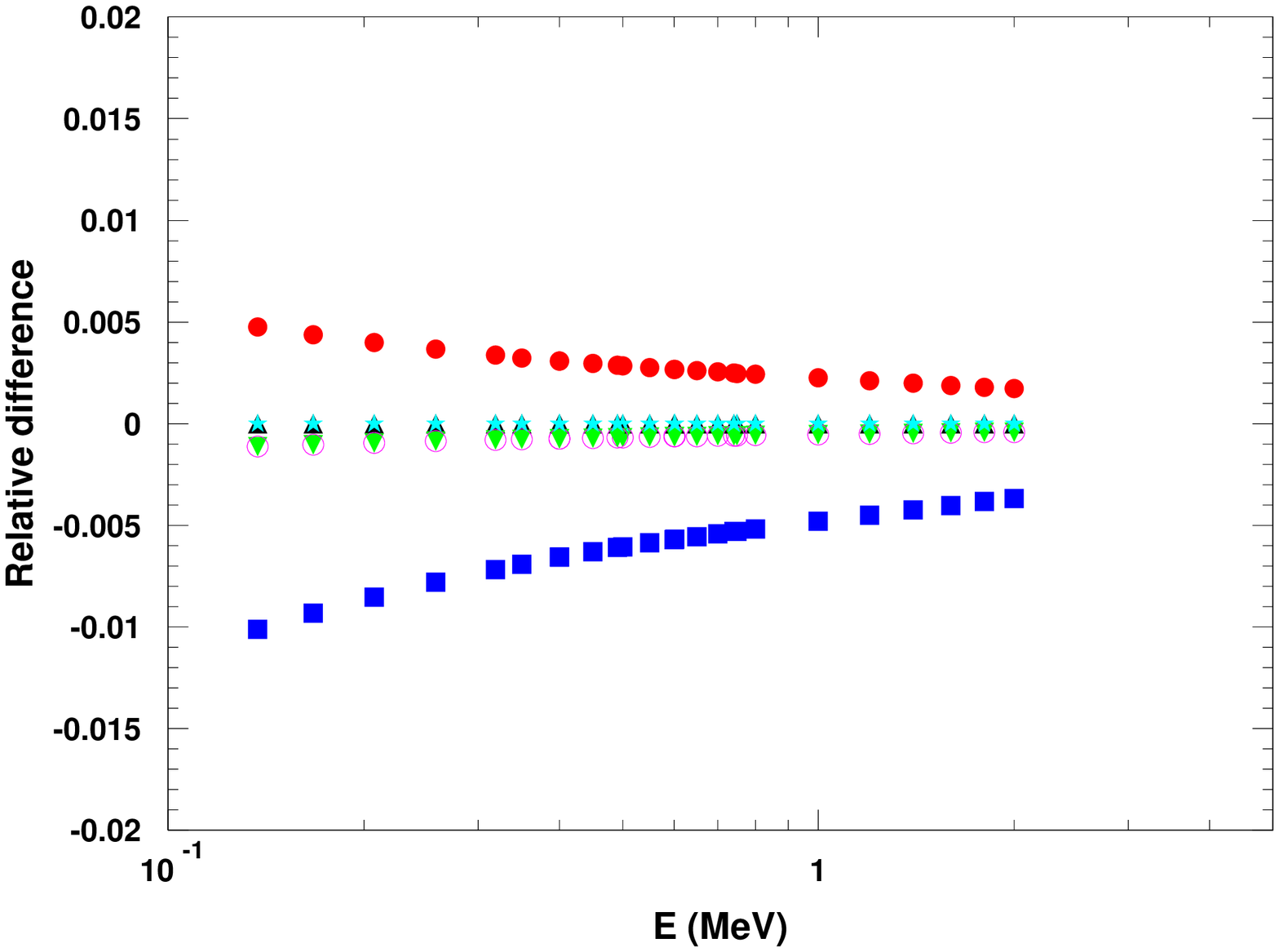}
        }%
    \end{center}
    \caption{%
       L sub-shell ionization cross sections by proton impact on cadmium calculated by the
ECPSSR model with different binding energies: the plot shows the relative
difference with respect to the values calculated with Bearden and Burr's binding
energies used by default by ISICS; the symbols identify cross sections calculated with EADL
(red circles), Carlson (blue squares), 1996 Table of Isotopes (black up
triangles), 1978 Table of Isotopes (green down triangles), Williams (pink empty
circles) and Sevier 1979 (turquoise stars) binding energies. Some symbols are
not visible in the plot due to the close values of some cross sections. 
     }%
    \label{fig_crossl_48}
\end{figure}

\begin{figure}[ht!]
    \begin{center}
        \subfigure[L$_1$]{%
            \label{fig_crossl1_74}
            \includegraphics[width=0.5\textwidth]{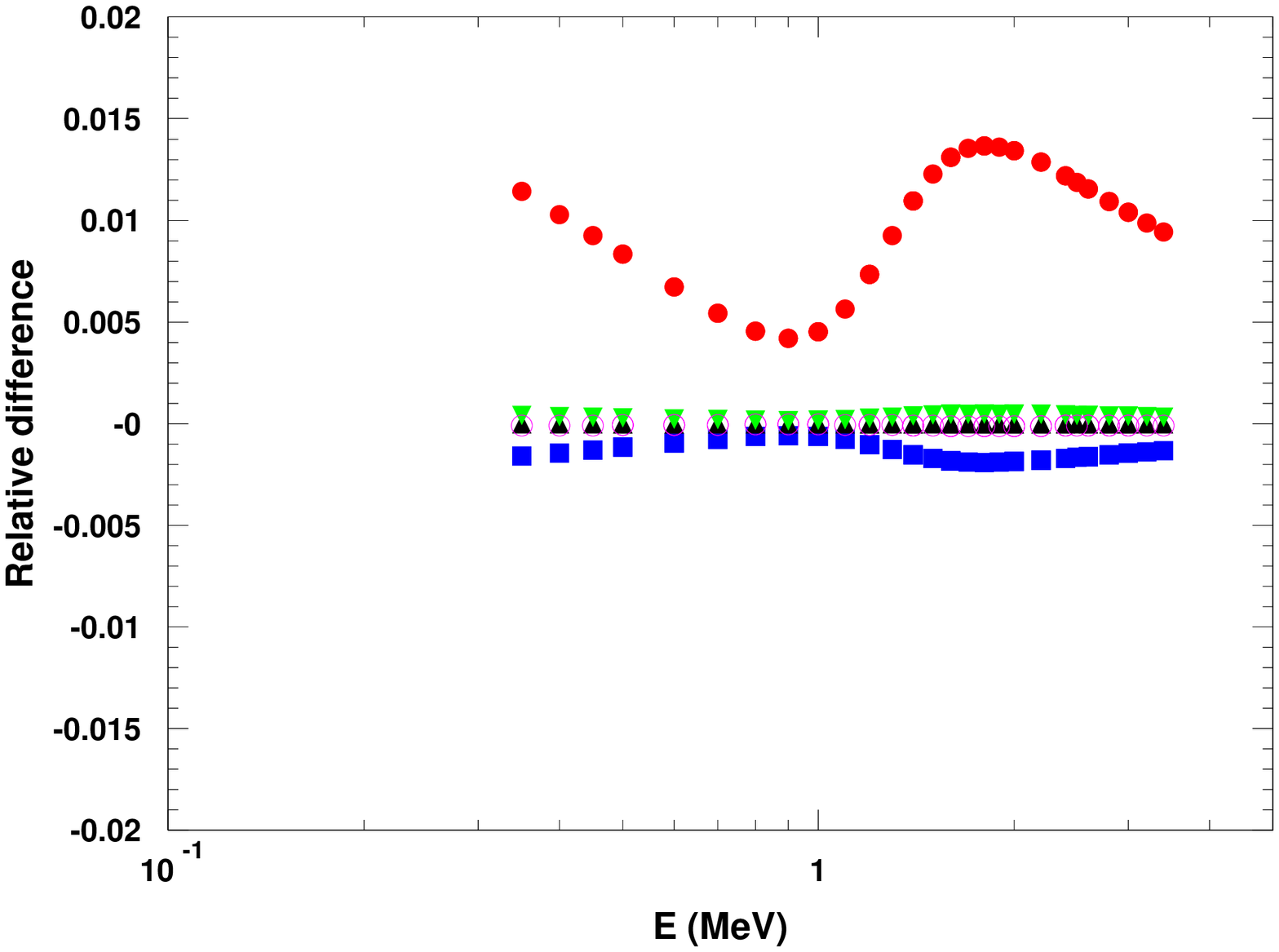}
        }%
        \subfigure[L$_2$]{%
           \label{fig_crossl2_74}
           \includegraphics[width=0.5\textwidth]{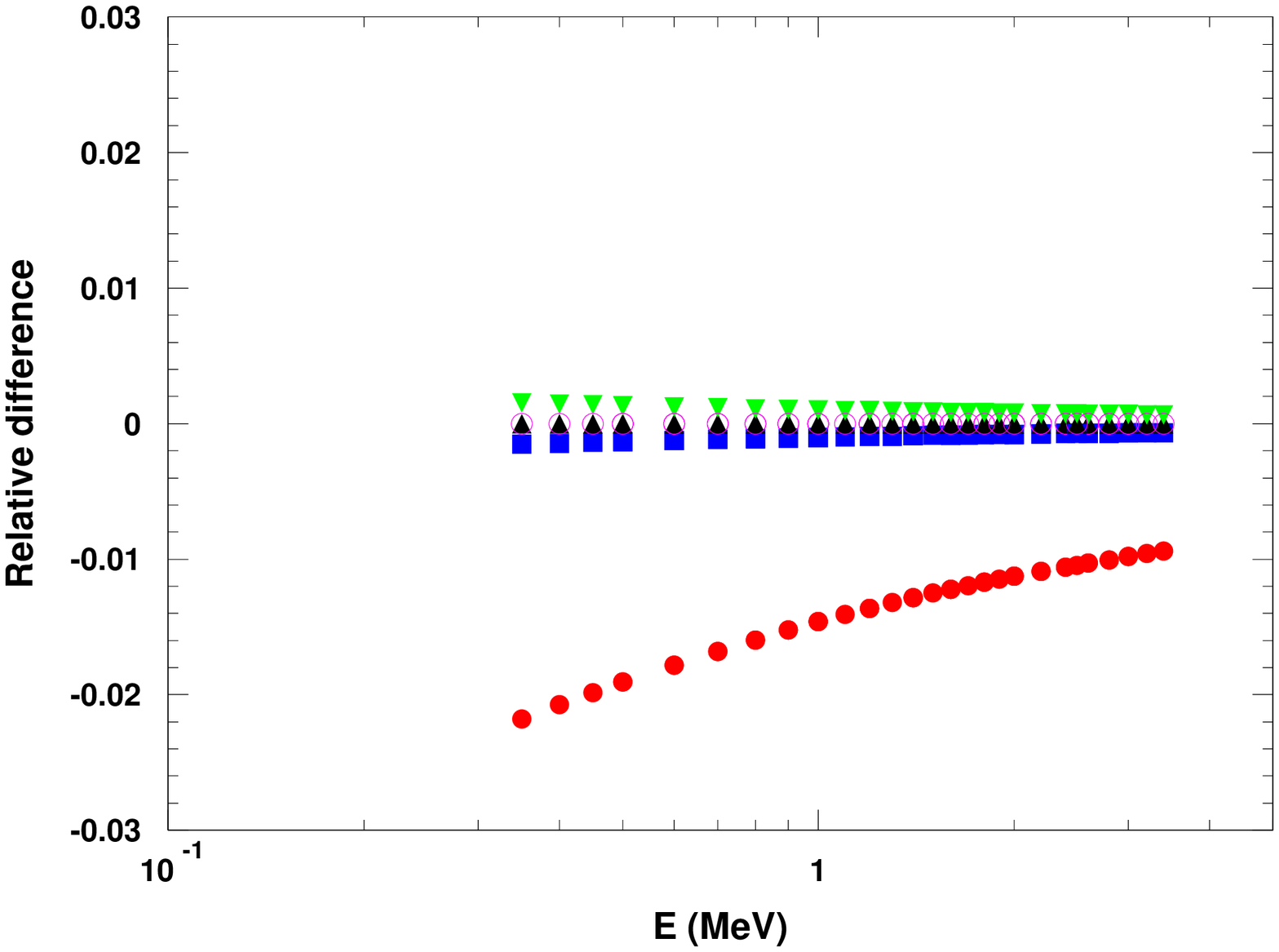}
        }\\ 
        \subfigure[L$_3$]{%
            \label{fig_crossl3_74}
            \includegraphics[width=0.5\textwidth]{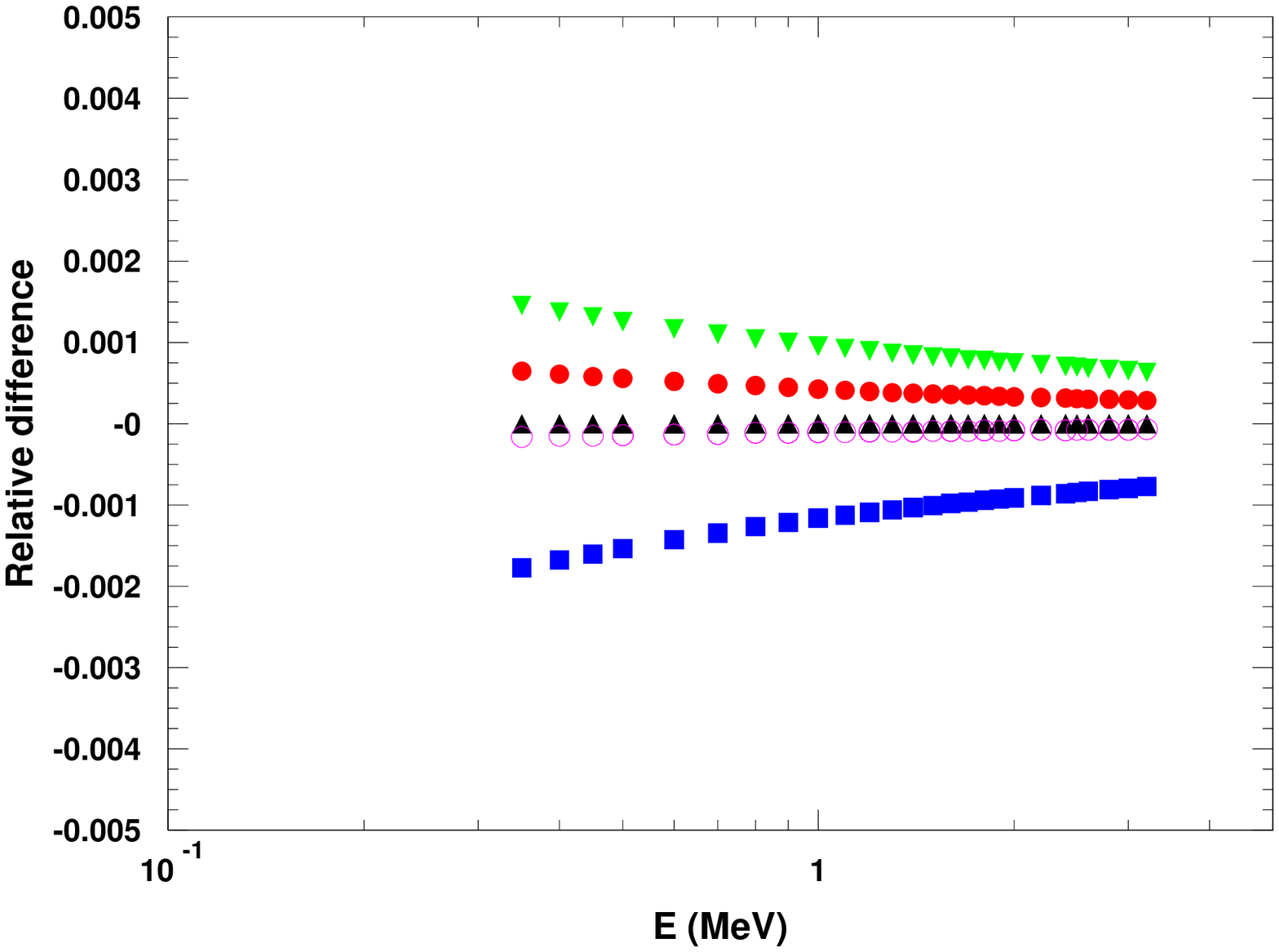}
        }%
    \end{center}
    \caption{%
       L sub-shell ionization cross sections by proton impact on tungsten calculated by the
ECPSSR model with different binding energies: the plot shows the relative
difference with respect to the values calculated with Bearden and Burr's binding
energies used by default by ISICS; the symbols identify cross sections calculated with EADL
(red circles), Carlson (blue squares), 1996 Table of Isotopes (black up
triangles), 1978 Table of Isotopes (green down triangles), Williams (pink empty
circles) and Sevier 1979 (turquoise stars) binding energies. Some symbols are
not visible in the plot due to the close values of some cross sections. 
     }%
    \label{fig_crossl_74}
\end{figure}

Cross sections calculated by ISICS 2011 version using different binding energy
compilations exhibit differences; some examples, concerning K and L shells, are
shown in figures \ref{fig_crossk}-\ref{fig_crossl_74}.
The differences appear larger for light elements and K shell.

%

The effects of different binding energies on the accuracy of proton ionization 
cross sections have been evaluated by comparing values based on  various 
binding energy collections with experimental data for K and L shells.
The experimental data derive from the reviews by Paul and Sacher
\cite{paul_sacher}, Sokhi and Crumpton \cite{sokhi} and Orlic et al.
\cite{orlic_exp}; the comparison process adopts the same strategy described in
\cite{tns_pixe} for the validation of the cross section models available in
Geant4.
For each element, the compatibility between calculated and experimental cross
sections is evaluated by means of the $\chi^2$ test; the significance of the
test for the rejection of the null hypothesis of equivalence of the compared
distributions is set to 0.05.

The analysis of the sensitivity to electron binding energies is reported here
for plain ECPSSR cross sections.
The fraction of tested elements for which K shell cross sections calculated with
various binding energies are compatible with experimental data is listed in
Table \ref{tab_ecpssrk}:
with the exception of EADL, all binding energy compilations appear to produce
equivalently accurate cross sections.
The use of EADL binding energies results in fewer test cases that are compatible
with measurements; the hypothesis of equivalent accuracy of cross sections based
on EADL with those based on other compilations is rejected with 0.05
significance for all the alternative binding energies.
As an example, the contingency table comparing the compatibility with
experimental data of cross sections based on EADL and Bearden and Burr binding
energies is reported in Table \ref{tab_contk}.


\begin{table}
\begin{center}
\caption{Fraction of tested elements for which ECPSSR K shell cross sections are
compatible with experimental data.}
\label{tab_ecpssrk}
\begin{tabular}{|l|c|}
\hline
{\bf Compilation} 	& {\bf Fraction}	\\
\hline
Bearden and Burr	& 0.76 $\pm$ 0.05	 \\
Carlson		& 0.76 $\pm$ 0.05	\\ 
EADL			& 0.58 $\pm$ 0.06	\\
Sevier 1979		& 0.76 $\pm$ 0.05	\\
ToI 1978		& 0.78 $\pm$ 0.05	\\
ToI 1996		& 0.76 $\pm$ 0.05	\\
Williams		& 0.78 $\pm$ 0.05	\\
\hline
\end{tabular}
\end{center}
\end{table}

\begin{table}
\begin{center}
\caption{Contingency table to estimate the equivalent accuracy of ECPSSR K shell
cross sections using EADL and Bearden and Burr's binding energies.}
\label{tab_contk}
\begin{tabular}{|l|c|c|c|c|}
\hline
{\bf $\chi^2$ test outcome}		&{\bf Bearden and Burr} &{\bf EADL}	\\
\hline
Pass					& 51			& 39 \\
Fail					& 16			& 28 \\
\hline
p-value Fisher test		& \multicolumn{2}{|c|}{0.042} \\
p-value Pearson $\chi^2$	& \multicolumn{2}{|c|}{0.027} \\
p-value Yates $\chi^2$		& \multicolumn{2}{|c|}{0.043} \\
\hline
\end{tabular}
\end{center}
\end{table}

The differences of cross sections associated with binding enegy compilations are
smaller for L shell than for K shell, as one can qualitatively observe in 
two examples, concerning cadmium and tungsten, shown in figures
\ref{fig_crossl_48}-\ref{fig_crossl_74}.
The comparison of L shell cross sections with experimental data does not identify
any significant differences associated with the use of different binding
energies; the fraction of tested elements for which L shell cross sections
calculated with various binding energies are compatible with experimental data
is listed in Table \ref{tab_ecpssrl}.

\begin{table}
\begin{center}
\caption{Fraction of tested elements for which ECPSSR L shell cross sections are
compatible with experimental data.}
\label{tab_ecpssrl}
\begin{tabular}{|l|c|c|c|}
\hline
{\bf Compilation} 	& {\bf Fraction, L$_1$}	& {\bf Fraction, L$_2$}	& {\bf Fraction, L$_3$}			\\
\hline
Bearden and Burr	& 0.68 $\pm$ 0.09 	& 0.68 $\pm$ 0.09	& 0.89 $\pm$ 0.06	\\
Carlson		& 0.68 $\pm$ 0.09	& 0.68 $\pm$ 0.09	& 0.86 $\pm$ 0.07	\\ 
EADL			& 0.64 $\pm$ 0.09	& 0.64 $\pm$ 0.09	& 0.89 $\pm$ 0.06	\\
Sevier 1979		& 0.68 $\pm$ 0.09	& 0.68 $\pm$ 0.09	& 0.89 $\pm$ 0.06	\\
ToI 1978		& 0.68 $\pm$ 0.09	& 0.68 $\pm$ 0.09	& 0.89 $\pm$ 0.06	\\
ToI 1996		& 0.68 $\pm$ 0.09	& 0.68 $\pm$ 0.09	& 0.89 $\pm$ 0.06	\\
Williams		& 0.68 $\pm$ 0.09	& 0.68 $\pm$ 0.09	& 0.89 $\pm$ 0.06	\\
\hline
\end{tabular}
\end{center}
\end{table}

The scarcity of experimental measurements prevents a similar analysis
on the effect of binding energies on M shell ionization cross sections.

Based on this analysis, the accuracy of proton ionization cross sections appears
statistically equivalent for all binding energy options but EADL.

\section{Effects on Compton scattering}

Doppler broadening of photon energy spectra arises from Compton scattering
between photons and moving electrons bound to atoms of the target medium.
Algorithms to account for Doppler broadening are implemented in widely used
Monte Carlo systems: those included in EGS \cite{namito}, MCNP
\cite{mcnp_doppler} and Geant4 \cite{nss_doppler} are based on
the method described in \cite{namito}; the algorithm implemented in Penelope 
produces equivalent results \cite{nss_doppler}.

A test was performed to ascertain if different binding energy compilations would
affect the calculated energy distributions of Compton scattering generated
in the simulation.
The test concerned a few target materials relevant to Compton telescopes
\cite{zoglauer}, silicon, germanium and xenon, which are characterized by
different experimental resolutions related to the effects of Doppler broadening.
For this investigation, the original implementation of Compton scattering with Doppler broadening in
Geant4 and associated unit test \cite{nss_doppler} was used.
The analysis compared the spectra deriving from two sets of binding energies:
those used in the simulation, which derive from EADL, and Carlson's compilation.
The latter was chosen as its binding energies for the considered elements
exhibit the largest difference with respect to EADL ones among the various
examined compilations.

\begin{figure}
\centerline{\includegraphics[angle=0,width=8.5cm]{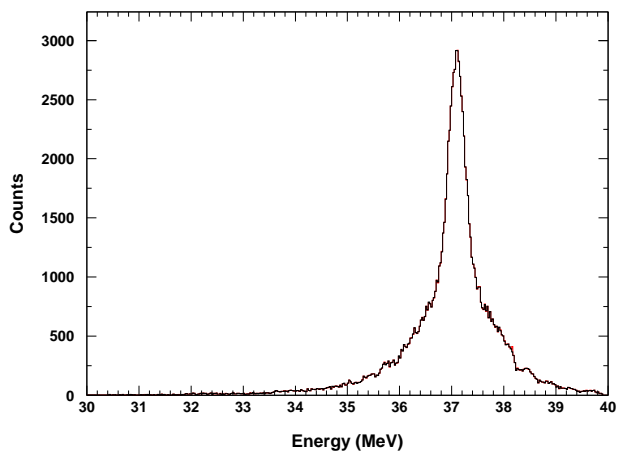}}
\caption{Energy distribution of photons between 89$^\circ$ and 91$^\circ$
resulting from Compton scattering of 40 keV photons orthogonally impinging onto
a silicon target, obtained using EADL (red) and Carlson's (black) binding
energies in the simulation. The two histograms associated with either binding
energy options are practically undistinguishable.}
\label{fig_doppler}
\end{figure}

No significant effect was visible in the spectra of the scattered photons as a
result of simulations using different binding energy compilations.
An example is illustrated in figure \ref{fig_doppler}, which shows the energy
spectrum of photons between 89$^\circ$ and 91$^\circ$ resulting from Compton
scattering of 40 keV photons orthogonally impinging onto a silicon target.

Pearson's $\chi^2$ test confirms the equivalence of the Doppler broadened photon
spectra based on EADL and Carlson's binding energies; the p-value resulting from
this test is 1 for all the three target materials. 
%

Therefore, based on this investigation, one can conclude that the choice
of binding energy compilation is not critical for the simulation of Compton scattering
accounting for Doppler broadening.

\section{Merged compilations}

The collection of binding energies in Geant4's \textit{G4AtomicShells} class has
been assembled specifically for Geant4, merging data from Carlson's compilation
with others from the 73rd edition of the CRC Handbook of Chemistry and Physics
\cite {crc73}.
The origin of the data for each element and shell is documented in the form of
comments in the code implementation.
The authors of this paper could not retrieve a copy of the
latter reference, which has been superseded by more recent editions (the most
recent one at the time of writing this paper is the 91st edition); nevertheless,
most of the values identified in the comments to the code as
originating from \cite{crc73} appear consistent with those published in the most
recent version of the Handbook, which includes Williams' compilation.
A few values in \textit{G4AtomicShells}, however, are consistent with neither Carlson's 
nor Williams' compilations.

\begin{figure}
\centerline{\includegraphics[angle=0,width=8.5cm]{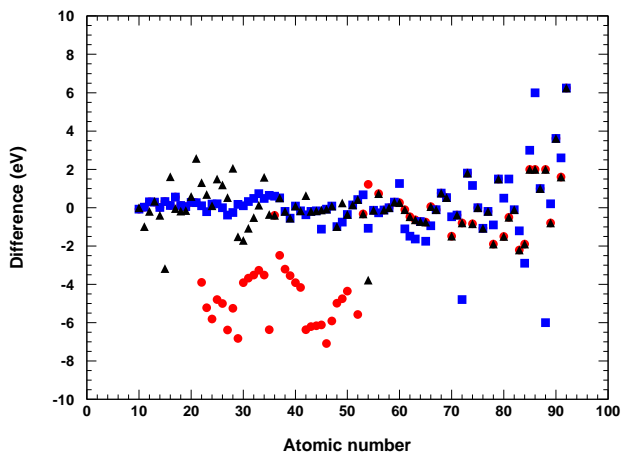}}
\caption{KL$_3$ transition, difference between X-ray energies calculated from
binding energies and experimental data from \cite{deslattes} versus atomic
number: binding energies from \textit{G4AtomicShells} (red circles), from
Carlson (blue squares) and Williams (black triangles).}
\label{fig_g4atokl3}
\end{figure}

\begin{figure}
\centerline{\includegraphics[angle=0,width=8.5cm]{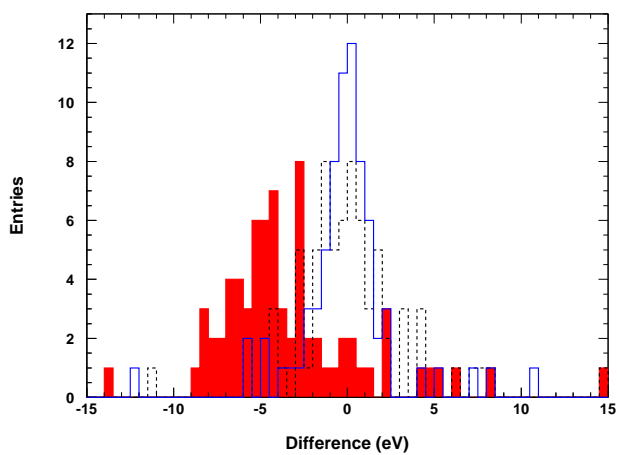}}
\caption{KM$_2$ transition, difference between X-ray energies calculated from
binding energies and experimental data from \cite{deslattes}: binding energies
from \textit{G4AtomicShells} (red shaded histogram), Carlson (blue solid
line histogram) and Williams (black dashed line histogram).}
\label{fig_g4atokm2}
\end{figure}

The two sources, Carlson's and Williams' compilations, report binding energies
based on different reference levels: the vacuum level for Carlson's data and the Fermi
level for Williams' data.
Data referring to different reference levels are associated with shells of the
same element in \textit{G4AtomicShells}.
The inconsistency of the data in the \textit{G4AtomicShells} class may generate
systematic effects in physics observables; some examples are illustrated in figures
\ref{fig_g4atokl3} and \ref{fig_g4atokm2}.
These plots show the differences between X-ray energies calculated from
\textit{G4AtomicShells} binding energies and the experimental data of Deslattes
et al. \cite{deslattes}, along with X-ray energies calculated from Carlson's and
Williams' compilations: the X-ray energies based on G4AtomicShells exhibit some
systematic shifts with respect to the experimental data, while the X-ray
energies based on Carlson's and Williams' compilations do not appear affected by
such systematic discrepancies with measurements.
The systematic effect is so large, that statistical tests appear redundant to 
identify its occurrence.


\section{Conclusion}

A survey of compilations of atomic binding energies compilations used by general purpose
Monte Carlo transport codes and other specialized software systems has been performed.
Most compilations are based on experimental data; the only exception among those
considered in this study is EADL, which is the result of theoretical
calculations.

The accuracy of these compilation has been evaluated through direct comparisons
with experimental data and through their effects on related physical quantities
used in particle transport and experimental observables.

The results of this study show that no single compilation is ideal for all applications.

Direct comparisons with reference experimental data, which concern a subset of
K, L, M and N shells, identify Williams' compilation, included in the X-ray Data
Booklet and CRC Handbook of Physics and Chemistry, 
as the one best agreeing with experimental values.
Regarding outer shells, ionization energies appear to be best reproduced by
Carlson's compilation; Lotz's ionization energies are identical to Carlson's for
elements with atomic number up to 92.

K and L shell X-ray energies are more accurately calculated based on the binding
energies reported in the 1996 Table of Isotopes. 
With respect to these binding energies, the X-ray energies derived from the
compilations by Bearden and Burr, Sevier and Williams do not exhibit any
statistically significant disagreement in compatibilility with average zero
difference from experiment.
Regarding the variance of such differences, the various compilations, with the
exception of EADL, are equivalent to the 1996 Table of Isotopes in 60\% to 79\% of
the transitions.
X-ray energies based on EADL, although less accurate than those produced by
other compilations, differ from the experimental references by less than 2\% for
most transitions: such an inaccuracy can be tolerable in some experimental
applications, while others, where accuracy of simulated X-ray energies is 
important, should utilize the 1996 Table of Isotopes or other compilations 
providing better accuracy than EADL in this domain.

Total cross sections for electron impact ionization addressing the energy range
below 1 keV (relevant to microdosimetry applications) are sensitive to the
values of ionization potentials, while they are marginally influenced by inner
shell binding energies.
Lotz-Carlson's compilation and EADL modified to include NIST experimental
ionization energies exhibit equivalent behaviour, while the use of EADL original
ionization energies 
substantially decreases the accuracy of the cross section calculation.

K shell ionization cross sections by proton impact are sensitive to the binding
energy values used in the ECPSSR calculation.
All the empirical binding energy compilations produce results compatible
with experimental measurements, while cross sections based on EADL show
statistically significant worse accuracy.
Binding energies also affect the calculation of ECPSSR cross sections for the L shell, but
the differences are smaller than for the K shell; the uncertainties of L shell experimental 
data are too large to appreciate the effect of different binding energies in terms of
modeling accuracy.

The simulation of Doppler broadening in Compton scattering appears insensitive
to the choice of binding energies among the examined compilations.

It is worthwhile to stress that these quantitative conclusions were not previously 
documented in the literature.

Simulation applications with high precision requirements may profit from the
results documented in this paper to identify the optimal set of binding energies
for specific scenarios (e.g. material analysis, microdosimetry, PIXE etc.).
Simulations applications not characterized by high precision requirements
may be satisfied by any of the compilations.
Monte Carlo kernel developers may profit form the results documented in this
paper to optimize the accuracy of the physics models they deliver to the 
experimental user community for applications.

As no single compilation is suitable for
all applications, it is highly desirable for simulation packages to
allow experimentalists to choose which compilation to use for their application.
This is much more easily possible in packages which read their binding
energy data from file at runtime, rather than hard-coding it into the
application binaries as in GEANT 3 and Geant4 \textit{materials}
package.



EADL binding energies appear consistently associated with worse accuracy in
all the test cases analysed in this paper.
An evolution of EADL to better reflect the state-of-the-art would be desirable;
it has already been advocated in \cite{relax_prob} regarding the improvement of
radiative transition probabilities.
However, modifications to EADL should preserve the consistency with two other
related data libraries, EEDL (Evaluated Electron Data Library) \cite{eedl} and
EPDL (Evaluated Photon Data Library)\cite{epdl97}, as these compilations are intended to
provide a consistent set of data for electron-photon trasport calculations.


\acknowledgments
The authors express their gratitude to CERN for support to the research
described in this paper.
The CERN Library, in particular Tullio Basaglia, has provided helpful assistance
and essential reference material for this study.
The authors thank Andy Buckley for proofreading the manuscript and valuable comments.



\begin{thebibliography}{199}

\bibitem{egs5}
H. Hirayama et al., 
\emph{The EGS5 Code System}, SLAC-R-730, Stanford, CA (2006).

\bibitem{egsnrc}
I. Kawrakow et al.,
\emph{The EGSnrc Code System: Monte Carlo
Simulation of Electron and Photon Transport},
NRCC PIRS-701, 5th printing, 2010.

\bibitem{g4nim} 
S.~Agostinelli et al., 
\emph{Geant4 - a simulation toolkit},
\emph{Nucl. Instrum. Meth. A} {\bf 506} (2003) 250.

\bibitem{g4tns}
J.~Allison et al., 
\emph{Geant4 Developments and Applications}, 
\emph{IEEE Trans. Nucl. Sci.} {\bf 53} (2006) 270. 

\bibitem{its5}
B. C. Franke et al.,
\emph{ITS5 theory manual}, 
Sandia Report SAND2004-4782, Albuquerque, 2004.

\bibitem{mcnp}
X-5 Monte Carlo Team, 
\emph{MCNP -- A General Monte Carlo N-Particle Transport Code, Version 5}, 
LLNL LA-UR-03-1987,  2003 revised 2005.

\bibitem{mcnpx}
J. S.~Hendricks et al., 
\emph{MCNPX, Version 26c}, 
LLNL  LA-UR-06-7991,  2006.

\bibitem{penelope}
J.~Baro et al.,
\emph{PENELOPE, an algorithm for Monte Carlo simulation of the
penetration and energy loss of electrons and positrons in
matter}, 
\emph{Nucl. Instrum. Meth. B} {\bf 100} (1995) 31. 

\bibitem{zschornack}
G. Zschornack,
\emph{Hadbook of X-Ray Data}, Springer, Berlin, 2007.

\bibitem{bearden}
J. A. Bearden et al.,
\emph{Reevaluation of X-Ray Atomic Energy Levels},
\emph{Rev. Mod. Phys.} {\bf 39} (1967) 125.

\bibitem{siegbahn}
M. Siegbahn et al.,
\emph{Phys. Zeitschr.} {\bf 17} (1916) 48.

\bibitem{sevier1979}
K. D. Sevier,
\emph{Atomic electron	binding energies}, 
\emph {Atom. Data Nucl. Data tables} {\bf 24} (1979) 323.

\bibitem{binding_nss2010}
H. Seo et al.,
\emph{Atomic parameters for Monte Carlo transport simulation: survey, validation and induced systematic effects},
IEEE Nucl. Sci. Symp. Conf. Rec., 2010 
\href{http://dx.doi.org/10.1109/NSSMIC.2010.5873948}{http://dx.doi.org/10.1109/NSSMIC.2010.5873948} 

\bibitem{slater1955}
J. C. Slater, 
\emph{One-Electron Energies of Atoms, Molecules, and Solids},
\emph{Phys. Rev.} {\bf 98} (1955) 1039.

\bibitem{crasemann}
B. Crasemann et al.,
\emph{Resource note: Theoretical atomic-electron binding energies},
\emph{Atom. Data Nucl. Data Tables} {\bf 36}  (1987) 355. 

\bibitem{carlson}
T. A. Carlson, 
\emph{Photoelectron and Auger spectroscopy}, Plenum, New York, 1975.

\bibitem{eadl}
S. T. Perkins et al., 
\emph{tables and Graphs of Atomic Subshell and
Relaxation Data Derived from the LLNL Evaluated Atomic Data Library (EADL)}, 
Z=1-100, UCRL-50400 Vol. 30, 1997.



\bibitem{toi1996}
R. B. Firestone et al.,
\emph{table of Isotopes, 8th ed.}, John Wiley \& Sons, New York, 1996.

\bibitem{toi1978}
M. Ledere et al.,
\emph{table of Isotopes, 7th ed.}, John Wiley \& Sons, New York, 1978.

\bibitem{xbook}
A. C. Thompson et al.,
\emph{X-ray Data Booklet}, 3rd ed., Lawrence Berkeley Natl. Lab., 2009.

\bibitem{crc90}
D. R. Lide ed.,
\emph{CRC Handbook of Chemistry and Physics}, 90th ed., Boca Raton, FL, 2009.

\bibitem{lotz}
W. Lotz, 
\emph{Electron binding energies in free atoms}, 
\emph{J. Opt. Soc. Am.} {\bf 60} (1970) 206.

\bibitem{shirley1977}
D. A. Shirley et al. 
\emph{Core-electron binding energies of the first thirty elements},
\emph{Phys. Rev. B} {\bf 15} (1977) 544. 

\bibitem{siegbahn_karlsson}
H. Siegbahn  et al., 
\emph{Photoelectron Spectroscopy} in
\emph{Encyclopedia of Physics} {\bf 31},  Springer, Berlin, 1982. 

\bibitem{larkins}
F. P. Larkins,
\emph{Semiempirical Auger-electron energies for elements $10 \leq Z \leq  100$},
\emph {Atom. Data Nucl. Data tables} {\bf 20}  (1977) 311.

\bibitem{sevier1972}
K. D. Sevier,
\emph{Appendix F}, 
in \emph{Low Energy Electron Spectrometry}, Wiley-Interscience, NY, 1972.

\bibitem{porter}
F. T. Porter et al.,
\emph{Recommended atomic electron binding energies, $1s$0 to $6p3/2$, for the heavy elements, 
Z=84 to 103},
\emph {J. Phys. Chem. Ref. Data} {\bf 7} (1978) 1267.

\bibitem{nist_ionipot}
W. C. Martinet al.,
\href{http://www.nist.gov/physlab/data/ion\_energy.cfm}{Ground Levels and Ionization Energies for the Neutral Atoms}.

\bibitem{scofield1969}
J. H. Scofield,
\emph{Radiative Decay Rates of Vacancies in the K and L Shells},
\emph{Phys. Rev. A} {\bf 18} (1969) 9.

\bibitem{scofield1974}
J. H. Scofield,
\emph{Relativistic Hartree-Slater values for K and L X-ray emission rates},
\emph {Atom. Data Nucl. Data tables} {\bf 14} (1974) 121.

\bibitem{scofield1978}
J. H. Scofield,
\emph{K- and L-shell ionization of  atoms by relativistic electrons},
\emph{Phys. Rev.} {\bf 179} (1978) 963. 

\bibitem{scofield1990}
J. H. Scofield,
\emph{Angular distribution of photoelectrons from polarized X-rays},
\emph{Phys. Scr.} {\bf 41} (1990) 59.

\bibitem{cardona_ley}
M. Cardona et al.,
\emph{Photoemission in Solids I: General Principles},
 Springer-Verlag, Berlin, 1978.

\bibitem{fuggle}
J. C. Fuggle et al.,
\emph{Core-Level Binding Energies in Metals}, 
\emph{J. Electron Spectrosc. Relat. Phenom.} {\bf 21} (1980) 275.


\bibitem{egs4}
W. R. Nelson et al., 
\emph{The EGS4 Code System},
SLAC-265, Stanford, CA, 1985.

\bibitem{crc73}
D. R. Lide ed.,
\emph{CRC Handbook of Chemistry and Physics}, 73rd ed., Boca Raton, FL, 1992.

\bibitem{lowe_chep}
S. Chauvie et al.,
\emph{Geant4 Low Energy Electromagnetic Physics},
in \emph{Proc. Computing in High Energy and Nuclear Physics}, 
Beijing, China, 2001.

\bibitem{lowe_nss}
S. Chauvie et al., 
\emph{Geant4 Low Energy Electromagnetic Physics},
in \emph{2004 IEEE Nucl. Sci. Symp. Conf. Rec.}  (2004) 1881.

\bibitem{eedl}
S. T. Perkins et al., 
\emph{tables and Graphs of Electron-Interaction Cross
Sections from 10 eV to 100 GeV Derived from the LLNL Evaluated
Electron Data Library (EEDL)}, 
UCRL-50400 Vol. 31, LLNL, 1997.

\bibitem{epdl97}	
D. Cullen et al., 
\emph{EPDL97, the Evaluated Photon Data Library}, 
UCRL-50400 {\bf 6}, Rev. 5, 1997.

\bibitem{haifa}
H. Abdelhwahed et al.,
\emph{New Geant4 Cross Section Models for PIXE Simulation},
\emph{Nucl. Instrum. Meth. B} {\bf 267} (2009) 37.

\bibitem{haifa2}
A. Mantero et al.,
\emph{PIXE Simulation in Geant4},
\emph{X-ray Spectrom} {\bf 40}  (2011) 135.

\bibitem{tns_pixe}
M. G. Pia et al.,
\emph{PIXE simulation with Geant4},
\emph{IEEE Trans. Nucl. Sci.} {\bf 56}   (2009) 3614.

\bibitem{isics}
Z.~Liu et al., 
\emph{ISICS: A program for calculating K-, L-, and M-shell cross sections from
ECPSSR theory using a personal computer},
\emph{Comp.\ Phys.\ Comm.} {\bf 97} (1996) 315.

\bibitem{geant3}
\emph{GEANT Detector Description and Simulation Tool}, 
CERN Program Library Long Writeup W5013, 1995.

\bibitem{grichine1991}
V. M. Grishin et al.,
\emph{Ionization energy loss in very thin absorbers },
\emph{Nucl. Instr. Meth.} {\bf 309}  (1991)  476.

\bibitem{gupix1}
J. A. Maxwell et al.,
\emph{The Guelph PIXE software package},
\emph{Nucl. Instrum. Meth. B} {\bf  43}  (1989) 218.


\bibitem{gupix3}
J. L. Campbell et al.,
\emph{The Guelph PIXE software package III: Alternative proton database},
\emph{Nucl. Instrum. Meth. B} {\bf 170} (2000) 193.

\bibitem{isics2011}
S. J. Cipolla, 
\emph{ISICS2011, an updated version of ISICS: A program for calculation K-, L-, and
M-shell cross sections from PWBA and ECPSSR theories using a personal
computer},
\emph{Comp.\ Phys.\ Comm.} {\bf 182}  (2011) 2439.


\bibitem{fluka1}
G. Battistoni et al.,
\emph{The FLUKA code: description and benchmarking},
\emph{AIP Conf. Proc.} {\bf 896} (2007) 31.

\bibitem{fluka2}
A.~Ferrari et al., 
\emph{Fluka: a multi-particle transport code}, 
Report CERN-2005-010, INFN/TC-05/11, SLAC-R-773, Geneva, 2005.


\bibitem{gof1}
G. A. P. Cirrone et al., 
\emph{A Goodness-of-Fit Statistical Toolkit}, 
\emph{IEEE Trans. Nucl. Sci.} {\bf 51} (2004)  2056.

\bibitem{gof2}
B. Mascialino et al., 
\emph{New developments of the Goodness-of-Fit Statistical Toolkit}, 
\emph{IEEE Trans. Nucl. Sci.} {\bf 53} (2006) 3834. 


\bibitem{halas}
S. Halas et al.,
\emph{Work functions of elements expressed in terms of the Fermi energy and the
density of free electrons},
\emph{J. Phys. Cond. Matter} {\bf 10}  (1998) 10815.

\bibitem{drummond}
T. J. Drummond,
\emph{Work Functions of the Transition Metals and Metal Silicides},
Sandia Report SAND99-0391J, Albuquerque, 1999.

\bibitem{powell1991}
C. J. Powell,
\emph{Formal databases for surface analysis: The current situation and future trends},
\emph{Surf. Interface Anal.} {\bf 17} (1991) 308.

\bibitem{powell1995}
C. J. Powell, 
\emph{Elemental binding energies for X-ray photoelectron spectroscopy},
\emph{Appl. Surf. Sci.} {\bf 89}  (1995) 141.

\bibitem{nist_xps}
J. R. Rumble Jr. et al.,
\emph{The NIST x-ray photoelectron spectroscopy database},
\emph{Surf. Interface Anal.} {\bf 19}  (1992) 241.

\bibitem{kolmogorov1933}
A.~N.~Kolmogorov, 
\emph{Sulla determinazione empirica di una legge di distribuzione}, 
\relax \emph{Gior. Ist. Ital. Attuari} {\bf 4} (1933) 83.

\bibitem{smirnov1939}
N.~V.~Smirnov, 
\emph{On the estimation of the discrepancy between empirical curves of distributions
for two independent samples},
\relax \emph{Bull. Math. Univ. Moscou}, 1939.

\bibitem{anderson1952}
T.~W.~Anderson et al., 
\emph{Asymptotic theory of certain goodness of fit criteria based on stochastic
processes},
\emph{Anls. Ma. St.} {\bf 23} (1952) 193.

\bibitem{anderson1954}
T.~W.~Anderson et al.,
\emph{A test of goodness of fit},
\emph{JASA} {\bf 49}  (1954) 765.

\bibitem{cramer1928}
H.~Cram\'er, 
\emph{On the composition of elementary errors. Second paper: statistical applications}, 
\emph{Skand. Aktuarietidskr.} {\bf 11}  (1928) 141.

\bibitem{vonmises1931}
R.~von Mises, 
\emph{Wahrscheinlichkeitsrechnung und ihre Anwendung in der Statistik und theoretischen Physik},
F. Duticke, Leipzig, 1931.

\bibitem{bock}
R. K. Bock et al.,
\emph{The Data Analysis BriefBook},
 Springer, Berlin, 1998.


\bibitem{nist_principal}
C. D. Wagneret al.,
\href{http://srdata.nist.gov/xps/RecomEnergy.aspx?EnergyType=PE}{NIST X-ray Photoelectron Spectroscopy Database}.


\bibitem{deslattes}
R. D. Deslattes et al., 
\emph{X-ray transition energies: new approach to a comprehensive evaluation}, 
\emph{Rev. Mod. Phys.} {\bf 75}  (2003) 35.

\bibitem{relax_nist}
S. Guatelli et al., 
\emph{Validation of Geant4 Atomic Relaxation against the NIST Physical Reference Data}, 
\emph{IEEE Trans. Nucl. Sci.} {\bf 54}  (2007) 594.

\bibitem{tns_relax}
S. Guatelli et al., 
\emph{Geant4 Atomic Relaxation}, 
\emph{IEEE Trans. Nucl. Sci.} {\bf 54}  (2007) 585.

\bibitem{fisher}
R. A. Fisher,
\emph{On the interpretation of  $\chi^2$ from contingency tables, 
and the calculation of P},
\emph{J. Royal Stat. Soc.} {\bf 85}  (1922) 87.

\bibitem{pearson}
K. Pearson,
\emph{On the $\chi^2$ test of Goodness of Fit},
\emph{Biometrika} {\bf 14}  (1922) 186.

\bibitem{yates}
F. Yates,
\emph{Contingency table involving small numbers and the $\chi^2$ test},
\emph{J. Royal Stat. Soc. Suppl.} {\bf 1} (1934) 217. 

\bibitem{beb1994} 
Y. K. Kim et al.,
\emph{Binary-encounter-dipole model for electron-impact ionization by electron impact},
\emph{Phys. Rev. A} {\bf 50}  (1994) 3954.

\bibitem{ecpssr}
W.~Brandt  et al.,
\emph{Energy-loss effect in inner-shell Coulomb
ionization by heavy charged particles}, 
\emph{ Phys. Rev. A} {\bf 23}  (1981) 1717.

\bibitem{dm1987} 
H. Deutsch et al.,
\emph{Calculation of absolute electron impact ionization cross-section functions for 
single ionization of He, Ne, Ar, Kr, Xe, N and F},
\emph{Int. J. Mass Spectrom. Ion Proc.} {\bf 79}  (1987) R1.

\bibitem{beb_mc2010}
H. Seo et al.,
\emph{Design, development and validation of electron ionization models for nano-scale simulation},
in \emph{Proc. SNA-Monte Carlo Conf.}, Tokyo, 2010.

\bibitem{beb_nss2010}
H. Seo et al.,
\emph{ionization models for nano-scale simulation},
in \emph{2010 IEEE Nucl. Sci. Symp.}, Knoxville, 2010.

\bibitem{tns_beb}
H. Seo et al.,
\emph{Ionization cross sections for low energy electron simulation},
submitted to \emph{JINST},  2011.


\bibitem{dmDeutsch2005} 
H. Deutsch et al., 
\emph{A detailed comparison of calculated and measured electron-impact ionization 
cross sections of atoms using the Deutsch-M\"ark (DM) formalism},
\emph{Int. J. Mass Spectrom.} {\bf 243}  (2005) 215.

\bibitem{dmDeutsch2005err} 
H. Deutsch et al., 
\emph{Erratum to A detailed comparison of calculated and measured electron-impact ionization 
cross sections of atoms using the Deutsch-M\"ark (DM) formalism},
\emph{Int. J. Mass Spectrom.} {\bf 246}  (2005) 113.

\bibitem{dmMarg1994} 
D. Margreiter et al., 
\emph{A semiclassical approach to the calculation of electron impact 
ionization cross-sections of atoms: from hydrogen to uranium},
\emph{Int. J. Mass Spectrom. Ion Processes} {\bf 139}  (1994) 127.

\bibitem{expKieffer} 
L. J. Kieffer  et al.,
\emph{Electron impact ionization cross-section data for atoms, atomic ions, and diatomic molecules: 
I. Experimental data},
\emph{Rev. Mod. Phys.} {\bf 38}  (1966) 1.

\bibitem{bebAli1997} 
M. A. Ali et al., 
\emph{Electron-impact total ionization cross sections of silicon and germanium hydrides},
\emph{J. Chem. Phys.} {\bf 106}  (1997) 9602.

\bibitem{bebKim1998} 
Y. K. Kim et al., 
\emph{Electron-impact ionization cross section of rubidium},
\emph{Phys. Rev. A} {\bf 57}  (1998) 246.

\bibitem{bebKim2000} 
Y. K. Kim et al., 
\emph{Cross sections for singly differential and total ionization of helium by electron impact},
\emph{Phys. Rev. A} {\bf 61}  (2000) 034702.

\bibitem{bebKim2000R} 
Y. K. Kim et al., 
\emph{Extension of the binary-encounter-dipole model to relativistic incident electrons},
\emph{Phys. Rev. A} {\bf 62}  (2000) 052710.

\bibitem{bebKim2001} 
Y. K. Kim  et al.,
\emph{Ionization of boron, aluminum, gallium, and indium by electron impact},
\emph{Phys. Rev. A} {\bf 64}  (2001) 052707.

\bibitem{bebKim2002} 
Y. K. Kim et al.,
\emph{Ionization of carbon, nitrogen, and oxygen by electron impact},
\emph{Phys. Rev. A} {\bf 66} (2002) 012708.

\bibitem{bebKim2007} 
Y. K. Kim et al.,
\emph{Ionization of silicon, germanium, tin and lead by electron impact},
\emph{J. Phys. B: At. Mol. Opt. Phys.} {\bf 40}  (2007) 1597.

\bibitem{bebAli2008} 
M. A. Ali et al.,
\emph{Ionization cross sections by electron impact on halogen atoms, diatomic halogen 
and hydrogen halide molecules},
\emph{J. Phys. B: At. Mol. Opt. Phys.} {\bf 41}  (2008) 145202.



\bibitem{expHshah1987} 
M. B. Shah et al., 
\emph{Pulsed crossed-beam study of the ionization of atomic hydrogen 
by electron impact},
\emph{J. Phys. B: At. Mol. Phys.} {\bf 20}  (1987) 3501.


\bibitem{fite}
W. L. Fite et al.,
\emph{Collisions of electrons with hydrogen atoms. I. Ionization},
\emph{Phys. Rev.} {\bf 112}  (1958) 1141.

\bibitem{rothe}
E. W. Rothe et al., 
\emph{Electron impact ionization of atomic hydrogen and atomic oxygen},
\emph{Phys. Rev.} {\bf 125}  (1962) 582.

\bibitem{expRejoub2002}
R. Rejoub et al., 
\emph{Determination of the absolute partial and total cross sections for 
electron-impact ionization of the rare gases},
\emph{Phys. Rev. A} {\bf 65}  (2002) 042713.

\bibitem{exp2S}
M. B. Shah et al., 
\emph{Single and double ionization of helium by electron impact},
\emph{J. Phys. B: At. Mol. Opt. Phys.} {\bf 21}  (1988) 2751.

\bibitem{rapp}
D. Rapp et al.,
\emph{Total cross sections for ionization and attachment in gases by electron impact},
\emph{Phys. Rev. A} {\bf 43}  (1965) 1464.

\bibitem{expHeNeSchram} 
B. L. Schram et al.,
\emph{Partial ionization cross sections of noble gases for electrons with energy 0)5-16 keV},
\emph{Physica} {\bf 32} (1966) 185. 

\bibitem{stephan}
K. Stephan et al., 
\emph{Mass spectrometric determination of partial electron impact ionization 
cross sections of He, Ne, Ar and Kr from threshold up to 180 eV},
\emph{J. Chem. Phys.} {\bf 73}  (1980) 3763.

\bibitem{krishnakumar}
E. Krishnakumar et al.,
\emph{ionization cross sections of rare-gas atoms by electron impact},
\emph{J. Phys. B: At. Mol. Opt. Phys.} {\bf 21}  (1988) 1055.

\bibitem{expMontague1984} 
R. G. Montague et al.,
\emph{A measurement of the cross section for ionization of helium by electron impact 
using a fast crossed beam technique},
\emph{J. Phys. B: At. Mol. Phys.} {\bf 17} (1984) 3295.

\bibitem{expNagy1980} 
P. Nagy et al.,
\emph{Absolute ionization cross sections for electron impact in rare gases},
\emph{J. Phys. B: At. Mol. Phys.} {\bf 13} (1980) 1249.

\bibitem{wetzel}
R. C. Wetzel et al., 
\emph{Absolute cross sections for electron-impact ionization of the rare-gas atoms 
by the fast neutral beam method},
\emph{Phys. Rev. A} {\bf 35} (1987)  559.

\bibitem{expM1965}
R. H. McFarland et al.,
\emph{Absolute cross sections of Li and other alkali metal atoms for ionization 
by electrons},
\emph{Phys. Rev.} {\bf 137}  (1965) 1058.

\bibitem{expZ1969}
I. P. Zapesochnyi et al.,
\emph{Ionization of alkali metal atoms by slow electrons},
\emph{Sov. Phys. JETP} {\bf 28}   (1969) 41.

\bibitem{jalin}
R. Jalin et al.,
\emph{Absolute electron impact ionization cross sections of Li in the energy range from 
100 to 2000 eV},
\emph{J. Chem. Phys.} {\bf 59} (1973)  952.

\bibitem{expBrook1978} 
E. Brook et al.,
\emph{Measurements of the electron impact ionization cross sections 
of He, C, O and N atoms},
\emph{J. Phys. B: Atom. Molec. Phys.} {\bf 11}   (1978) 3115.

\bibitem{smith}
A. C. H. Smith et al.,
\emph{Electron impact ionization of atomic nitrogen},
\emph{Phys. Rev.} {\bf 127}  (1962)  1647.

\bibitem{expPeterson} 
J. R. Peterson, 
\emph{Atomic Collision Processes}, 
North-Holland Publishing, Amsterdam, 1964.

\bibitem{exp8}
W. R. Thompson et al.,
\emph{Single and double ionization of atomic oxygen by electron impact},
\emph{J. Phys. B: At. Mol. Opt. Phys.} {\bf 28}  (1995) 1321.

\bibitem{zipf}
E. C. Zipf, 
\emph{The ionization of atomic oxygen by electron impact},
\emph{Planet. Space Sci.} {\bf 33} (1985) 1303.

\bibitem{expOfite1959} 
W. L. Fite et al.,
\emph{Ionization of atomic oxygen on electron impact},
\emph{Phys. Rev.} {\bf 113} (1959) 815.

\bibitem{expHayes} 
T. R. Hayes et al.,
\emph{Absolute electron-impact-ionization cross-section measurements of the 
halogen atoms},
\emph{Phys. Rev. A} {\bf 35} (1987) 578. 

\bibitem{adamczyk}
B. Adamczyk et al.,
\emph{Partial ionization cross sections of He, Ne, H$_2$, and CH$_4$ for electrons from 20 to 500 eV},
\emph{J. Chem. Phys.} {\bf 44}  (1966) 4640.

\bibitem{almeida}
D. P. Almeida et al.,
\emph{Electron-impact ionization cross section of neon ($\sigma$$_{n+}$, n = 1-5)},
\emph{J. Phys. B: At. Mol. Opt. Phys.} {\bf 28} (1995) 3335.

\bibitem{fletcher}
J. Fletcher et al.,
\emph{Electron impact ionization of neon and argon},
\emph{J. Phys. B: At. Mol. Phys.} {\bf 6} (1973) L258.

\bibitem{sorokin}
A. A. Sorokin et al.,
\emph{Measurements of electron-impact ionization cross sections of neon by 
comparison with photoionization},
\emph{Phys. Rev. A} {\bf 58}  (1998) 2900.

\bibitem{brink}
G. O. Brink, 
\emph{Absolute ionization cross sections of the alkali metals},
\emph{Phys. Rev.} {\bf 134}  (1964)  A345.

\bibitem{fujii}
K. Fujii et al.,
\emph{A measurement of the electron-impact ionization cross section of sodium},
\emph{J. Phys. B} {\bf 28} (1995) L559.

\bibitem{johnston}
A. R. Johnston et al.,
\emph{Electron-impact ionization of Na},
\emph{Phys. Rev. A} {\bf 51}   (1995) R1735.

\bibitem{tan}
W. S. Tan et al.,
\emph{Electron-impact ionization of laser-excited sodium atom},
\emph{Phys. Rev. A} {\bf 54}  (1996)  R3710.

\bibitem{expFreund} 
R. S. Freund et al.,
\emph{Cross-section measurements for electron-impact ionization of atoms},
\emph{Phys. Rev. A} {\bf 41}  (1990) 3575.

\bibitem{boivin}
R. F. Boivin et al.,
\emph{Electron-impact ionization of Mg},
\emph{J. Phys. B: At. Mol. Opt. Phys.} {\bf 31}  (1998) 2381.

\bibitem{karstensen}
F. Karstensen et al.,
\emph{Absolute cross sections for single and double ionization of Mg atoms by electron impact},
\emph{J. Phys. B: At. Mol. Phys.} {\bf 11}   (1978) 167.

\bibitem{expMgMcCallion} 
P. McCallion et al., 
\emph{Multiple ionization of magnesium by electron impact},
\emph{J. Phys. B: At. Mol. Opt. Phys.} {\bf 25}  (1992) 1051.

\bibitem{expVainsh} 
L. A. Vainshtein et al.,
\emph{Absolute values of electron impact ionization cross sections for magnesium, 
calcium, strontium and barium},
\emph{Sov. Phys. JETP} {\bf 34}  (1972)  271.

\bibitem{okunoMg}
Y. Okuno et al.,
\emph{Absolute measurement of total ionization cross section of Mg by electron impact},
\emph{J. Phys. Soc. Japan} {\bf 29}  (1970) 164.

\bibitem{golovach}
D. G. Golovach et al.,
\emph{Measurment of the ionization cross section of aluminum atoms by electronic impact},
\emph{Meas. Tech. (USSR)} {\bf 30} (1987) 587.

\bibitem{shimonAlInTl}
L. L. Shimon et al.,
\emph{Effective total electron-impact ionization cross sections for aluminum, gallium, indium and thallium},
\emph{Sov. Phys. Tech. Phys.} {\bf 20}  (1975) 434.

\bibitem{ziegler}
D. L. Ziegler et al.,
\emph{Single and multiple ionization of sulfur atoms by electron impact},
\emph{Planet. Space Sci.} {\bf 30}  (1982)  1269.

\bibitem{exp18S} 
H. C. Straub et al.,
\emph{Absolute partial and total cross sections for electron-impact ionization of argon
from threshold to 1000 eV},
\emph{Phys. Rev. A} {\bf 52}  (1995) 1115.

\bibitem{expArKrXeSchram} 
B. L. Schram, 
\emph{Partial ionization cross sections of noble gases for electrons with energy 0)5-18 keV},
\emph{Physica} {\bf 32} (1966) 197.

\bibitem{expArMcCallion} 
P. McCallion et al.,
\emph{A crossed beam study of the multiple ionization of argon by electron impact},
\emph{J. Phys. B: At. Mol. Opt. Phys.} {\bf 25}  (1992)  1061.

\bibitem{ma}
C. Ma et al.,
\emph{A pulsed electron beam time of flight apparatus for measuring absolute electron impact 
ionization and dissociative ionization cross sections},
\emph{Rev. Sci. Inst.} {\bf 62}  (1991) 909.

\bibitem{korchevoi}
Yu. P. Korchevoi et al.,
\emph{Effective electron impact excitation and ionization cross sections for cesium, 
rubidium, and potassium atoms in the pre-threshold region},
\emph{Sov. Phys. JETP} {\bf 24}   (1967) 1089.

\bibitem{expKNygaard} 
K. J. Nygaard, 
\emph{Electron impact autoionization in heavy alkali metals},
\emph{Phys. Rev. A} {\bf 11}   (1975) 1475.

\bibitem{expM1967} 
R. H. McFarland, 
\emph{Electron-impact ionization measurements of surface-ionizable atoms},
\emph{Phys. Rev.} {\bf 159}  (1967) 20.

\bibitem{okuno}
Y. Okuno,
\emph{Ionization cross sections of Ca, Sr and Ba by electron impact},
\emph{J. Phys. Soc. Japan} {\bf 31}  (1971)  1189.

\bibitem{schneider}
M. Schneider, 
\emph{Measurement of absolute ionization cross sections for electron impact},
\emph{J. Phys. D: Appl. Phys.} {\bf 7}  (1974) L83.

\bibitem{rakhovskii}
V. J. Rakhovski et al.,
\emph{Absolute values of the apparent cross section for calcium ionization by electron collision},
\emph{High Temp.} {\bf 7}  (1969) 1001.


\bibitem{expFeShah1993} 
M. B. Shah et al., 
\emph{Multiple ionization of iron by electron impact},
\emph{J. Phys. B: At. Mol. Opt. Phys.} {\bf 26} (1993) 2393.

\bibitem{bolorizadeh}
M. A. Bolorizadeh et al.,
\emph{Multiple ionization of copper by electron impact},
\emph{J. Phys. B: At. Mol. Opt. Phys.} {\bf 27} (1994) 175.

\bibitem{pavlovCuAg}
S. I. Pavlov et al.,
\emph{Measurement of cross sections for ionization by electron impact at low vapor pressures},
\emph{Sov. Phys. JETP} {\bf 25}  (1967) 12.

\bibitem{schroeer}
J. M. Schroeer et al.,
\emph{Electron impact ionization cross sections of Cu and Au between 40 and 250 eV, and 
the velocity of evaporated atoms},
\emph{J. Chem. Phys.} {\bf 58}  (1973) 5135.

\bibitem{expZnCd} 
R. F. Pottie, 
\emph{Cross sections for ionization by electrons. I. Absolute ionization cross sections 
of Zn, Cd, and Te$_2$. II. Comparison of theoretical with experimental values for 
atoms and molecules},
\emph{J. Chem. Phys.} {\bf 44}  (1966)  916.

\bibitem{expShul1989} 
R. J. Shul et al.,
\emph{Electron-impact-ionization cross section of the Ga and In atoms},
\emph{Phys. Rev. A} {\bf 39} (1989)  5588.

\bibitem{expGaInVainsh} 
L. A. Vainshtein et al.,
\emph{Cross sections for ionization of gallium and indium by electrons},
\emph{Sov. Phys. JETP} {\bf 66}  (1987) 36.
 
\bibitem{patton}
C. J. Patton et al.,
\emph{Multiple ionization of gallium by electron impact},
\emph{J. Phys. B: At. Mol. Opt. Phys.} {\bf 29} (1996) 1409.

\bibitem{expRbNygaard} 
K. J. Nygaard et al.,
\emph{Total electron impact ionization cross section in rubidium from threshold
to 250 eV},
\emph{J. Chem. Phys.} {\bf 58}  (1973)  3493.

\bibitem{schappe}
R. S. Schappe et al.,
\emph{Absolute electron-impact ionization cross section measurements 
using a magneto-optical trap},
\emph{Phys. Rev. Lett.} {\bf 76}  (1996)  4328.

\bibitem{crawford}
C. K. Crawford et al.,
\emph{Electron-impact ionization cross sections for silver},
\emph{J. Chem. Phys.} {\bf 47}  (1967) 4667.

\bibitem{franzreb}
K. Franzreb et al.,
\emph{Absolute cross sections for electron impact ionization of Ag$_2$},
\emph{Z. Phys. D} {\bf 19}  (1991) 77.

\bibitem{lin}
S. S. Lin et al.,
\emph{Electron-impact ionization cross sections. IV. group IVb atoms},
\emph{J. Chem. Phys.} {\bf 47}  (1967) 4664.

\bibitem{lyubimov}
A. P. Lyubimov et al.,
\emph{Procedure for measuring the ionization cross sections and ionization coefficients of metal atoms},
\emph{Bull. Acad. USSR. Phys. Ser.} {\bf 17}  (1963) 1033.

\bibitem{mathur}
D. Mathur  et al.,
\emph{Ionization of xenon by electrons: Partial cross sections for single, double, and triple ionization},
\emph{Phys. Rev. A} {\bf 35}   (1987) 1033.

\bibitem{expStephan1984} 
K. Stephan et al.,
\emph{Absolute partial electron impact ionization cross sections of Xe from threshold up to 180 eV},
\emph{J. Chem. Phys.} {\bf 81}  (1984)  3116.

\bibitem{heil}
H. Heil et al., 
\emph{Cesium ionization cross section from threshold to 50 eV},
\emph{Phys. Rev.} {\bf 145}   (1966) 279.

\bibitem{expCsNygaard} 
K. J. Nygaard, 
\emph{Electron-impact ionization cross section in cesium},
\emph{J. Chem. Phys.} {\bf 49}  (1968)  1995.

\bibitem{dettmann}
J. M. Dettmann  et al.,
\emph{Absolute ionization functions for electron impact with barium},
\emph{J. Phys. B: At. Mol. Phys.} {\bf 15} (1982) 287.

\bibitem{expYagi2000} 
S. Yagi et al.,
\emph{Absolute total and partial cross-sections for ionization of Ba and Eu atoms by electron impact},
\emph{J. Phys. Soc. Japan} {\bf 69}  (2000)  1374.

\bibitem{golovachBaPb}
D. G. Golovach et al.,
\emph{Apparatus for measurement of electronic-ionization cross sections of metal atoms},
\emph{Instr. Exp. Tech.} {\bf 29} (1987) 1396.

\bibitem{expYagi2001} 
S. Yagi et al.,
\emph{Absolute total and partial cross sections for ionization of free lanthanide atoms by 
electron impact},
\emph{J. Phys. Soc. Japan} {\bf 70}  (2001)  2559.

\bibitem{shimon}
L. L. Shimon et al.,
\emph{Multiple ionization of samarium, europium, thulium, and ytterbium atoms by electrons},
\emph{Sov. Phys. Tech. Phys.} {\bf 34}   (1989) 1264.

\bibitem{exp80}
W. Bleakney,
\emph{Probability and critical potentials for the formation of multiply charged ions in Hg 
vapor by electron impact},
\emph{Phys. Rev.} {\bf 35}  (1930)  139.

\bibitem{exp82M}
P. C. E. McCartney et al.,
\emph{Multiple ionization of lead by electron impact},
\emph{J. Phys. B: At. Mol. Opt. Phys.} {\bf 31}  (1998) 4821.

\bibitem{pavlov}
S. I. Pavlov et al.,
\emph{Single and multiple ionization of lead atoms by electrons},
\emph{Sov. Phys. JETP} {\bf 31}  (1970)  61.

\bibitem{beilina}
G. M. Beilina et al.,
\emph{Measurement of electron impact ionization functions for metal atoms},
\emph{J. Appl. Mechan. Tech. Phys.} {\bf 2}  (1965) 86.

\bibitem{wareing}
J. B. Wareing et al.,
\emph{A measurement of the cross section for ionization of Li$^+$ to Li$^{2+}$ by electron impact},
\emph{Proc. Phys. Soc.} {\bf 91}  (1967) 887.

\bibitem{exp92}
J. C. Halle et al.,
\emph{Ionization of uranium atoms by electron impact},
\emph{Phys. Rev. A} {\bf 23}   (1981) 1708.

\bibitem{pwba}
E. Merzbacher et al.,
\emph{Handbuch der Physik} {\bf 34} Springer, Berlin, 1958.

\bibitem{isics_linux}
M. Batic et al.,
\emph{ISICSoo: a class for the calculation of ionisation cross sections from PWBA and ECPSSR theory},
\emph{submitted to Comp. Phys. Comm.} (2011)


\bibitem{paul_sacher}
H.~Paul et al.,
\emph{Fitted empirical reference cross sections for K-shell ionization 
by protons}, 
\emph{At.\ Data Nucl.\ Data Tab.} {\bf 42} (1989) 105.

\bibitem{sokhi}
R. S. Sokhi et al.,
\emph{Experimental L-Shell X-Ray Production and Ionization Cross
Sections for Proton Impact}, 
\emph{At. Data Nucl. Data tables} {\bf 30}  (1984) 49.

\bibitem{orlic_exp}
I.~Orlic et al., 
\emph{Experimental L-shell X-ray
production and ionization cross sections for proton impact}, 
\emph{At. Data Nucl.  Data tables} {\bf 56}  (1994) 159.

\bibitem{namito}
Y. Namito et al.,
\emph{Implementation of the Doppler broadening of a Compton-scattered photon 
into the EGS4 code}, 
\emph{Nucl. Instrum. Meth. A} {\bf 349}  (1994) 489.

\bibitem{mcnp_doppler}
A. Sood,
\emph{Doppler Energy Broadening for Incoherent Scattering in 
MCNP5, Part I},
Los Alamos Report LA-UR-04-0487, 2004.

\bibitem{nss_doppler}
F. Longo et al.,
\emph{New Geant4 developments for Doppler broadening simulation in Compton
scattering - development of charge transfer simulation models in Geant4},
in \emph{IEEE Nucl. Sci. Symp. Conf. Rec.} 2008.
%
\bibitem{zoglauer}
A. Zoglauer et al., 
\emph{Doppler Broadening as a Lower Limit to the Angular Resolution of Next
Generation Compton Telescopes},
in \emph{Proc. SPIE 4581} (2003) 1302.

\bibitem{relax_prob}
M. G. Pia et al.,
\emph{Validation of K and L Shell Radiative Transition Probability Calculations},
\emph{IEEE Trans. Nucl. Sci.} {\bf 56} (2009) 3650.


%

%


%
%

%

%
%




%



%










%
%
%
%
%
%
%
%

%
%
%
%
%
%
%
%
%
%
%
%
%
%
%




\end{thebibliography}
\end{document}